\UseRawInputEncoding

\documentclass[twocolumn]{aastex62}

\usepackage{amsmath}

\newcommand\msun{M$_{\odot}$}

\graphicspath{{./}{figures/}}

\received{--}
\revised{--}
\accepted{--}

\submitjournal{ApJ}

\shorttitle{Magnetic fields in L1689}
\shortauthors{Pattle et al.}

\begin{document}

\title{JCMT POL-2 and BISTRO Survey observations of magnetic fields in the L1689 molecular cloud}

\correspondingauthor{Kate Pattle}

\author[0000-0002-8557-3582]{Kate Pattle}
\email{katherine.pattle@nuigalway.ie}
\affil{Centre for Astronomy, National University of Ireland Galway, University Road, Galway, Ireland}
\affil{Institute for Astronomy and Department of Physics, National Tsing Hua University, No. 101, Sec. 2, Guangfu Road, Hsinchu 30013, Taiwan}

\author[0000-0001-5522-486X]{Shih-Ping Lai}
\affil{Institute for Astronomy and Department of Physics, National Tsing Hua University, No. 101, Sec. 2, Guangfu Road, Hsinchu 30013, Taiwan}
\affil{Academia Sinica Institute of Astronomy and Astrophysics, No. 1, Sec. 4., Roosevelt Road, Taipei 10617, Taiwan}

\author[0000-0002-9289-2450]{James Di Francesco}
\affil{NRC Herzberg Astronomy and Astrophysics, 5071 West Saanich Road, Victoria, BC V9E 2E7, Canada}
\affil{Department of Physics and Astronomy, University of Victoria, Victoria, BC V8W 2Y2, Canada}

\author[0000-0001-7474-6874]{Sarah Sadavoy}
\affil{Department for Physics, Engineering Physics and Astrophysics, Queen's University, Kingston, ON, K7L 3N6, Canada}

\author[0000-0003-1140-2761]{Derek Ward-Thompson}
\affil{Jeremiah Horrocks Institute, University of Central Lancashire, Preston PR1 2HE, United Kingdom}

\author[0000-0002-6773-459X]{Doug Johnstone}
\affil{NRC Herzberg Astronomy and Astrophysics, 5071 West Saanich Road, Victoria, BC V9E 2E7, Canada}
\affil{Department of Physics and Astronomy, University of Victoria, Victoria, BC V8W 2Y2, Canada}

\author[0000-0003-2017-0982]{Thiem Hoang}
\affil{Korea Astronomy and Space Science Institute, 776 Daedeokdae-ro, Yuseong-gu, Daejeon 34055, Republic of Korea}
\affil{University of Science and Technology, Korea, 217 Gajeong-ro, Yuseong-gu, Daejeon 34113, Republic of Korea}

\author[0000-0002-1959-7201]{Doris Arzoumanian}
\affil{Instituto de Astrof\'isica e Ci{\^e}ncias do Espa\c{c}o, Universidade do Porto, CAUP, Rua das Estrelas, PT4150-762 Porto, Portugal}

\author[0000-0002-0794-3859]{Pierre Bastien}
\affil{Centre de recherche en astrophysique du Qu\'{e}bec \& d\'{e}partement de physique, Universit\'{e} de Montr\'{e}al, C.P. 6128 Succ. Centre-ville, Montr\'{e}al, QC, H3C 3J7, Canada}

\author[0000-0001-7491-0048]{Tyler L. Bourke}
\affil{SKA Organisation, Jodrell Bank, Lower Withington, Macclesfield, SK11 9FT, UK}
\affil{Jodrell Bank Centre for Astrophysics, School of Physics and Astronomy, University of Manchester, Manchester, M13 9PL, UK}

\author[0000-0002-0859-0805]{Simon Coud\'{e}}
\affil{SOFIA Science Center, Universities Space Research Association, NASA Ames Research Center, Moffett Field, California 94035, USA}

\author[0000-0001-8746-6548]{Yasuo Doi}
\affil{Department of Earth Science and Astronomy, Graduate School of Arts and Sciences, The University of Tokyo, 3-8-1 Komaba, Meguro, Tokyo 153-8902, Japan}

\author[0000-0003-4761-6139]{Chakali Eswaraiah}
\affil{CAS Key Laboratory of FAST, National Astronomical Observatories, Chinese Academy of Sciences, Peopleʼs Republic of China}
\affil{National Astronomical Observatories, Chinese Academy of Sciences, A20 Datun Road, Chaoyang District, Beijing 100012, People's Republic of China}

\author[0000-0001-9930-9240]{Lapo Fanciullo}
\affil{Academia Sinica Institute of Astronomy and Astrophysics, No. 1, Sec. 4., Roosevelt Road, Taipei 10617, Taiwan}

\author[0000-0003-0646-8782]{Ray S. Furuya}
\affil{Institute of Liberal Arts and Sciences Tokushima University, Minami Jousanajima-machi 1-1, Tokushima 770-8502, Japan}

\author[0000-0001-7866-2686]{Jihye Hwang}
\affil{Korea Astronomy and Space Science Institute, 776 Daedeokdae-ro, Yuseong-gu, Daejeon 34055, Republic of Korea}
\affil{University of Science and Technology, Korea, 217 Gajeong-ro, Yuseong-gu, Daejeon 34113, Republic of Korea}

\author[0000-0002-8975-7573]{Charles L. H. Hull}
\affil{National Astronomical Observatory of Japan, NAOJ Chile, Alonso de C\'ordova 3788, Office 61B, 7630422, Vitacura, Santiago, Chile }
\affil{Joint ALMA Observatory, Alonso de C\'ordova 3107, Vitacura, Santiago, Chile}
\affil{NAOJ Fellow}

\author[0000-0001-7379-6263]{Jihyun Kang}
\affil{Korea Astronomy and Space Science Institute, 776 Daedeokdae-ro, Yuseong-gu, Daejeon 34055, Republic of Korea}

\author[0000-0003-2412-7092]{Kee-Tae Kim}
\affil{Korea Astronomy and Space Science Institute, 776 Daedeokdae-ro, Yuseong-gu, Daejeon 34055, Republic of Korea}
\affil{University of Science and Technology, Korea, 217 Gajeong-ro, Yuseong-gu, Daejeon 34113, Republic of Korea}

\author[0000-0002-3036-0184]{Florian Kirchschlager}
\affil{Department of Physics and Astronomy, University College London, WC1E 6BT London, UK}

\author[0000-0003-2815-7774]{Jungmi Kwon}
\affil{Department of Astronomy, University of Tokyo, 7-3-1 Hongo, Bunkyo-ku, Tokyo 113-0033, Japan}

\author[0000-0003-4022-4132]{Woojin Kwon}
\affil{Department of Earth Science Education, Seoul National University, 1 Gwanak-ro, Gwanak-gu, Seoul 08826, Republic of Korea}
\affil{Korea Astronomy and Space Science Institute, 776 Daedeokdae-ro, Yuseong-gu, Daejeon 34055, Republic of Korea}

\author[0000-0002-3179-6334]{Chang Won Lee}
\affil{Korea Astronomy and Space Science Institute, 776 Daedeokdae-ro, Yuseong-gu, Daejeon 34055, Republic of Korea}
\affil{University of Science and Technology, Korea, 217 Gajeong-ro, Yuseong-gu, Daejeon 34113, Republic of Korea}

\author[0000-0002-5286-2564]{Tie Liu}
\affil{Key Laboratory for Research in Galaxies and Cosmology, Shanghai Astronomical Observatory, Chinese Academy of Sciences, 80 Nandan Road, Shanghai 200030, People’s Republic of China}

\author[0000-0002-1021-9343]{Matt Redman}
\affil{Centre for Astronomy, National University of Ireland Galway, University Road, Galway, Ireland}

\author[0000-0002-6386-2906]{Archana Soam}
\affil{SOFIA Science Center, Universities Space Research Association, NASA Ames Research Center, Moffett Field, California 94035, USA}

\author[0000-0001-8749-1436]{Mehrnoosh Tahani}
\affil{Dominion Radio Astrophysical Observatory, Herzberg Astronomy and Astrophysics Research Centre, National Research Council Canada, P. O. Box 248, Penticton, BC V2A 6J9 Canada}

\author[0000-0002-6510-0681]{Motohide Tamura}
\affil{Department of Astronomy, University of Tokyo, 7-3-1 Hongo, Bunkyo-ku, Tokyo 113-0033, Japan}
\affil{Astrobiology Center, 2-21-1 Osawa, Mitaka-shi, Tokyo 181-8588, Japan}
\affil{National Astronomical Observatory, 2-21-1 Osawa, Mitaka-shi, Tokyo 181-8588, Japan}

\author[0000-0002-4154-4309]{Xindi Tang}
\affil{Xinjiang Astronomical Observatory, Chinese Academy of Sciences, 830011 Urumqi, Peopleʼs Republic of China}

\begin{abstract}

We present 850$\mu$m polarization observations of the L1689 molecular cloud, part of the nearby Ophiuchus molecular cloud complex, taken with the POL-2 polarimeter on the James Clerk Maxwell Telescope (JCMT).  We observe three regions of L1689: the clump L1689N which houses the IRAS 16293-2433 protostellar system, the starless clump SMM-16, and the starless core L1689B.  We use the Davis-Chandrasekhar-Fermi method to estimate plane-of-sky field strengths of $366\pm 55$\,$\mu$G in L1689N, $284\pm 34$\,$\mu$G in SMM-16, and $72\pm 33$\,$\mu$G in L1689B, for our fiducial value of dust opacity.  These values indicate that all three regions are likely to be magnetically trans-critical with sub-Alfv\'{e}nic turbulence.
In all three regions, the inferred {mean} magnetic field direction is {approximately} perpendicular to the local filament direction identified in $Herschel$ Space Telescope observations.  The core-scale field morphologies for L1689N and L1689B are consistent with the cloud-scale field morphology measured by the $Planck$ Space Observatory, suggesting that material can flow freely from large to small scales for these sources.  Based on these magnetic field measurements, we posit that accretion from the cloud onto L1689N and L1689B may be magnetically regulated.  However, in SMM-16, the clump-scale field is nearly perpendicular to the field seen on cloud scales by $Planck$, suggesting that it may be unable to efficiently accrete further material from its surroundings.
\end{abstract}


\section{Introduction} \label{sec:intro}

The role of magnetic fields in the process of star formation, and in particular the dynamic importance of magnetic fields in the later stages of the star formation process -- the collapse of prestellar cores to form protostellar systems -- remains poorly constrained.  Magnetic fields are generally considered to act to resist the gravitational collapse of starless cores to form protostars \citep[e.g.][]{mouschovias1976}, although debate continues over whether magnetic fields mediate the star formation process, or are subdominant or negligible in comparison to turbulence in the interstellar medium (ISM) \citep[e.g.][]{hennebelle2019,krumholz2019}.  In most ISM environments, dust grains are expected to be preferentially aligned with their major axes perpendicular to the local magnetic field direction \citep{davis1951, lazarian2007, andersson2015}, and so dust emission polarimetry is a key tool for investigating magnetic fields in star-forming regions.

Magnetic fields in molecular clouds are preferentially aligned either parallel or perpendicular to dense elongated or filamentary structures \citep{soler2013, planck2016a}. These filaments are ubiquitous in star-forming regions, and may be the main site for the formation of Sun-like stars \citep{konyves2010}.  This has led to the suggestion that material flows onto filaments along magnetic field lines, until they have accreted sufficient mass to collapse under gravity to form a series of prestellar cores \citep{palmeirim2013,soler2013, andre2014}.

The role of magnetic fields in the evolution of prestellar cores -- gravitationally bound overdensities which will go on to form a single protostellar system \citep{wardthompson1994} -- is not yet well-characterized.  A dynamically important magnetic field is broadly expected to support a prestellar core against, and to impose a preferred direction on, gravitational collapse \citep{mouschovias1976}.  Magnetically-dominated systems might be expected to show the classical `hourglass' magnetic field indicative of ambipolar-diffusion-mediated gravitational collapse \citep[e.g.][]{fiedler1993}.  Observations of the polarization geometry of prestellar cores typically show linear magnetic field geometries, oriented within $\sim 30^{\circ}$ of the minor axis of the core \citep{wardthompson2000,kirk2006,liu2019,coude2019}. Hourglass fields have not been definitively observed in prestellar cores; however, such fields have been observed by interferometers in some protostellar sources, most famously in NGC1333 IRAS 4A \citep{girart2006}. 

Recent interferometric observations of protostellar sources have shown that the majority have outflow direction uncorrelated with their overall magnetic field direction, while a minority may be strongly correlated \citep{hull2013}.  This has led to the suggestion that while magnetic fields are dynamically subdominant or even negligible in the majority of protostellar systems, there are a minority of magnetically dominated systems \citep{hull2019}.  

In this work, we define clumps as subregions of a molecular cloud that contain sufficient mass to form multiple and distinct stellar systems.  These stellar systems form from smaller substructures within the clumps, which we refer to as cores.  Those substructures which contain no protostellar sources are referred to as starless cores, while those with embedded protostellar sources are referred to as protostellar cores.   The gravitationally bound subset of starless cores, we term prestellar cores.  We note that cores need not be embedded in clumps; they can also exist within filamentary structures of the larger molecular cloud, or in isolation.

In this work we present observations of three regions in the Ophiuchus L1689 molecular cloud with the POL-2 polarimeter on the James Clerk Maxwell Telescope (JCMT): the L1689N star-forming clump, the SMM-16 starless clump, and the L1689B starless core, all of which are embedded in a larger filamentary network.  This paper is structured as follows: in Section~\ref{sec:intro_oph} we briefly review the L1689 cloud and discuss its distance.  In Section~\ref{sec:obs} we present our POL-2 observations of L1689.  In Section~\ref{sec:mag_fields} we estimate the magnetic field properties in L1689.  In Section~\ref{sec:grains} we discuss grain alignment.  In Section~\ref{sec:discussion} we interpret our results.  Section~\ref{sec:summary} summarizes this work.

\section{The L1689 molecular cloud}
\label{sec:intro_oph}

\begin{figure*}
    \centering
    \includegraphics[width=0.75\textwidth]{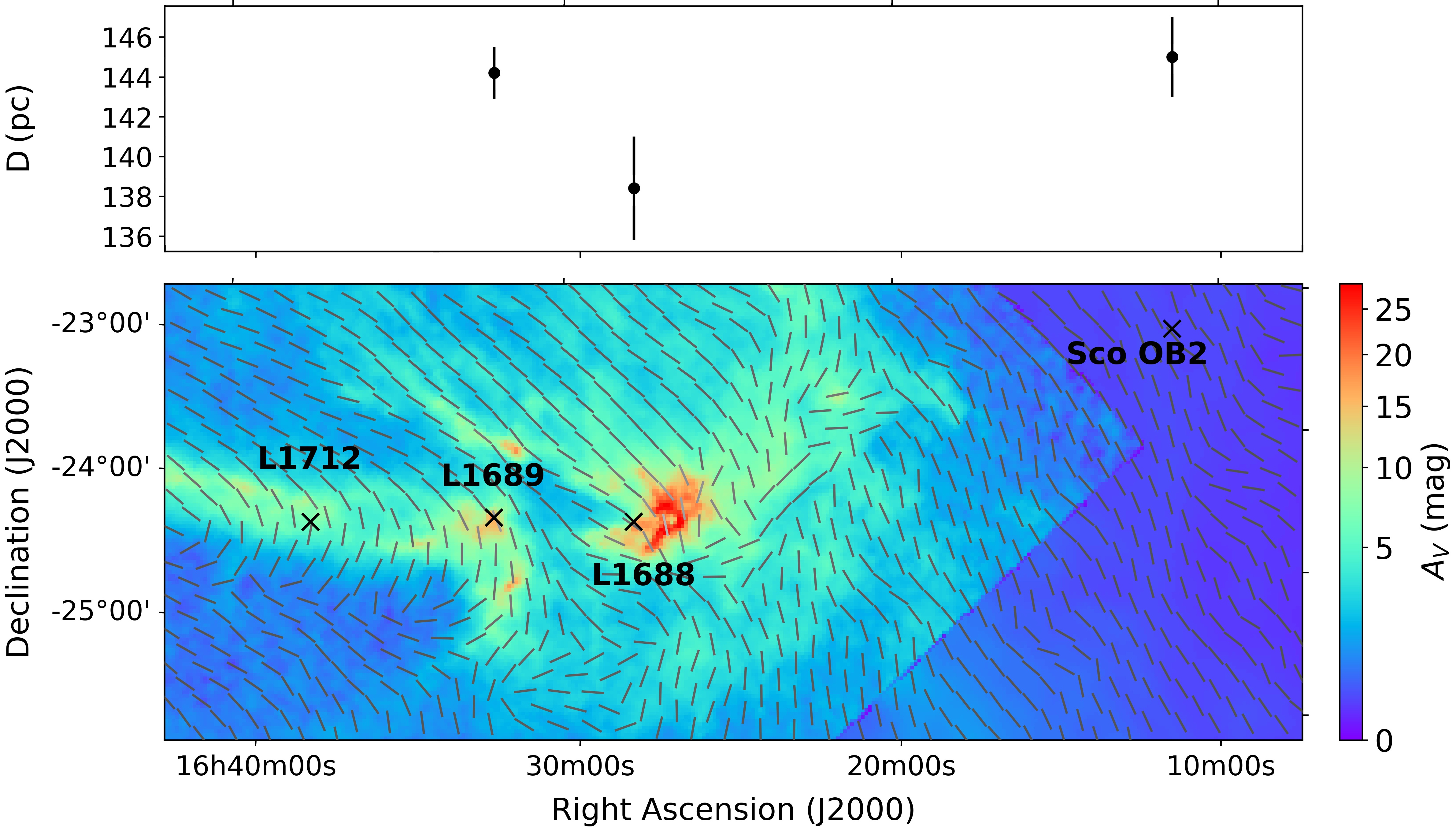}
    \caption{Main (lower) panel: The Ophiuchus molecular cloud, observed in magnitudes of visual extinction ($A_{V}$).  High-resolution $A_{V}$ mapping of L1688, L1689 and part of their filamentary streamers is taken from 2MASS imaging made as part of the COMPLETE Survey \citep{ridge2006}; extinction values in areas not covered by COMPLETE are taken from the NASA/IPAC Infrared Science Archive \citep{schlegel1998, schlafly2011}.  {The L1689/L1712 filamentary streamer runs approximately east-west across the northern part of L1689, turning to run approximately north-east/south-west, towards and beyond the L1712 region, on the eastern side of L1689 \citep{loren1989a}.}  Grey half-vectors show magnetic field direction inferred from $Planck$ 353\,GHz polarization angle measurements, with 10\arcmin\ spacing, drawn at a constant length and rotated by 90$^{\circ}$ to trace the large-scale magnetic field direction.  Crosses mark the positions used to calculate distances between the clouds and Sco OB2, as described in the text.  Upper panel: line-of-sight distances to L1688, L1689 and Sco OB2 \citep{dezeeuw1999, ortizleon2018}.}
    \label{fig:finder}
\end{figure*}

The Ophiuchus molecular cloud is a nearby, well-studied region of low-to-intermediate-mass star formation \citep{wilking2008}, located at a distance $\sim 140$\,pc from the Solar System \citep{ortizleon2018}.  The region is made up of two central dark clouds, L1688 and L1689 \citep{lynds1962}, both of which have extensive filamentary streamers to their north-east, but are sharp-edged on their south-western side \citep{vrba1977}.  The strong asymmetry of the region is thought to be due to the influence of the nearby Sco OB2 association \citep{vrba1977,loren1989a}, located to the west of and behind Ophiuchus \citep{mamajek2008}.  The two main clouds and their filamentary streamers are shown in Figure~\ref{fig:finder}, along with the central position of Sco OB2.  Ophiuchus is threaded by a large-scale magnetic field, preferentially oriented $\sim 50^{\circ}$ east of north, as shown in Figure~\ref{fig:finder} \citep{vrba1976,planck2015, kwon2015}.

L1688 contains a number of dense clumps, including clumps Oph A, B and C, which were recently observed as part of the JCMT BISTRO (B-Fields in Star-Forming Regions Observations) Survey \citep{wardthompson2017,kwon2018,soam2018,liu2019}.  The Oph A region has also recently been observed in polarized far-infrared emission \citep{santos2019}.  These observations have shown that the dense clumps within L1688 have a mean magnetic field direction consistent with the large-scale NE/SW field, but with significant deviations, both ordered and disordered, from the mean field direction.

L1688 contains more than 20 embedded protostars and two B stars, whereas L1689 contains only five embedded protostellar systems \citep{enoch2009}. \citet{nutter2006} proposed that there is a star formation gradient across Ophiuchus driven by the global influence from the Sco OB2 association \citep{loren1989a}, under the assumption that L1689 is located further from Sco OB2 than L1688.  However, the evolution of dense gas within the two clouds also appears to be strongly influenced by local effects, without clear evidence for a west-to-east star formation gradient within L1688 \citep{pattle2015}.

Although L1688 (R.A., Dec$=16^{h}28^{m},-24^{\circ}32^{\prime}$; $l,b=353.22^{\circ},16.53^{\circ}$) and L1689 (R.A., Dec.=$16^{h}32^{m}$, $-24^{\circ}28^{\prime}$; $l,b=353.94^{\circ},15.84^{\circ}$) have previously been treated as being at the same distance, revised distance estimates, combining Gaia DR2 and VLBA parallaxes, place L1689 5.8\,pc behind L1688, with the two clouds located at distances of $144.2\pm 1.3$\,pc and $138.4\pm 2.6$\,pc respectively \citep{ortizleon2018}.  The Sco OB2 association is located behind the Ophiuchus molecular cloud, at R.A.,\,Dec.$=16^{h}11^{m},-23^{\circ}18^{\prime}$ ($l,b=351.5^{\circ},20.2^{\circ}$) and at a distance of $145\pm 2$\,pc \citep{dezeeuw1999}.  The plane-of-sky and line-of-sight locations of the clouds and of Sco OB2 are shown in Figure~\ref{fig:finder}.  Coordinates for each object are taken from the Simbad database \citep{simbad}.  We note that these revised distance estimates suggest that L1688 and L1689 are located at similar distances to Sco OB2, with the L1688-to-Sco OB2 distance being $11.5\pm 3.3$\,pc, the L1689-to-Sco OB2 distance being $12.2\pm 2.4$\,pc, and L1688 and L1689 being located $6.3\pm 2.9$\,pc from one another.  If this is the case, the differences in star formation history between L1688 and L1689 may not be attributable to their relative proximity to Sco OB2, and alternative explanations for the relatively lackluster star formation in L1689 must be sought.

\begin{figure*}
    \centering
    \includegraphics[width=0.75\textwidth]{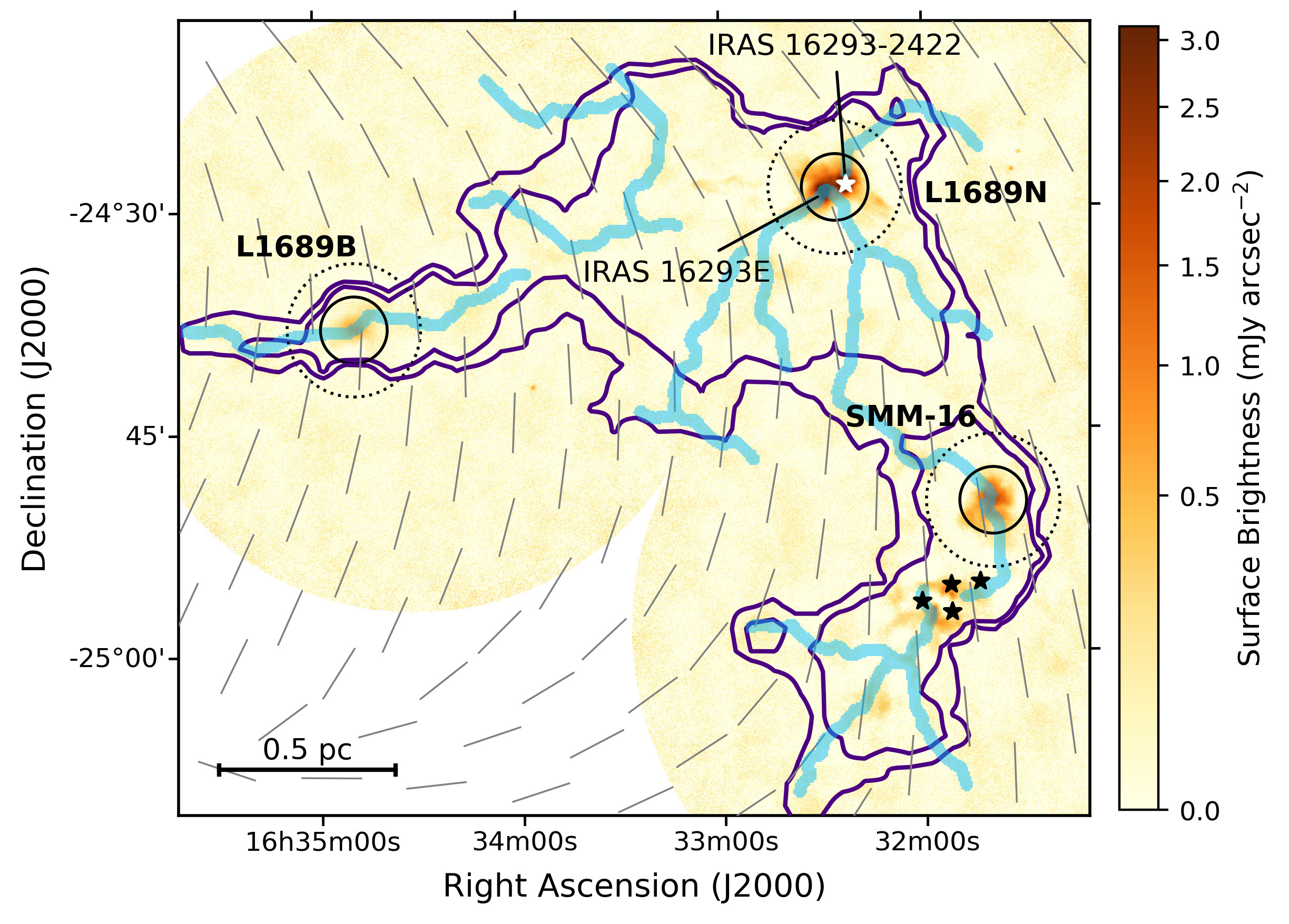}
    \caption{The L1689 molecular cloud observed in 850$\mu$m emission with SCUBA-2 \citep{pattle2015} {(white areas are beyond the extent of the SCUBA-2 map)}.  Black circles mark the extent of our POL-2 observations: solid circles show the central 6\arcmin\ region of useful coverage; dotted circles show the full extent of the observation.  Grey half-vectors show $Planck$ 353\,GHz polarization angles, drawn at a constant length and rotated by 90$^{\circ}$ to trace the large-scale magnetic field direction.  Purple {lines mark the $A_{V}=7$ contour \citep{ridge2006}, approximately delineating the extent of the L1689 cloud, and the $A_{V}=8$ contour enclosing both L1689N and L1689B, as part of the L1689/L1712 filamentary streamer \citep[cf.][]{loren1989a}}.  Blue lines mark the filamentary network identified in \emph{Herschel} observations by \citet{ladjelate2020}.  Stars mark embedded protostars in the region, as identified by \citet{enoch2009}.  The L1689N region, the IRAS 16293-2422 protostellar core, the IRAS 16293E (SMM 19) core, the SMM 16 clump, and the L1689B core are also labelled.  Physical scale is shown in the lower left-hand corner, for an assumed distance of 144.2\,pc.}
    \label{fig:finding_chart}
\end{figure*}

The L1689 cloud, shown in Figure~\ref{fig:finding_chart}, contains two significant clumps.  The northern clump, hereafter referred to as L1689N, contains the well-studied protostellar system IRAS 16293-2422, a multiple system of Class 0 protostars \citep{wootten1989,mundy1992}, with a quadrupolar set of outflows \citep{walker1988,mizuno1990}.  The system contains two main sources, IRAS 16293A (south) and 16293B (north), separated by $\sim 5$\arcsec\ \citep{chandler2005} but joined by a bridge of emission of length $\sim 700$\,AU \citep{pineda2012}.
The region also contains the starless core IRAS 16293E (SMM-19 in the nomenclature of \citealt{nutter2006}), which is a candidate for gravitational collapse \citep{sadavoy2010a}.  See \citet{jorgensen2016} for a detailed review of the IRAS 16293 system. 

The southern part of L1689, known as L1689S, contains several structures, including SMM-16 \citep{nutter2006}, a strong candidate for being a gravitationally bound prestellar clump or core.  \citet{chitsazzadeh2014} identified SMM-16 as a starless core based on its high degree of deuterium fractionation and lack of associated infrared emission, and found it to be virially bound.  However, they found no evidence for infall, instead finding SMM-16 to be oscillating.  However, \citet{pattle2015} identified three fragments, SMM-16a, b, and c, within SMM-16, suggesting that the region is a starless clump rather than a single core.  Similarly, \citet{ladjelate2020} identify three dense fragments in the center of SMM-16, and three further fragments in the periphery of the clump.

L1689 also contains the L1689B prestellar core candidate, embedded in a filamentary structure to the east of the main body of the cloud \citep{jessop2000,kirk2007,steinacker2016}.  The core, which is generally considered to be undergoing large-scale infall \citep[e.g.][]{lee2001}, has been extensively studied in terms of its chemistry \citep[e.g.][]{redman2002,crapsi2005,bacmann2016,kim2020}, internal structure and dynamics \citep[e.g.][]{lee1999a,lee2001,redman2004,seo2013,roy2014} due to its relative isolation and simple morphology.

\section{Observations} \label{sec:obs}

\begin{figure}
    \centering
    \includegraphics[width=0.47\textwidth]{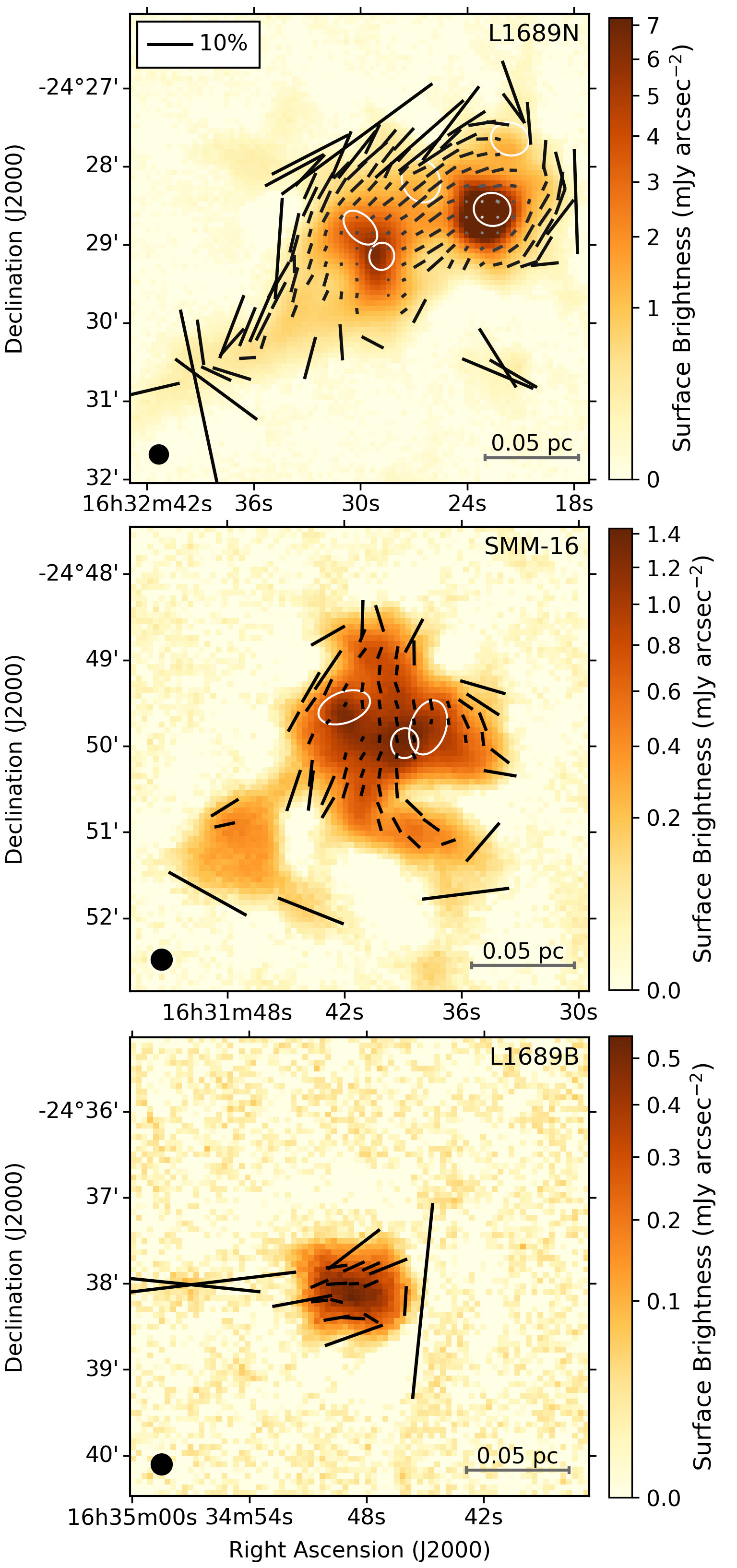}
    \caption{Debiased polarization half-vectors in L1689N (top), SMM-16 (middle) and L1689B (bottom), overlaid on POL-2 Stokes $I$ data.  POL-2 data are shown as black half-vectors (except in L1689N, where half-vectors are colored by total intensity for contrast), with selection criteria $p_{{\rm db}}/\delta p>3$, $I/\delta I > 5$ and $\delta \theta < 10^{\circ}$.  Sources identified by \citet{pattle2015} are marked as white circles: {in L1689N, their sources SMM-19 and -22 overlap in the IRAS 16293E core, their SMM-20 corresponds to IRAS 16293-2422, and their SMM-23 and -24 are in the clump periphery.  Their sources SMM-16a, -16b and -16c are shown in SMM-16}.  The JCMT beam size is shown in the lower left-hand corner, and a physical scale bar is shown in the lower right-hand corner, of each plot.}
    \label{fig:presentation_p}
\end{figure}
.

\begin{figure}
    \centering
    \includegraphics[width=0.47\textwidth]{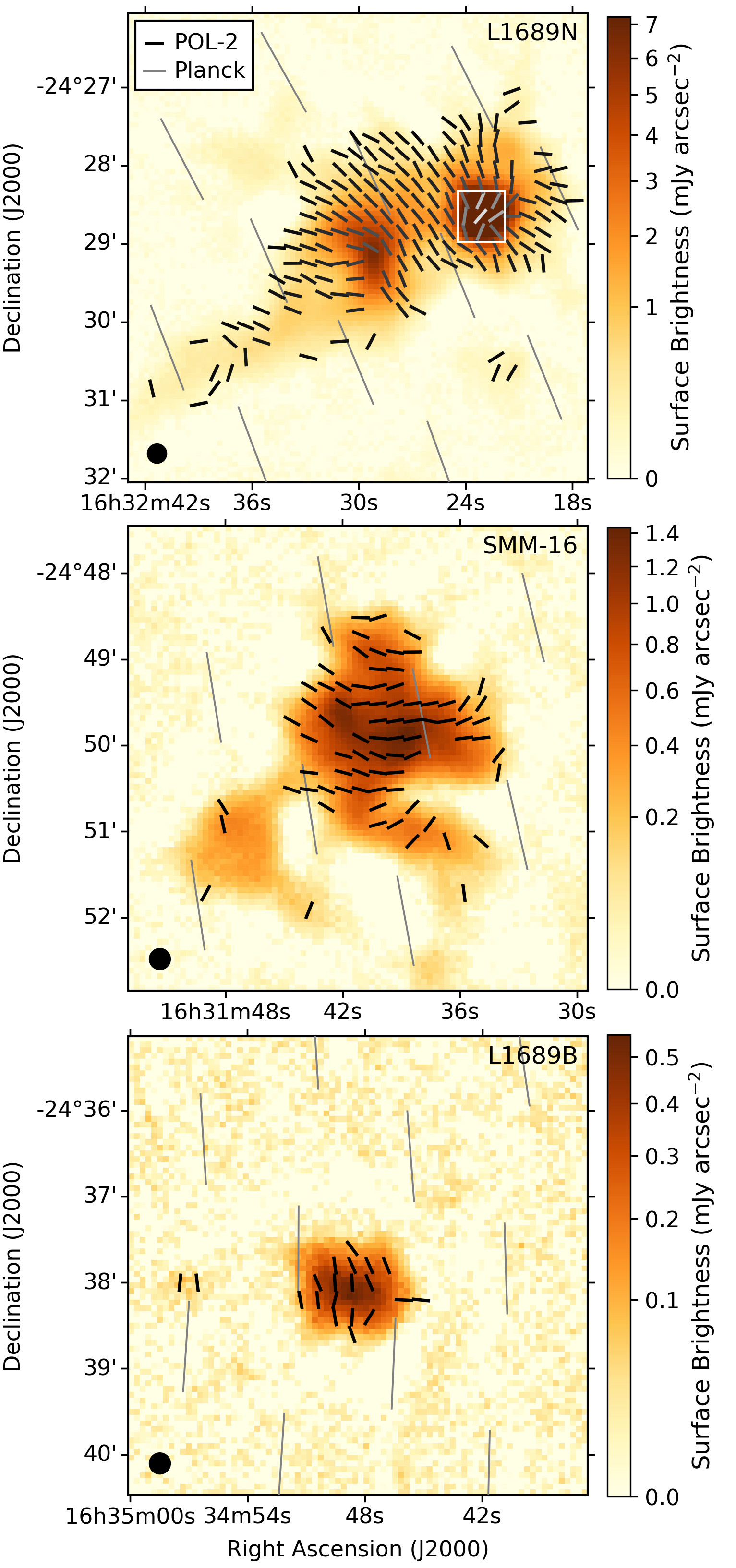}
    \caption{Magnetic field half-vectors (polarization half-vectors rotated by 90$^{\circ}$) in L1689N (top), SMM-16 (middle) and L1689B (bottom), overlaid on POL-2 Stokes $I$ data.  POL-2 data are shown as uniform-length black half-vectors (except L1689N, where half-vectors are colored by total intensity for contrast) with selection criteria as in Figure~\ref{fig:presentation_p}.  Planck data are shown as grey half-vectors.  The JCMT beam size is shown in the lower left-hand corner of each plot.  (Note that the $Planck$ half-vectors shown in this image are oversampled.)  {The half-vectors in L1689N associated with IRAS 16293-2422 are enclosed in a white box.}}
    \label{fig:presentation}
\end{figure}

We observed L1689 using the POL-2 polarimeter on the SCUBA-2 camera on the JCMT.  L1689N and SMM-16 were observed under project code M19AP038, and L1689B was observed as part of the JCMT BISTRO Survey \citep{wardthompson2017}, under project code M16AL004.  Each field was observed 20 times in Band 2 weather ($0.05<\tau_{225\,{\rm GHz}}<0.08$), giving a total integration time of 14 hours per field.  A single POL-2 observation consists of 40 minutes of observing time using the POL-2-DAISY scan pattern \citep{friberg2016},  which produces a 12\arcmin-diameter output map, of which the central 3\arcmin\ has approximately uniform noise, and the central 6\arcmin\ has a useful level of coverage.  The data were reduced in a two-stage process using the $pol2map$ routine\footnote{http://starlink.eao.hawaii.edu/docs/sun258.htx/sun258ss73.html} recently added to \textsc{Smurf}\footnote{http://starlink.eao.hawaii.edu/docs/sun258.htx/sun258.html}. 

In the first stage, the raw bolometer timestreams for each observation were converted into separate Stokes $Q$, $U$, and $I$ timestreams.  An initial Stokes $I$ map was then created from the $I$ timestream from each observation using the iterative map-making routine $makemap$ \citep{chapin2013}. For each reduction, areas of astrophysical emission were defined using a signal-to-noise-based mask determined iteratively by $makemap$.  Areas outside this masked region were set to zero until the final iteration of $makemap$ (see \citealt{mairs2015} for a detailed discussion of the role of masking in SCUBA-2 data reduction).  Each map was compared to the first map in the sequence to determine a set of relative pointing corrections. The individual $I$ maps were then coadded to produce an initial $I$ map of the region.

In the second stage, an improved Stokes $I$ map was created from the $I$ timestreams of each observation using $makemap$, and Stokes $Q$ and $U$ maps were created from their respective timestreams. The initial $I$ map (described above) was used to generate a fixed signal-to-noise-based mask for all iterations of $makemap$. In the second stage, the $skyloop$\footnote{http://starlink.eao.hawaii.edu/docs/sun258.htx/sun258ss72.html} routine is used, in which one iteration of $makemap$ is performed on each of the observations in the set in turn, with the set being averaged together at the end of each iteration, rather than each observation being reduced consecutively.
The pointing corrections determined in the first stage were applied to the Stokes $Q$, $U$ and $I$ maps during the map-making process.  Correction for instrumental polarization in the Stokes $Q$ and $U$ maps was performed based on the final output $I$ map, using the `January 2018' instrumental polarization (IP) model \citep{friberg2018}.  Variances in the final coadded maps were calculated according to the standard deviation of the measured values in each pixel across the 20 observations, and in the final coadded maps, each observation was weighted according to the mean of its associated variance values.  The output $Q$, $U$, and $I$ maps were gridded to 4$^{\prime\prime}$ pixels and calibrated in mJy\,beam$^{-1}$ using a flux conversion factor (FCF) of 725 Jy\,pW$^{-1}$ (the standard SCUBA-2 850$\mu$m FCF of 537 Jy\,pW$^{-1}$ multiplied by a factor of 1.35 to account for additional losses from POL-2; \citealt{dempsey2013,friberg2016}).

Half-vector catalogues were created from the final $I$, $Q$ and $U$ maps, on a 12\arcsec\ pixel grid (the primary beam size of the JCMT at 850\,$\mu$m; \citealt{dempsey2013}).  The term `half-vector' refers to the $\pm180^{\circ}$ ambiguity in
magnetic field direction.  These catalogues list the derived polarization properties of each pixel: polarized intensity ($P$), debiased polarized intensity ($P_{{\rm db}}$), polarization fraction ($p$), debiased polarization fraction ($p_{{\rm db}}$), and polarization angle, ($\theta_{P}$), and uncertainties on each quantity.  The formulae for these derived quantities are given in Appendix~\ref{sec:appendix_pol}.  The average RMS noise in Stokes $Q$ and $U$ on 12\arcsec\ pixels over the central 3\arcmin\ of the map is 0.83\,mJy\,beam$^{-1}$ in L1689N, 0.81 mJy\,beam$^{-1}$ in SMM-16, and 0.75\,mJy\,beam$^{-1}$ in L1689B.  {The maps and half-vector catalogues used in this work are available at \url{http://dx.doi.org/10.11570/20.0013}.}

\begin{figure}
    \centering
    \includegraphics[width=0.47\textwidth]{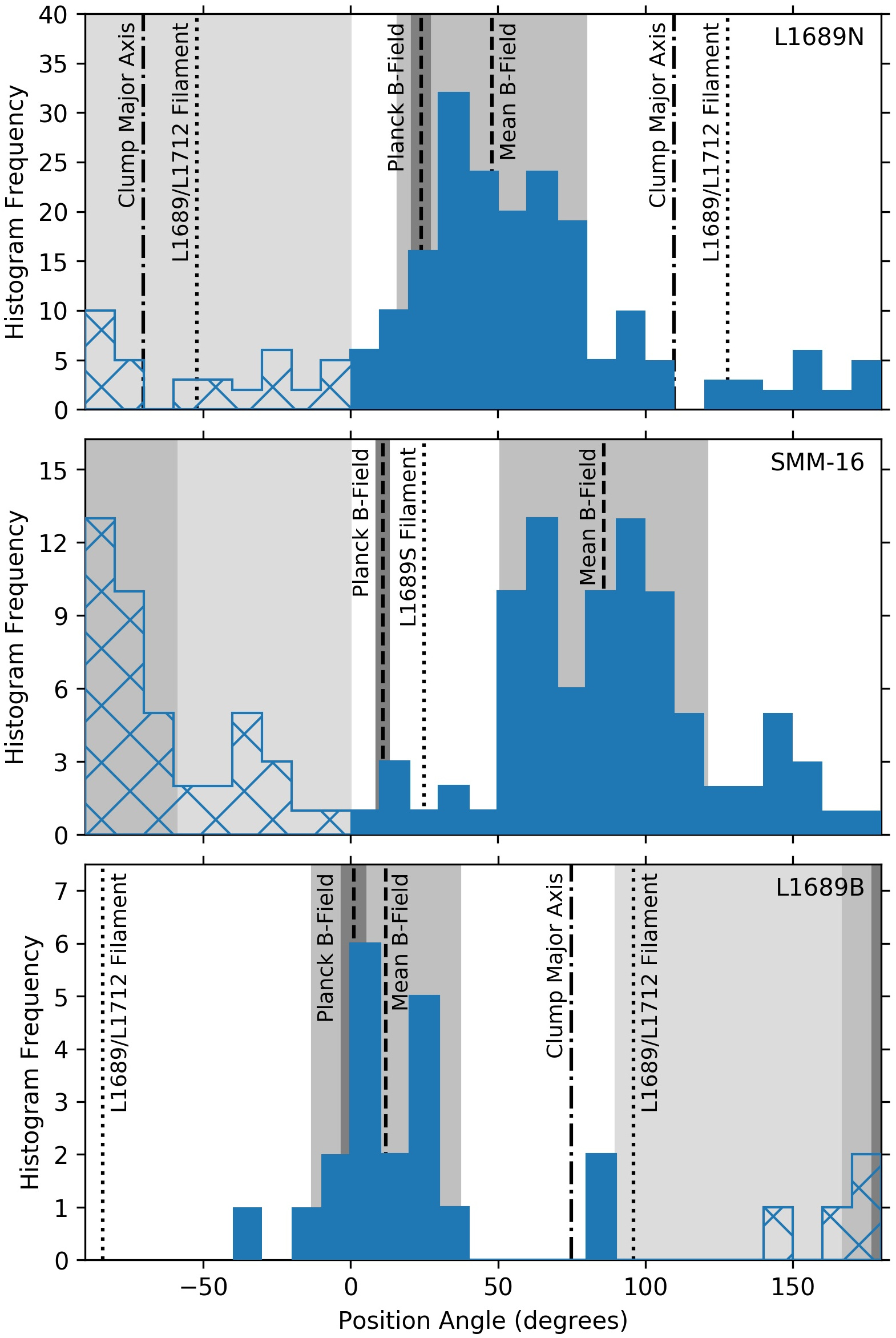}
    \caption{Distribution of {magnetic field} position angles in L1689N (top), SMM-16 (middle) and L1689B (bottom).   Position angles are {given} in degrees E of N, {in the range $-90 \leq \theta < +180^{\circ}$.  The ranges $-90^{\circ}-0^{\circ}$ and $+90^{\circ}-+180^{\circ}$ are identical.  For L1689N and SMM-16, we emphasize the range $0-+180^{\circ}$, while for L1689B we emphasize the range $-90 - +90^{\circ}$, in each case chosen to best illustrate the typical magnetic field direction and the distribution of the magnetic field angles.  The duplicated range is in each panel shaded in light grey, with position angles plotted in hatched histogram bins.  In each panel, black dashed lines mark the mean POL-2 and $Planck$ position angles; shaded grey areas show their standard deviations.  Black dot-dashed lines mark the fitted major axis position angle of the clump/core as listed in Table~\ref{tab:gaussians}, with the exception of SMM-16, which does not have a well-defined major axis.  Dotted lines mark} the approximate major axis position angle of the filament in which the clump/core is embedded, {as determined from $Herschel$ measurements \citep{ladjelate2020}.}}
    \label{fig:angles}
\end{figure}

In the case of optically thin polarized emission from dust grains aligned with their minor axes parallel to the magnetic field direction, polarization angles must be rotated by 90$^{\circ}$ in order to trace the plane-of-sky magnetic field direction.  In the following discussion we denote magnetic field angle as $\theta$.  POL-2 polarization half-vector maps for {L1689N, SMM-16 and L1689B} are shown in Figure~\ref{fig:presentation_p}, while magnetic field half-vector maps are shown in Figure~\ref{fig:presentation}.  Histograms of magnetic field angle are shown in Figure~\ref{fig:angles} {(position angles are given in the range $-90 \leq \theta < +180^{\circ}$, where the ranges $-90^{\circ}-0^{\circ}$ and $+90^{\circ}-+180^{\circ}$ are identical)}.  In these figures and in all subsequent analysis {(except where noted in Section~\ref{sec:grains})}, we select half-vectors where $I/\delta I > 5$, $p_{{\rm db}}/\delta p > 3$ and $\delta \theta < 10^{\circ}$, noting that Serkowski's approximation ($\delta\theta \approx 28.65^{\circ}\times(\delta p/p)$; \citealt{serkowski1962}) makes the latter two criteria approximately equivalent to one another.

\section{Magnetic field properties in L1689} \label{sec:mag_fields}

\begin{figure}
    \centering
    \includegraphics[width=0.47\textwidth]{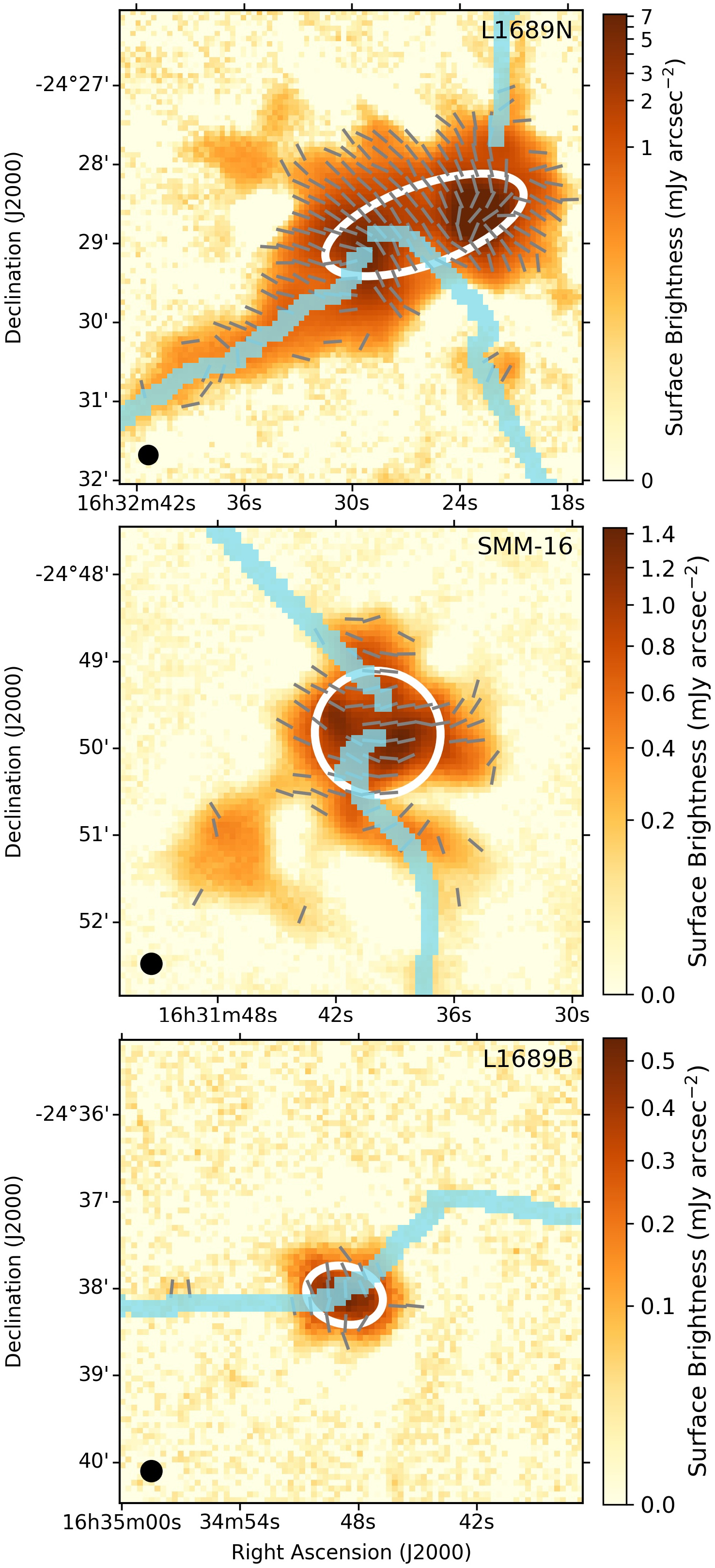}
    \caption{{POL-2 magnetic field half-vectors in L1689N (top), SMM-16 (middle) and L1689B (bottom), overlaid on POL-2 Stokes $I$ data, as in Figure~\ref{fig:presentation_p}.  The FWHM best-fit ellipses to the Stokes $I$ emission are shown as white ellipses.  The filaments identified in $Herschel$ observations by \citet{ladjelate2020} are shown as blue lines.  The L1689N Stokes $I$ data are plotted using a logarithmic color scale, to emphasize the faint larger-scale structure.}}
    \label{fig:presentation_ef}
\end{figure}

\subsection{Magnetic field morphology}

\subsubsection{L1689N}

L1689N has a well-ordered magnetic field which is aligned {broadly NE/SW, with a mean direction $48^{\circ}\pm32^{\circ}$ E of N (calculated over the circular range $0 - 180^{\circ}$)}.  The exception to this is on the position of the IRAS 16293-2422 protostar itself, where the inferred magnetic field direction {is $166^{\circ}\pm 31^{\circ}$ E of N, measured over the 9 independent pixels covering a $36^{\prime\prime}\times36^{\prime\prime}$ area centered on the protostar.
Excluding IRAS 16293-2422, the mean field direction is $50^{\circ}\pm90^{\circ}$ E of N, so that the polarisation half-vectors of IRAS 16293-2422 are $64^{\circ}\pm43^{\circ}$ offset from the mean elsewhere in L1689N.} The half-vectors associated with IRAS 16293-2422 are {marked on Figure \ref{fig:presentation}.}

Away from IRAS 16293-2422, the magnetic field inferred from POL-2 observations is ordered and generally linear, with a mean direction {$50^{\circ}\pm 30^{\circ}$ E of N.  We see some deviations from the mean field direction in the southeastern and northwestern edges of the clump, particularly in the vicinity of IRAS 16293E.}

{The mean field direction in L1689N is similar to} the field observed on arcminute scales by $Planck$ ($\langle\theta_{Planck}\rangle = 24^{\circ}\pm3^{\circ}$), and approximately perpendicular to {($\sim 78^{\circ}$ offset from)} the major axis of the L1689/L1712 filament of which the clump is a part, which runs $\sim 128^{\circ}$ E of N (cf. \citealt{ladjelate2020}, see also Figures~\ref{fig:finding_chart}, \ref{fig:angles}).

{Although the L1689N clump is at the junction of two filaments, one running approximately SE/NW and the other approximately NE/SW, as shown in Figures~\ref{fig:finding_chart} and \ref{fig:presentation_ef}, we consider L1689N to be principally associated with the SE/NW filament, which extends over several degrees as part of the L1689/L1712 filamentary streamer, as shown in Figure~\ref{fig:finder}.  L1689N and L1689E are embedded in the same larger structure, with both enclosed in the same $A_{V}=8$ contour, as shown in Figure~\ref{fig:finding_chart}, and this structure extends west beyond the extent of L1689N which we observe.  This filament is also detectable in our Stokes $I$ observations, unlike the lower-column-density NE/SW filament, as shown in Figure~\ref{fig:presentation_ef}.  Moreover, the major axis of L1689N is well-aligned with the major axis of the SE/NW filament, as shown in Figures~\ref{fig:angles} and \ref{fig:presentation_ef} (clump orientation is determined as described in Section~\ref{sec:dcf_main}, below).  However, an alternative interpretation is that L1689N is located at the point that the main L1689/L1712 filamentary streamer turns to become the short extension to L1689S.  In this case, the major axis of the filament at the position of L1689N could be defined as the average of the SE/NW and NE/SW filament directions.  The NE/SW filament is oriented $\sim 27^{\circ}$ E of N, and so the average orientation of the filaments in the junction of which L1689N is embedded in this interpretation is $\sim 78^{\circ}$ E of N.  This is again significantly offset from the plane-of-sky magnetic field direction which we infer.}

On the position of IRAS 16293-2422, the magnetic field inferred from POL-2 observations flips by {$\sim 64^{\circ}$} relative to the large-scale field, appearing to run {approximately} NW/SE.  ALMA polarization observations of protostellar discs have shown abrupt changes in polarization angle \citep[e.g.][]{ko2020}.  Mechanisms proposed to explain this change include scattering from large dust grains \citep{kataoka2015, sadavoy2019}, intrinsic polarization due to thermal emission from non-spherical dust grains \citep{kirchschlager2019,guillet2020}, and the effects of optically thick emission \citep{ko2020}. However, in IRAS 16293-2422 the JCMT beam encompasses not only two differently-inclined protostellar discs but also the protobinary envelope and its internal dust structures \citep{pineda2012}.  The magnetic field direction seen in our observations corresponds well with the magnetic field direction seen by ALMA in the bridge of dense gas which connects the IRAS 16293A and B discs \citep{sadavoy2018}, as discussed in Section~\ref{sec:l1689_discuss} below.  This good agreement with ALMA observations suggests that our observations of IRAS 16293-2422 traces dust grains that are both optically thin and aligned with respect to the magnetic field, and so we consider our observations to be tracing the magnetic field throughout L1689N.

\subsubsection{SMM-16}

SMM-16 has a fairly well-ordered magnetic field which is aligned approximately E/W in the densest parts of the clump, but which shows significant deviations in its outer regions {(mean direction: $86^{\circ}\pm 35^{\circ}$)}.  The inferred field direction tends towards a NE/SW direction in the eastern side of the clump, and to a SE/NW direction in the western side of the clump.  The average magnetic field direction is approximately perpendicular both to the $Planck$-scale magnetic field direction, which runs north-south across the region ($\langle\theta_{Planck}\rangle = 11^{\circ}\pm2^{\circ}$), and to the orientation of the filament in which the clump is embedded, which runs $\sim 25^{\circ}$ E of N \citep[cf.][see also Figures~\ref{fig:finding_chart}, \ref{fig:angles}]{ladjelate2020}.  {Similar discrepancies between POL-2 and $Planck$-scale fields have recently been seen in BISTRO survey observations of Perseus NGC 1333 \citep{doi2020}.}  The distribution of POL-2 magnetic field half-vectors which we observe is thus suggestive of a field which is perpendicular to the large-scale field in the dense center of the clump, {with significant excursions in the lower-density periphery}.  {This perpendicular component is represented by the main peak in Figure~\ref{fig:angles}, and is more clearly apparent in Figures~\ref{fig:ellipses} and \ref{fig:histograms} in Appendix~\ref{sec:appendix_dcf}.}  We further tentatively suggest that in the {lower}-density periphery of the eastern and western sides of the clump, the field may curve to join the large-scale field in the lower-density surrounding region, {as is suggested by the peaks at 65$^{\circ}$ and 150$^{\circ}$ E of N in Figure~\ref{fig:angles}, representing ordered field structure on the eastern and western sides of the core respectively.}.

\subsubsection{L1689B}

L1689B is considerably smaller and fainter than the other two clumps, and consequently fewer reliable polarization half-vectors are detected.  Nonetheless, we observe a consistent magnetic field running {broadly N/S} across the core, with the exception of two half-vectors on the western edge of the core.  {The mean field direction is $12^{\circ}\pm25^{\circ}$.}  The Planck-scale field in the region also runs N/S ($\langle\theta_{Planck}\rangle = 1^{\circ}\pm4^{\circ}$), similar to the mean field direction which we observe.  The field which we observe with POL-2 is again approximately perpendicular to the filament in which the core is embedded, which runs $\sim 96^{\circ}$ E of N, approximately E/W, \citep[cf.][see also Figures~\ref{fig:finding_chart}, \ref{fig:angles}]{kirk2007, ladjelate2020}.

\subsubsection{Comparison between regions}

The discrepant plane-of-sky POL-2 and $Planck$-scale fields in SMM-16 contrast notably with L1689N and L1689B, in which the POL-2- and $Planck$-scale fields agree well in projection on the plane of the sky. However, in all three clumps the POL-2-scale magnetic field is perpendicular to the local filament direction.

\begin{table*}[]
    \centering
    \begin{tabular}{c c c c}
    \hline
         & L1689N & SMM-16 & L1689B \\
    \hline
    Peak value (mJy\,arcsec$^{-2}$) & 3.00 & 0.95 & 0.53  \\
    Center R.A. (hh:mm:ss.ss) & 16:32:25.96 & 16:31:39.87 & 16:34:48.70 \\
    Center Dec. (dd:mm:ss.s) & $-24$:28:46.2 & $-24$:49:49.7 & $-24$:38:04.7 \\
    $a$ (major std. dev.) (arcsec) & 67.4 & 37.1 & 22.9 \\
    $b$ (minor std. dev.) (arcsec) & 25.5 & 36.4 & 16.9 \\
    P.A. (deg E of N) & 109.8 & 56.3 & 74.9 \\
    $R$ (pc) & 0.056 & 0.030 & 0.019 \\
    \hline
    \end{tabular}
    \caption{Gaussian fits to Stokes $I$ emission from L1689N, SMM-16 and L1689B. Note that IRAS 16293-2422 was masked as shown in Figure~\ref{fig:ellipses}.}
    \label{tab:gaussians}
\end{table*}

\begin{table*}[]
    \centering
    \begin{tabular}{c cc c c}
    \hline
         & \multicolumn{2}{c}{L1689N} & SMM-16 & L1689B\\ \cline{2-3}
         & All & With masking & & \\
    \hline
    $F_{\nu}$ (Jy) & 45.14 & 23.62 & 5.41 & 0.64 \\
    $M$ (M$_{\odot}$) & $17.2 \pm 14.4$ & $9.0 \pm 7.5$ & $2.1\pm 1.4$ & $0.24\pm 0.17$ \\
    $N({\rm H}_{2})$ (cm$^{-2}$) & $(7.9\pm 6.6)\times 10^{22}$ & $(4.2\pm 3.5) \times 10^{22}$ & $(3.3 \pm 2.2)\times 10^{22}$ & $(9.8\pm 7.7)\times10^{21}$ \\
    $n({\rm H}_{2})$ (cm$^{-3}$) & $(3.5\pm 2.9)\times 10^{5}$ & $(1.8\pm 1.5)\times 10^{5}$ & $(2.7\pm 1.8) \times 10^{5}$ & $(1.3\pm 1.0)\times 10^{5}$\\
    $\Delta v_{\textsc{nt}}$ (km\,s$^{-1}$) & \multicolumn{2}{c}{$0.61 \pm 0.04$} & $0.57\pm 0.04$ & $0.29 \pm 0.03$ \\
    \hline
    $\langle\theta\rangle$ (deg)  & --- & $34.2 \pm 0.8$ & $95.9 \pm 0.4$ & 8.4\\
    $\sigma_{\theta}$ (deg) & --- & $6.6\pm 0.6$ & $9.7 \pm 0.4$ & $13.3\pm 4.7$  \\
    $\mathcal{M}_{A}$ & --- & $0.23 \pm 0.05$ & $0.34 \pm 0.06$ & $0.47\pm 0.26$ \\
    $v_{A}$ (km\,s$^{-1}$) & --- & $1.12\pm 0.17$ & $0.72\pm0.08$ & $0.26\pm 0.12$ \\
    $B_{{\rm pos}}$ (fiducial $\kappa_{\nu}$) ($\mu$G) & --- & $366 \pm 55$ & $284 \pm 31$ & $72\pm 33$ \\
    $B_{{\rm pos}}$ (full range) ($\mu$G) & --- & 157---575 & 160---408 & 12---132 \\
    $\lambda$ (fiducial $\kappa_{\nu}$) & $1.52\pm 0.23$ & $0.79\pm 0.12$ &  $0.81\pm0.09$ & $0.95\pm 0.44$ \\
    $\lambda$ (full range) & 0.66---2.38 & 0.34---1.24 &  0.47---1.15 & 0.14---1.76 \\
    \hline
    \end{tabular}
    \caption{Quantities relevant to magnetic field strength calculations: flux density ($F_{\nu}$), mass ($M$), column density ($N({\rm H}_{2})$), volume density ($n({\rm H}_{2})$), FWHM non-thermal velocity dispersion ($v_{\textsc {fwhm}}$), mean magnetic field angle ($\langle\theta\rangle$), dispersion in magnetic field angle ($\sigma_{\theta}$), Alfv\'{e}n Mach number ($\mathcal{M}_{A}$), Alfv\'{e}n velocity ($v_{A}$), plane-of-sky magnetic field strength ($B_{\rm pos}$), and mass-to-flux ratio ($\lambda$), in the latter two cases listing both the fiducial-$\kappa_{\nu}$ value and its statistical uncertainty, and the full range of results including systematic uncertainty on density. When calculating masked values in L1689N, $F_{\nu}$, $M$, $N({\rm H}_{2})$ and $n({\rm H}_{2})$ were calculated assuming a constant flux density over the masked area covering IRAS 16293-2422 as described in Appendix~\ref{sec:appendix_dcf}, and all magnetic field half-vectors associated with the masked area were excluded from determination of $\langle\theta\rangle$ and $\sigma_{\theta}$. $\Delta v_{\textsc{nt}}$ values are taken from \citet{pan2017} for L1689N and SMM-16, and from \citet{lee2001} for L1689B, and are corrected for the thermal component of the linewidth.  Mean position angle and dispersion in position angle are taken from Gaussian fitting for L1689N and SMM-16, and calculated directly for L1689B.}
    \label{tab:properties}
\end{table*}

\subsection{Davis-Chandrasekhar-Fermi analysis}
\label{sec:dcf_main}

We estimated magnetic field strengths in L1689 using the Davis-Chandrasekhar-Fermi (DCF) method \citep{davis1951a,chandrasekhar1953}, which assumes that perturbations in the magnetic field are Alfv\'{e}nic; i.e. deviation in angle from the mean field direction is due to distortion by small-scale non-thermal motions, and the Alfv\'en Mach number of the gas is given by
\begin{equation}
\mathcal{M}_{A} = \frac{\sigma_{\textsc{nt}}}{v_{A}} = \frac{\sigma_{\theta}}{Q},
\end{equation}
where $\sigma_{\theta}$ is the dispersion in magnetic field angle in radians, $Q$ is a correction factor for line-of-sight and sub-beam integration effects, such that $0 < Q < 1$ \citep{ostriker2001}, $\sigma_{\textsc{nt}}$ is the non-thermal velocity dispersion of the gas and $v_{A}$ is the Alfv\'{e}n velocity of the magnetic field, which is then given in cgs units by
\begin{equation}
    v_{A} = \frac{B}{\sqrt{4\pi\rho}} = Q\,\frac{\sigma_{\textsc{nt}}}{\sigma_{\theta}},
    \label{eq:va_theory}
\end{equation}
where $B$ is magnetic field strength and $\rho$ is gas density.  This equation can then be rearranged to give an expression for plane-of-sky magnetic field strength $B_{{\rm pos}}$, noting that the value of $\sigma_{\theta}$ which we measure represents only the plane-of-sky component of the angle dispersion,
\begin{equation}
    B_{{\rm pos}} \approx Q\sqrt{4\pi\rho}\,\frac{\sigma_{\textsc{nt}}}{\sigma_{\theta}}.
    \label{eq:cf}
\end{equation}
\citet{crutcher2004} note that on average $B_{\rm pos}/B \approx \pi/4$.  Using plane-of-sky angular dispersion will also cause $v_{A}$ to be correspondingly underestimated, and $\mathcal{M}_{A}$ to be overestimated.  However, as the \citet{crutcher2004} correction factor can be meaningfully applied only to a statistical ensemble of measurements, we do not apply it in this work, noting that its inclusion would not alter any of our conclusions.

We use the formulation of Equation~\ref{eq:cf} given by \citet{crutcher2004},
\begin{equation}
    B_{{\rm pos}}(\mu{\rm G}) \approx 9.3\sqrt{n({\rm H}_{2})({\rm cm}^{-3})}\,\frac{\Delta v_{\textsc{nt}}({\rm km\,s}^{-1})}{\sigma_{\theta}({\rm deg})},
    \label{eq:cf_c04}
\end{equation}
where number density $n({\rm H}_{2}) = \rho/\mu m_{\textsc{h}}$, and FWHM non-thermal velocity dispersion $\Delta v_{\textsc{nt}} = \sigma_{\textsc{nt}}\sqrt{8\ln2}$.  Note that this formulation takes $Q=0.5$ (cf. \citealt{heitsch2001,ostriker2001}) and assumes a mean molecular weight $\mu = 2.8$, which we adopt throughout this work.  Equation~\ref{eq:cf_c04} is equivalent to the following relations for the Alfv\'{e}n Mach number,
\begin{equation}
    \mathcal{M}_{A} \approx 3.5\times 10^{-2}\,\sigma_{\theta}({\rm deg}),
    \label{eq:ma_emp}
\end{equation}
and for Alfv\'{en} velocity,
\begin{equation}
    v_{A}({\rm km\,s}^{-1}) \approx 12.1 \,\frac{\Delta v_{\textsc{nt}}({\rm km\,s}^{-1})}{\sigma_{\theta}({\rm deg})}.
    \label{eq:va_emp}
\end{equation}

We outline here our approach to DCF analysis in L1689.  A detailed description of how the values which we use are derived is given in Appendix~\ref{sec:appendix_dcf}.

We estimated the size of each of the clumps by fitting a 2D Gaussian distribution to the emission associated with the clump.  In the case of L1689N, we masked emission in the 40$^{\prime\prime}$-diameter region surrounding the IRAS 16293-2422 protostar, as we aim to investigate the magnetic field strength in the larger clump, rather than the behavior of IRAS 16293-2422 itself, the magnetic field of which appears to be behaving quite differently to that in the clump in which it is embedded (see Section~\ref{sec:l1689_discuss}).  Polarization half-vectors in this region are excluded from the determination of angle dispersions, and flux from this region is included in mass and density calculations only where noted.  The best-fit Gaussian distributions are listed in Table~\ref{tab:gaussians}, {and marked on Figure~\ref{fig:presentation_ef}.}

We measured flux densities $F_{\nu}$ via aperture photometry over a 1-FWHM-diameter elliptical aperture for each region.  We measured dispersion in polarization angle $\sigma_{\theta}$ over the same areas, by fitting a Gaussian model in L1689N and SMM-16, and through direct calculation in L1689B.  The values for each region are listed in Table~\ref{tab:properties}.

We further defined a representative HWHM radius for each source,
\begin{equation}
    R = D\sqrt{2\ln2 \tan a \tan b},
    \label{eq:R}
\end{equation}
where $a$ is the major axis Gaussian width, $b$ is the minor axis Gaussian width, and $D=144.2$\,pc is the distance to L1689.  This radius, also listed in Table~\ref{tab:gaussians}, is used in calculations of column and volume density and in Section~\ref{sec:energetics} below.

Masses for each source were calculated using the \citet{hildebrand1983} relation,
\begin{equation}
    M = \frac{F_{\nu}D^{2}}{\kappa_{\nu}B_{\nu}(T)},
    \label{eq:mass_main}
\end{equation}
where dust opacity $\kappa_{\nu} = 0.0125$\,cm$^{2}$g$^{-1}$, and $B_{\nu}(T)$ is the Planck function.  We take temperature $T=12$\,K for all of the sources, and conservatively assume a systematic uncertainty of $50$\% on $\kappa_{\nu}$ \citep[cf.][]{roy2014}.  Column density is calculated as
\begin{equation}
    N({\rm H}_{2}) = \frac{M}{\mu m_{\textsc{h}}}\frac{1}{\pi R^{2}},
    \label{eq:coldens_main}
\end{equation}
and volume density as
\begin{equation}
    n({\rm H}_{2}) = \frac{M}{\mu m_{\textsc{h}}}\frac{3}{4\pi R^{3}}.
    \label{eq:density_main}
\end{equation}

Gas FWHM velocity dispersion values $\Delta v$ were taken from N$_{2}$H$^{+}$ $J=1\to0$ measurements made by \citet{pan2017} in the case of L1689N and SMM-16, and by \citet{lee2001} in the case of L1689B, and corrected for the thermal linewidth component to give $\Delta v_{\textsc{nt}}$.  The flux densities ($F_{\nu}$), masses ($M$), column and volume densities ($N({\rm H}_{2})$ and $n({\rm H}_{2})$ respectively), FWHM velocity dispersions ($\Delta v_{\textsc{nt}}$), mean magnetic field angle ($\langle\theta\rangle$) and magnetic field angle dispersion ($\sigma_{\theta}$) determined for each clump are listed in Table~\ref{tab:properties}.

\subsubsection{Alfv\'en Mach number}

Using equation~\ref{eq:ma_emp}, we find that in L1689N, $\mathcal{M}_{A} = 0.23 \pm 0.05$, while in SMM-16, $\mathcal{M}_{A} = 0.34 \pm 0.06$, and in L1689B, $\mathcal{M}_{A} = 0.47 \pm 0.26$.  All three clumps have sub-Alfv\'{e}nic non-thermal motions, suggesting that magnetic fields are more important to clump stability than are non-thermal motions.  This result follows from the ordered field morphologies seen in the clumps.

\subsubsection{Alfv\'en velocity}

Using equation~\ref{eq:va_emp}, we find that in L1689N, $v_{A}=1.12\pm 0.17$\,km\,s$^{-1}$, while in SMM-16, $v_{A}=0.72\pm 0.06$\,km\,s$^{-1}$, and in L1689B, $v_{A}=0.26\pm 0.12$\,km\,s$^{-1}$. 

\subsubsection{Magnetic field strength}

Combining these measurements using equation~\ref{eq:cf_c04}, we estimated plane-of-sky magnetic field strengths of $366\pm 209$\,$\mu$G in L1689N, $284\pm 127$\,$\mu$G in SMM-16, and $72\pm 61$\,$\mu$G in L1689B.  We emphasize that the uncertainties on these values are predominantly systematic, being dominated by systematic uncertainty on $n({\rm H}_{2})$, itself predominantly caused by uncertainty on $\kappa_{\nu}$ (cf. equations~\ref{eq:mass_main} and \ref{eq:density_main}).  These measurements are more meaningfully described as being in the range $\sim 160 - 580$\,$\mu$G in L1689N, $\sim 160 - 410$\,$\mu$G in SMM-16, and $\sim 10 - 130$\,$\mu$G in L1689B.  The magnetic field strengths at our fiducial value of $\kappa_{\nu}$ are $366\pm 55$\,$\mu$G, $284\pm 31$\,$\mu$G and $72\pm33$\,$\mu$G in L1689N, SMM-16 and L1689B respectively.

The magnetic field strength values in L1689N and SMM-16 are quite large, indicating a significant enhancement of the magnetic field strength over that of the diffuse ISM  \citep[median strength $6.0\pm 1.8\,\mu$G;][]{heiles2005},
but comparable to some previous measurements of magnetic field strength in nearby low-mass star-forming and starless clumps and cores inferred from JCMT observations; \citet{crutcher2004} measured plane-of-sky magnetic field strengths of 140\,$\mu$G, 80\,$\mu$G and 160\,$\mu$G in the starless cores L1544, L183, and L43, respectively.  These values are also comparable to recent measurements in L1688 made using POL-2; \citet{kwon2018} measured field strengths in the range $200-5000$\,$\mu$G in the star-forming clump Oph A, while \citet{soam2018} measured $220 - 1040$\,$\mu$G in the star-forming clump Oph B, and \citet{liu2019} measured $103-213$\,$\mu$G in the starless clump/core Oph C.  The highly ordered magnetic field morphologies seen in Figure~\ref{fig:presentation} are also consistent with a relatively strong magnetic field.  The magnetic field strength of L1689B is somewhat weaker, comparable to the magnetic field strengths measured in starless cores by \citet{crutcher2004}, but also to those measured in the isolated starless cores L1498 and L1517B, which were found by \citet{kirk2006} to have magnetic field strengths of $\sim 10$\,$\mu$G and $\sim 30$\,$\mu$G respectively.

\subsubsection{Mass-to-flux ratio}

The relative importance of magnetic fields and gravity in a clump or core can be characterized using the mass-to-flux ratio $\lambda$, the ratio of measured mass to the maximum mass which could be supported against collapse under self-gravity by the measured magnetic flux.  A magnetically subcritical object, $\lambda< 1$, is magnetically supported, while a magnetically supercritical object, $\lambda >1$, is gravitationally unstable.

We estimated the mass-to-flux ratio $\lambda$ using the \citet{crutcher2004} formulation,
\begin{equation}
    \lambda = 7.6\times 10^{-21}\frac{N({\rm H}_{2})\,({\rm cm}^{-2})}{B_{{\rm pos}}\,(\mu{\rm G})}.
\end{equation}
In L1689N we measured $\lambda = 0.79\pm 0.45$ (for fiducial $\kappa_{\nu}$, $\lambda = 0.79\pm 0.12$) when IRAS 16293-2422 is masked, and $\lambda = 1.52 \pm 0.86$ (fiducial $\kappa_{\nu}$: $\lambda = 1.52\pm 0.23$) using the total mass, while in SMM-16 we measured $\lambda = 0.81 \pm 0.34$ (fiducial $\kappa_{\nu}$: $\lambda = 0.81\pm 0.09$), and in L1689B, $\lambda = 0.95 \pm 0.81$ (fiducial $\kappa_{\nu}$: $\lambda = 0.95\pm 0.44$).  These values suggest that all of the clumps are magnetically trans-critical 
-- i.e. they {are either marginally} supported against collapse by their internal magnetic fields, or they are marginally gravitationally unstable, but magnetic fields are dynamically important in the {clumps'} evolution.  However, we note that the validity of a comparison of plane-of-sky magnetic field strength to line-of-sight column density is uncertain; our values of $\lambda$ could be overestimated by a factor of $\sim 1.33 - 3$ depending on the three-dimensional geometry of the clumps \citep{crutcher2004,planck2016a}. 
We discuss the interpretation of these values further in Section~\ref{sec:discussion}, but note that the values of $\lambda$ determined here are best considered as a qualitative indicator that these clumps are approximately magnetically critical rather than as a precise measure of their stability against collapse.

\subsection{Energy balance}
\label{sec:energetics}

\begin{table*}[]
    \centering
    \begin{tabular}{l cc c c}
    \hline
    Energy ($\times 10^{35}$ J) & \multicolumn{2}{c}{L1689N} & SMM-16 & L1689B\\ \cline{2-3}
     & All & With masking & & \\
    \hline
    Gravitational ($E_{G}$) & $-274$ & $-75$ & $-7.3$ & $-0.16$ \\
    Magnetic ($E_{B}$) & --- & 113 & 11 & 0.17 \\ 
    Thermal Kinetic	($E_{K,\textsc{t}}$) & 18 & 9.5 & 2.2 & 0.25 \\
    Non-thermal Kinetic ($E_{K, \textsc{nt}}$) & 11 & 5.9 & 1.2 & 0.04 \\
    Rotational ($E_{R}$) & --- & --- & 0.03 & --- \\
    \hline
    \end{tabular}
    \caption{Energetics of L1689N, SMM-16 and L1689B.}
    \label{tab:energy}
\end{table*}

We calculated the energy balance for each of our clumps using the properties derived above.

\subsubsection{Gravitational potential energy}

The gravitational potential energy (GPE) of a spherically symmetric density distribution is given by 
\begin{equation}
    E_{G} = -\eta\frac{GM^{2}}{R}.
\end{equation}
The value of the constant $\eta$, and so of $E_{G}$, depends on the density profile of the sphere.  For consistency with our previous analysis, we choose to model the core as a uniform sphere of radius $R$, for which $\eta = 3/5$.

\subsubsection{Magnetic energy}

Magnetic energy is given, in SI units, by
\begin{equation}
    E_{B} = \frac{B^{2}V}{2\mu_{0}},
\end{equation}
where $B$ is total magnetic field strength, and volume $V = (4\pi/3)R^{3}$.  We again take $B\approx B_{\rm pos}$, noting that this will produce an underestimate of the true magnetic energy.  If the \citet{crutcher2004} statistical correction applies to our results, the true magnetic energies will typically be $\sim 1.6\times$ those that we derive.  We give our values without correction, noting that they are, as with the other energies listed, accurate to order of magnitude.

\subsubsection{Kinetic energy}

Thermal kinetic energy is given by
\begin{equation}
    E_{K,\textsc{t}} = \frac{3}{2}M\frac{k_{\textsc{b}}T}{\mu m_{\textsc{h}}},
\end{equation}
and non-thermal kinetic energy is given by
\begin{equation}
    E_{K,\textsc{nt}} = \frac{1}{2}M\sigma_{\textsc{nt}}^{2},
\end{equation}
For L1689N we again calculate these values both including and excluding the mass of IRAS 16293-2422.

\subsubsection{Rotational energy}

{\citet{loren1990} inferred SMM-16 to be rotating with an angular velocity of $\omega = 1.85\pm 0.25$ km\,s$^{-1}$\,pc$^{-1}$, with a rotation axis of $68^{\circ}$ east of north, using 2.4-arcmin resolution DCO$^{+}$ data, and so} for SMM-16, we also calculate a value for rotational energy, 
\begin{equation}
    E_{R} = \frac{1}{2}\mathcal{I}\omega^{2},
\end{equation}
where $\mathcal{I}$, the moment of inertia, is
\begin{equation}
    \mathcal{I} = \frac{2}{5}MR^{2},
\end{equation}
again taking SMM-16 to be a uniform-density sphere.

\subsubsection{Comparison of values}

The energy values for each clump are listed in Table~\ref{tab:energy}.  We note that these values have significant uncertainties, and thus that the energy values which we derive are accurate only to order of magnitude.  However, all of the derived energy values depend linearly on dust opacity $\kappa_{\nu}$, our dominant and systematic source of uncertainty, except for GPE, which goes as $\kappa_{\nu}^{2}$.  Thus we consider only fiducial-$\kappa_{\nu}$ values here, as we can meaningfully compare different energy terms to one another despite the large uncertainties on their absolute values.

We note that the energy values which we list are largely a restatement of the mass-to-flux ratios and Alfv\'{e}n Mach numbers previously calculated, with $E_{G}/E_{B}\propto \lambda^{2}$, and $E_{K,\textsc{nt}}/E_{B} \propto \mathcal{M}_{A}^{2}$.  Nonetheless, calculation of energy terms is a helpful exercise, allowing the various forces determining the evolution of the clump/core to be compared to one another in the same terms.  We broadly expect any relatively long-lived object in the ISM to be near equipartition, and in L1689 we particularly expect gravitational and magnetic energies to be comparable, as all of our cores have mass-to-flux ratios consistent with unity.

We find that in L1689N, when IRAS 16293-2422 is excluded, the gravitational and magnetic energies are similar, with the gravitational energy slightly smaller than the magnetic energy.  When the mass of IRAS 16293-2422 is included, the gravitational potential energy of L1689N is larger than, but still comparable to, the magnetic energy, consistent with the fragmentation across field lines and ongoing star formation in the clump, and with the $\lambda > 1$ which we infer in this case.  In L1689N in particular, we note that a uniform sphere model is a crude approximation to the true geometry of the clump, which has fragmented into two significant cores, and we stress that the GPE values which we measure are approximate.

In SMM-16 and L1689B, the gravitational, magnetic and kinetic energies are comparable to one another, as we expect.  The rotational energy of SMM-16 is two orders of magnitude smaller than any of the other terms.  SMM-16 and L1689B being in approximate equipartition is consistent with their lack of ongoing star formation.

\section{Grain alignment in L1689}
\label{sec:grains}

\begin{figure}
    \centering
    \includegraphics[width=0.47\textwidth]{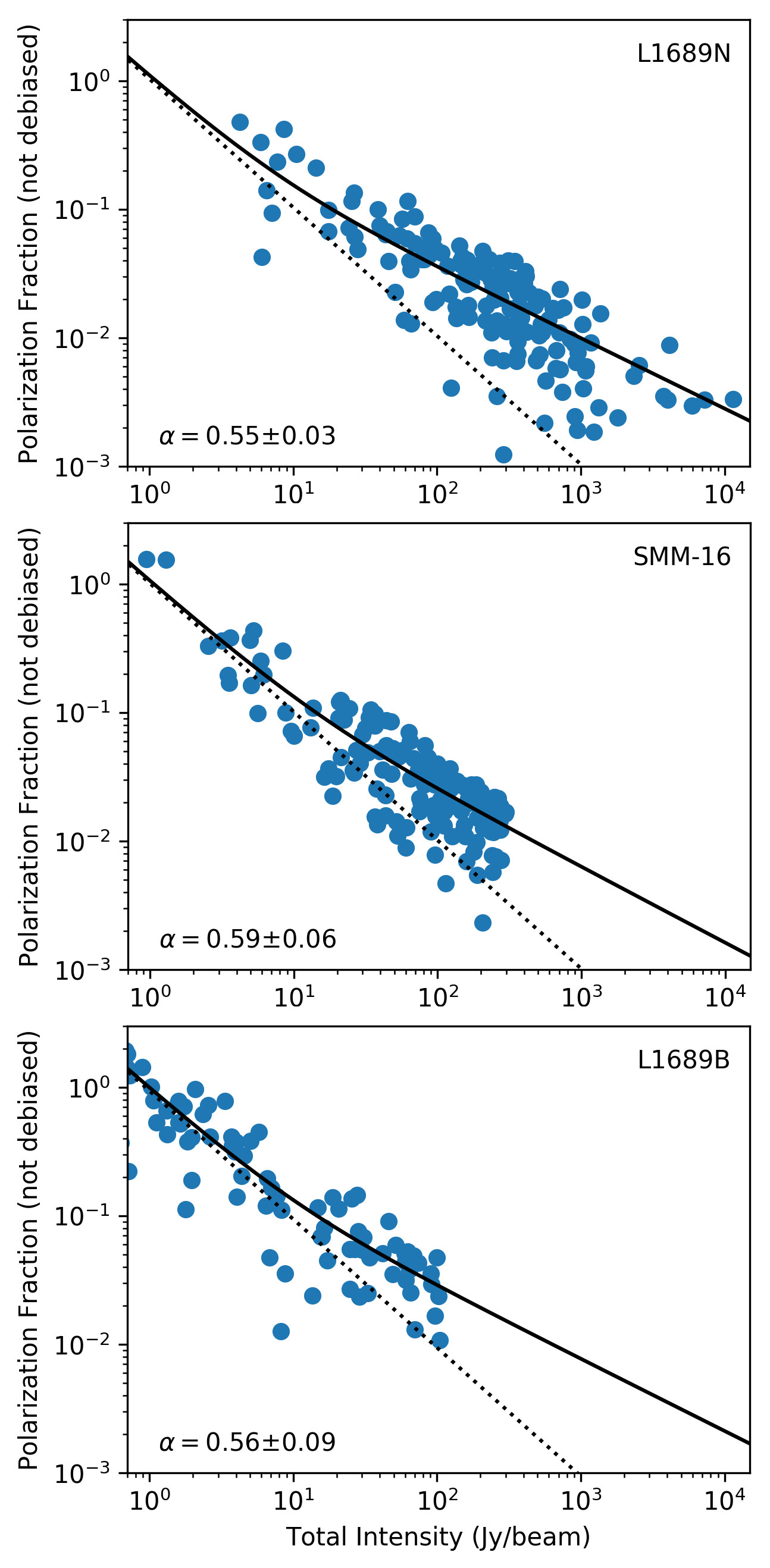}
    \caption{Polarization fraction $p$ (not debiased) vs. total intensity, fitted with the mean of the Ricean distribution of $p$ associated with a $p\propto I^{-\alpha}$ model (see Appendix~\ref{sec:appendix_grains} for details).  Solid line shows best-fit model; dashed line shows expected behavior of non-aligned grains.}
    \label{fig:polfrac}
\end{figure}

There is ongoing debate over the degree to which grains remain aligned relative to their local magnetic field direction at high $A_{V}$ \citep{whittet2008,alves2014,jones2015,wang2019,pattle2019a}.  In the diffuse ISM, dust grains are expected to be aligned with their minor axes parallel to the local magnetic field direction \citep{davis1951}. However, at high optical depths, grains are expected to become less effectively aligned with the magnetic field \citep[e.g][]{andersson2015}. This is a corollary of the radiative alignment torques (RATs) model of grain alignment, in which irregularly-shaped grains are spun up by incident photons from a non-isotropic radiation field \citep{dolginov1976,lazarian2007}.  In this paradigm, high extinction will perforce lead to a lack of grain alignment.

\subsection{Measurements}

In order to confirm the validity of our preceding use of polarization observations to analyse the properties of the magnetic fields in L1689, we assess the degree of grain alignment in each clump.  Grain alignment can be quantified through examination of the relationship between polarization efficiency and visual extinction \citep[e.g][]{whittet2008,jones2015}.  In submillimeter emission polarimetry, this is commonly treated as a relationship between $p$ and $I$ (e.g. \citealt{alves2014}).  Polarization efficiency is identical to polarization fraction for optically thin emission \citep{alves2015}, and for optically thin, isothermal dust emission, total intensity is proportional to visual extinction \citep{jones2015,santos2017}.  Observations of polarized dust emission typically show a power-law dependence, $p\propto I^{-\alpha}$, where $0\leq \alpha \leq 1$.  A steeper index (higher $\alpha$) indicates poorer grain alignment: $\alpha = 0$ indicates that grains are equally well aligned at all depths, while $\alpha = 1$ indicates either a total absence of aligned grains, or that all observed polarized emission is produced in a thin layer at the surface of the cloud \citep{pattle2019a}.

We apply the method of fitting the $p-I$ relationship described by \citet{pattle2019a} to our observations of L1689.  The details of this analysis are described in Appendix~\ref{sec:appendix_grains}.  We find that the fitting results for all regions agree with one another within error.  In L1689N we find $\alpha=0.55\pm0.03$ while in SMM-16 we find $\alpha=0.59\pm0.06$, and in L1689B, $\alpha = 0.56\pm 0.09$.  These results are shown in Figure~\ref{fig:polfrac}.  This suggests that in all three regions, grains become less well-aligned with the magnetic field as density increases, but some degree of alignment persists to the highest densities which we observe, justifying our use of our polarization observations to analyse the properties of the magnetic fields which permeate the clumps.

\subsection{Comparison with L1688}

The degree of grain alignment in L1689 appears to be broadly consistent with that recently measured in the neighbouring L1688 region \citep{pattle2019a}.  The $\alpha$ values observed in L1689 are larger than that of the externally-illuminated Oph A region ($\alpha=0.34$), but comparable to those of Oph B and C ($\alpha\approx 0.6-0.7$) \citep{pattle2019a}, and, in the cases of L1689N and SMM-16, better-characterized than those in Oph B and C.  The global interstellar radiation field (ISRF) on L1689 is likely to be similar to that on L1688, thanks to their comparable proximity to Sco OB2, as discussed in Section~\ref{sec:intro_oph}.

The agreement between the fitting results for L1689N and the starless sources suggests that the presence of the L1689N protostellar system is not significantly affecting the grain alignment in the clump as a whole, although it may be aligning grains on scales smaller than can be resolved by the JCMT.  The only clump in Ophiuchus showing a significant difference in behavior is Oph A, which is illuminated by the two B stars of L1688, in which grains appear to be significantly better-aligned with the magnetic field \citep{pattle2019a}.  This provides further evidence suggesting that grain alignment within the dense clumps embedded within L1688 and L1689 is driven by the local radiation field on the clump.

\subsection{Grain growth in L1689}

{Our observations show that grains in L1689 retain some degree of alignment up to the highest gas densities which we observe.}  In RAT alignment theory, grains are efficiently aligned when they can be spun up to suprathermal rotation by an anisotropic radiation field, with there being a critical grain size $s_{{\rm crit}}$ above which grains can become aligned, which increases with increasing gas density \citep{lazarian2007,hoang2008}.  We calculate this critical grain size for the clumps of L1689 using equation 3 of \citet{lee2020}, assuming $T_{{\rm gas}} = T_{\rm{dust}} = 12$\,K, and taking grain mass density $\rho=3$\,g\,cm$^{-3}$, radiation field anisotropy $\gamma = 0.1$ \citep{draine1996}, and mean incident wavelength $\bar{\lambda}=1.2\,\mu$m \citep{mathis1983,draine2007}.  We further take the radiation strength (the ratio of local radiation energy density to that of the interstellar radiation field) to be $U = (T_{{\rm dust}}/16.4\,{\rm K})^{6} \simeq 0.1535$ \citep{draine2011}.  With these assumptions, and in the limit where hydrogen number density $n({\rm H}) = 2n({\rm H}_{2}) \gg 30$\,cm$^{-3}$, the critical grain size is given by
\begin{equation}
    s_{{\rm crit}} \approx 3.88\times 10^{-6}\,{\rm cm}\times \left(\frac{n(\rm{H})}{30\,{\rm cm}^{-3}}\right)^{\frac{5}{16}}.
    \label{eq:scrit}
\end{equation}
Taking the range of $n({\rm H})$ values corresponding to the average densities listed in Table~\ref{tab:properties}, we find that in L1689N, for $n({\rm H}) = 7\times 10^{5}$\,cm$^{-3}$, $s_{\rm crit} \approx 0.90\,\mu$m, while in the lowest-density source, L1689B, for $n({\rm H}) = 2.6\times 10^{5}$\,cm$^{-3}$, $s_{\rm crit} = 0.66\,\mu$m.  We note that in equation~\ref{eq:scrit}, we have conservatively adopted $\gamma = 0.1$.  \citet{draine1996} found $\gamma=0.1$ in the diffuse ISM and $\gamma=0.7$ in molecular clouds, while \citet{bethell2007} found $\gamma\sim 0.34$ in clumpy molecular clouds.  If we take $\gamma = 0.7$, our values of $s_{crit}$ will be smaller by a factor 0.55, with $s_{{\rm crit}, \gamma=0.7} = 0.49$\,$\mu$m and 0.36\,$\mu$m in L1689N and L1689B respectively.  The maximum grain size in the diffuse ISM is $\sim 0.25 - 0.3\,\mu$m \citep{mathis1977,draine2007}.  Thus, {as our observations show that} a significant population of dust grains remain aligned at high densities in L1689, grain growth must have occurred.  {That such grain growth and evolution takes place is suggested by recent studies of $Herschel$ Space Observatory data, which find that dust opacity increases with column density in nearby molecular clouds \citep[e.g.][]{roy2013, ysard2013,juvela2015}.  Particularly, \citet{schirmer2020} find that dust emission modelling of the Horsehead Nebula requires a minimum grain size $2-2.5\times$ that in the diffuse ISM to reproduce $Herschel$ and $Spitzer$ observations.   Polarization observation such as ours provide independent confirmation that grain growth occurs in these dense environments.}

\section{Discussion} \label{sec:discussion}

L1689N, SMM-16 and L1689B all appear to be, on the scales probed by POL-2, magnetically trans-critical environments with sub-Alfv\'{e}nic turbulence.  However, the three regions appear to have evolved in quite different manners.

$Planck$ measurements (\citealt{planck2015}; Figure~\ref{fig:finding_chart}) show that in L1689 the magnetic field goes from running NE/SW in the north, similar to the overall magnetic field direction across the complex ($\sim 50^{\circ}$ east of north; \citealt{vrba1976}), to running approximately N/S in the south.
In our data, we see agreement between POL-2 and $Planck$ magnetic fields in L1689N and L1689B, but an approximately 80$^{\circ}$ disagreement at the highest column densities in SMM-16, as shown in Figure~\ref{fig:angles}.

\subsection{L1689N}
\label{sec:l1689_discuss}

\begin{figure}
    \centering
    \includegraphics[width=0.47\textwidth]{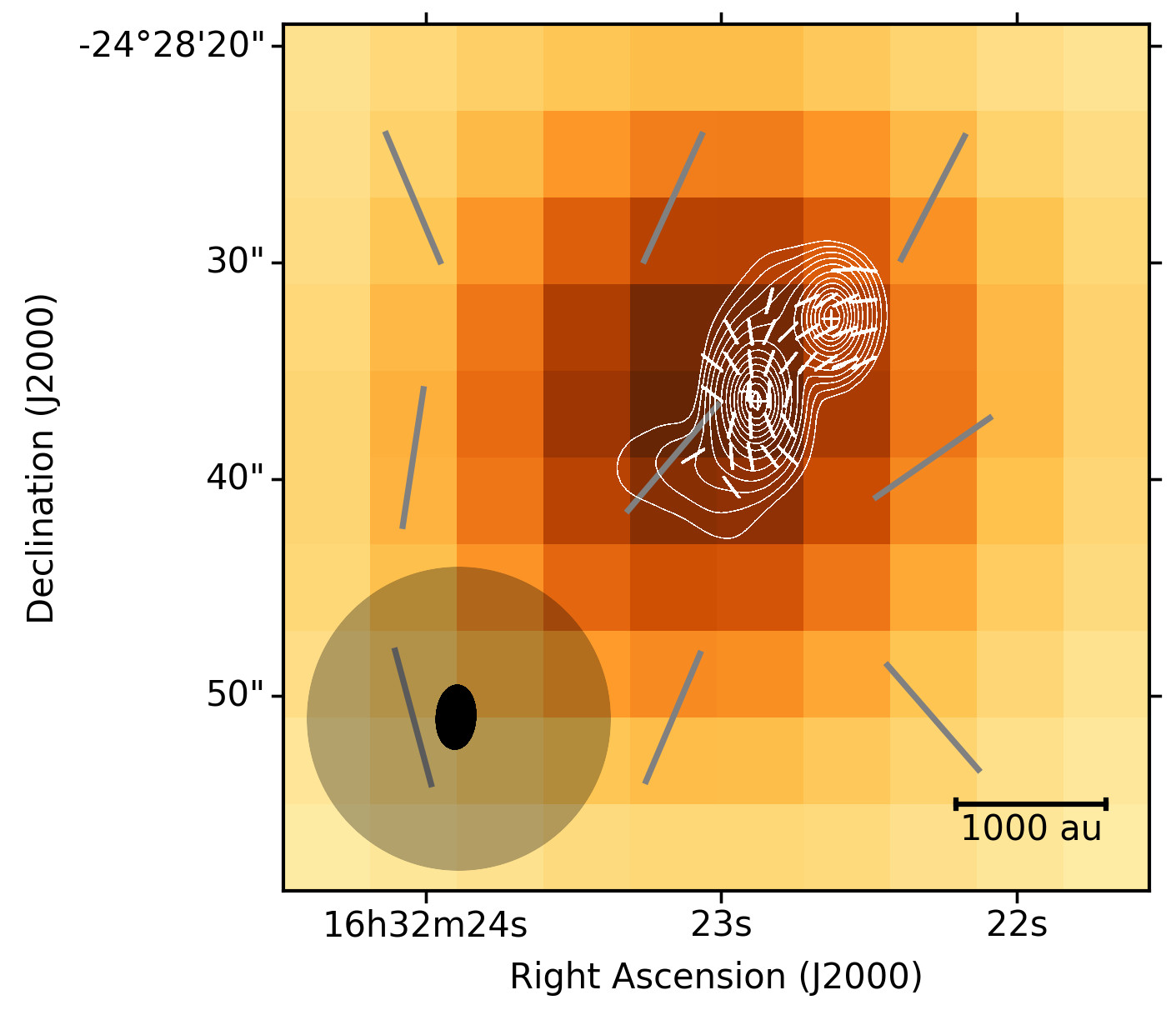}
    \caption{A comparison of POL-2 and SMA observations of IRAS 16293-2422.  Underlying image is of POL-2 Stokes $I$ emission in which IRAS 16293-2422 is a point source.  Grey half-vectors show the POL-2 magnetic field direction.  White contours show SMA Stokes I emission, while white half-vectors show the SMA-inferred magnetic field direction \citep{rao2009}.  The JCMT and SMA beam sizes are shown in the lower left-hand corner as a translucent grey circle and solid black ellipse respectively.  Note that the registration of the two images is based on the nominal coordinates of each observation, and that the JCMT has a typical pointing uncertainty of $2^{\prime\prime}-6^{\prime\prime}$ (0.5--1.5 4\arcsec pixels) \citep{mairs2017}, comparable to or larger than the SMA beam.}
    \label{fig:iras16293}
\end{figure}

On cloud-to-clump scales, L1689N appears to be undergoing magnetically-mediated {evolution}.  The average magnetic field direction is uniform from $Planck$ to JCMT scales.  The magnetic field which we infer from our POL-2 observations is magnetically trans-critical and uniform in the clump's center, {albeit with some deviation from linearity} in its north-western and south-eastern edges. 
Moreover, the clump itself is significantly elongated perpendicular to the plane-of-sky magnetic field direction, consistent with having formed through collection of material along magnetic field lines.  L1689N thus, {on clump scales,} appears consistent with {having formed} in a magnetically dominated environment.

{Despite the properties of the L1689N clump being broadly consistent with having formed in a strongly magnetized environment, the internal structure of the clump suggests that the magnetic field does not control the evolution of the star-forming cores embedded within it.   Our energetics analysis suggests that the clump is sufficiently massive to be gravitationally bound and unstable to fragmentation.  The fragmentation across magnetic field lines within the L1689N clump, which has created the IRAS 16293-2422 and IRAS 16293E cores, and the ongoing star formation in the clump, indicated by the presence of the IRAS 16293-2422 protostellar system, confirm} that the clump must be magnetically supercritical on some scales. Moreover, the magnetic field direction which we infer in IRAS 16293-2422 is perpendicular to that in the surrounding clump, indicating that the magnetic field direction is not consistent on all scales.  {Thus even if the magnetic field has been instrumental in forming the L1689N clump, as is suggested but not confirmed by our measurements, it appears not to be the dominant influence on the stars forming within it.}

\subsubsection{IRAS 16293-2422}

The IRAS 16293-2422 system has been observed in 850$\mu$m polarized light using both the SMA \citep{rao2009} and ALMA \citep{sadavoy2018}.  The A and B protostars are linked by a `bridge' of emission oriented approximately SE/NW \citep[e.g.][]{pineda2012,jorgensen2016}.  
The magnetic field in the IRAS 16293-2422 system runs along this bridge \citep{rao2009,sadavoy2018}, with a magnetic field strength of $23-78$\,mG, at a density of $5.6\times10^{8}$\,cm$^{-3}$ \citep{sadavoy2018}.

In Figure~\ref{fig:iras16293} we compare the magnetic field direction which we infer on the position of IRAS 16293-2422 with that inferred by \citet{rao2009} from SMA observations.  {In our POL-2 observations we see an average magnetic field direction across the core of $166^{\circ}\pm31^{\circ}$,} consistent with the SE/NW field seen in interferometric observations.  {The average field direction in the \citet{rao2009} observations is $152^{\circ}\pm66^{\circ}$, although this value includes substantial contributions from the IRAS16293A and B protostellar discs, polarized emission from which is likely to be dominated by scattering \citep{sadavoy2019}.  The average field direction in the \citet{sadavoy2019} observations is $176^{\circ}\pm54^{\circ}$, again including both of the protostellar systems, while in the Bridge feature the average value is $130^{\circ}\pm14^{\circ}$.  All of these values are consistent with our measurement.}

If the dust emission features observed with ALMA formed from flux-frozen collapse under gravity, we can estimate what the magnetic field strength would have been in the gas from which the core formed.  Taking $B = (23 - 78\,{\rm mG})\times\sqrt{n/(5.6\times10^{8}\,{\rm cm}^{-3})}$ (assuming that in the high-density ISM, $B\propto n^{0.5}$; cf. \citealt{crutcher2010}), for our estimated density in the L1689N clump, $n=1.8\times 10^{5}$\,cm$^{-3}$, we infer a field strength of $412 - 1398\,\mu$G, consistent with our inferred field strength of $160-580\,\mu$G.  This suggests that the features observed on interferometric scales could have evolved from the larger-scale field which we observe.

\citet{jacobsen2018} proposed a model of IRAS 16293-2422 in which material is accreting along the bridge of emission onto the two protostars.  \citet{sadavoy2018} posit that if the bridge is not a transient structure, it must itself be accreting material from the surrounding envelope, and so that the field in the surrounding envelope ought to be perpendicular to that in the bridge (cf. \citealt{gomez2018}).  This is what we see in our observations (Figure~\ref{fig:presentation}), supporting the suggestion that accretion onto the IRAS 16293-2422 protostars is magnetically regulated.

\subsubsection{IRAS 16293E}

{As IRAS 16293E is a starless core, without a central hydrostatic object \citep{kirk2017}, its geometry is not consistently defined between studies.  \citet{pattle2015} split the core into two components (their SMM-19 and SMM-22), the former approximately circular with an aspect ratio 1.1, the latter with an aspect ratio of 1.6, oriented $44.7^{\circ}$ E of N (these sources are marked on Figure~\ref{fig:presentation_p}).  \citet{ladjelate2020} class IRAS 16293E as a single core (their source 464), with an aspect ratio 1.7, oriented $2^{\circ}$ E of N.  \citet{kirk2017}, observing with ALMA, find an aspect ratio 2.0, oriented $4.8^{\circ}$ E of N (their source 37).  The core is thus consistently found to be elongated broadly NE/SW, in a direction similar to the local average magnetic field direction.  This is not consistent with the behavior predicted for a strongly magnetized starless core \citep{fiedler1993}, or with the larger-scale behavior of the clump, again suggesting that the magnetic field does not control the evolution of the cores embedded within the L1689N clump.}

\subsection{SMM-16}

The magnetic field which we infer from POL-2 data in SMM-16 is, in the center of the  clump, oriented approximately 90$^{\circ}$ E of N, in contrast to the magnetic field at the same location inferred from the $Planck$ data, which is oriented approximately $10^{\circ}$ E of N.  This significant difference in magnetic field direction between cloud and clump scales suggests that a large-scale reordering of the magnetic field has taken place during the formation of the clump.

Despite evidence for a large-scale rotation gradient across the region \citep{loren1990}, our energetics analysis suggests that the rotational energy of SMM-16 itself is two orders of magnitude smaller than the gravitational, magnetic and kinetic energies.  This suggests that, whatever the cause of the reordering of the magnetic field in the clump, its dynamics are not presently significantly affected by rotation, as expected from previous studies of rotation in dense cores \citep{caselli2002a, tafalla2004, xu2020}.
The similarity between POL-2 field lines in the periphery of the {clump} and the field direction inferred from $Planck$ data suggests that the magnetic field in the clump may remain connected to the larger-scale field, despite its reordering in the center.

The density structure of the clump -- in particular the multiple dense cores identified by \citet{pattle2015} and \citet{ladjelate2020} -- and the complex velocity structure \citep{chitsazzadeh2014}, suggest that the kinematics of the clump itself are more complex than that of simple solid-body rotation.
Our energetics analysis suggests that the clump is not sufficiently dense to be undergoing gravitational fragmentation and collapse.  If the sources identified by \citet{pattle2015} and \citet{ladjelate2020} represent distinct cores, rather than being transient features in an oscillating clump structure \citep{chitsazzadeh2014}, they are unlikely to be gravitationally unstable.

In the dense center of SMM-16, the magnetic field appears to be strong enough to provide significant support against gravitational collapse.  We posit that the misalignment between the magnetic field in the clump and that in the surrounding cloud means that, if material is accreted onto clumps along magnetic field lines, SMM-16 may not be able to efficiently acquire further mass from its surroundings, as even if the magnetic field in the clump does remain connected to the larger-scale field, infalling material would have to lose a significant amount of momentum in order to flow along the twisted magnetic field lines onto the core.  SMM-16 may thus have been prevented from yet becoming sufficiently massive to form stars. 

\subsection{L1689B}

The magnetic field which we infer from POL-2 data in L1689B runs approximately N/S, perpendicular to the filament in which the core is embedded, and approximately parallel to the $Planck$-scale magnetic field.

L1689B is a candidate gravitationally bound prestellar core.  Its stability and age remain uncertain, with a number of studies identifying the core as contracting or showing signs of infall \citep{redman2004, sohn2007, lee2011, seo2013}, while \citet{schnee2013} found it to be static.  \citet{redman2002} found L1689B to be relatively long-lived, with an age of at least one freefall time inferred from CO freezeout in the core's center; however, \citet{lee2003} argued that the core is chemically young, with a lack of freezeout of HCO$+$ and also a potential lack of CO freezeout.
The trans-critical mass-to-flux ratio which we measure suggests a dynamically important magnetic field in the core.  While this does not provide information on the core's age, it does suggest magnetic support as a means by which the core could be long-lived \citep[e.g.][]{jessop2000}.

\citet{redman2004} inferred a SW/NE ($\sim 45^{\circ}$ E of N) rotation axis in L1689B, different from both the POL-2 and $Planck$ magnetic field directions.  This rotation axis is similar to the rotation axis in SMM-16, and the core is rotating in the same sense.  \citet{seo2013} found infall velocities in L1689B greater than could be caused by gravitational collapse, and so inferred that core collapse has been instigated by some sort of external perturbation, suggesting that turbulence or a sudden increase in external pressure might be responsible.

Our results suggest that the magnetic field is likely to be dynamically important in the evolution of the core, and that the core may be accreting material along magnetic field lines.  Despite this, we do not see any suggestion of an hourglass field morphology in the core \citep[cf.][]{fiedler1993}, but note that, due to the relatively low SNR of our observations of L1689B, we do not detect polarization in the periphery of the core, where such a field structure would be most apparent.

\subsection{Magnetic field orientation with respect to the L1689 cloud and its substructures}

We next consider the orientation of the magnetic field with respect to the L1689 cloud, the filaments embedded within the cloud, and the clumps/cores which we observe.

\subsubsection{Cloud/magnetic field alignment}

Molecular clouds typically have a large-scale magnetic field direction which is either parallel or perpendicular to the major axis of the molecular cloud \citep{li2013}.  In both NIR and $Planck$ observations, the magnetic field of L1689 is parallel to the cloud major axis \citep[See Figures~\ref{fig:finder}, \ref{fig:finding_chart};][]{li2013,planck2015}. 
\citet{li2017} further suggest that clouds formed parallel to their magnetic field direction have higher SFRs, as the field does not hinder gravitational fragmentation, taking Ophiuchus as an example of a parallel-field cloud.

$Planck$ observations of Ophiuchus have found that at lower column densities the magnetic field is parallel to density structures, with some hint of a transition to perpendicularity at high column densities \citep{planck2016a}.  \citet{soler2019} identified this transition as occurring at $N({\rm H}) \sim 10^{21.8}$\,cm$^{-2}$ in the L1688/L1689 region.  However, this study did not distinguish between L1688 and L1689, and we note that {the mass distribution of Ophiuchus is dominated by L1688, which has approximately 2.5 times the mass of L1689 \citep{loren1989a}.  While L1688 has a clearly-defined major axis \citep[e.g.][]{ladjelate2020}, L1689 has a more complex geometry, and its preferred orientation with respect to the large-scale magnetic field is less well-defined.}
\citet{planck2016a} identify an average magnetic field strength in the Ophiuchus molecular cloud of $B_{pos} \sim 13 - 25\,\mu$G, and a mass-to-flux ratio $\lambda \sim 0.2-0.4$, suggesting that on large scales, the cloud is significantly magnetically sub-critical.  {We note, however, that} interpretation of the large-scale magnetic field in Ophiuchus is complicated by feedback effects from the Sco OB2 association.  The large-scale dust emission and magnetic field structures of L1689 in particular are clearly bowed, indicative of the large-scale west-to-east influence on the cloud \citep[Figures~\ref{fig:finder}, \ref{fig:finding_chart},][]{vrba1976, loren1989a}.

\subsubsection{Filament/magnetic field alignment}

$Herschel$ observations \citep{arzoumanian2019,ladjelate2020} have shown filaments in Ophiuchus to be preferentially aligned either parallel or perpendicular to the large-scale NE/SW streamers \citep{loren1989a}, and so to the large-scale magnetic field direction (\citealt{vrba1976}, \citealt{planck2015}; see Figure~\ref{fig:finder}).  Figure~\ref{fig:finding_chart} shows that the filaments in which L1689N and L1689B are embedded are {part of, and approximately parallel to the overall direction of, the L1689/L1712 filamentary streamer and perpendicular to the $Planck$-scale magnetic field direction}, while SMM-16 is embedded in a filament which is approximately parallel to the {local} $Planck$-scale field and {perpendicular to the major axis of the L1689/L1712 filamentary streamer}.

\subsubsection{Filament stability}

\citet{arzoumanian2019} produced a catalogue of filamentary structures across all regions observed by the $Herschel$ Gould Belt Survey \citep{andre2010}.  We took their values for the masses per unit length and temperatures of the principal filaments on which our three clumps are located.  \citet{arzoumanian2019} find three principal filaments meeting in L1689N; two, running approximately north-south, meeting in SMM-16; and one running approximately east-west through L1689B\footnote{L1689N filament indices: 8, 10, 35; SMM-16 filament indices: 29, 54; L1689B filament index: 6, in the nomenclature of \citet{arzoumanian2019}.  Filament skeleton maps are available from the $Herschel$ Gould Belt Survey archive at \url{http://www.herschel.fr/cea/gouldbelt/en/}}.  These filaments are very similar to those found by \citet{ladjelate2020} in the same region, shown in Figure \ref{fig:finding_chart}.  

The \citet{ostriker1964} critical mass per unit length for a uniform, unmagnetized, isothermal filament is given by 
\begin{equation}
    \left(\frac{M}{L}\right)_{crit} = \frac{2 c_{s}^{2}}{G},
    \label{eq:crit_line_mass}
\end{equation}
where $c_{s}$ is the sound speed in the filament, $c_{s} = (k_{\textsc{b}}T/\mu m_{\textsc{h}})^{0.5}$.  \citet{arzoumanian2019} give integrated line masses ($M/L$) of 12.7, 20.6 and 36.8 \msun\,pc$^{-1}$ and temperatures of 16.1, 15.1 and 15.8\,K for the three filaments in L1689N.
Neglecting any contribution from non-thermal sources of support, these values result in critical line mass ratios $(M/L)/(M/L)_{crit}$ of 0.57, 1.00 and 1.70, respectively.  
Similarly, the two filaments which meet at SMM-16 have integrated line masses of 5.5 and 15.7 \msun\,pc$^{-1}$, and temperatures of 15.8 and 16.2\,K, resulting in thermal critical line masses of 0.25 and 0.72, respectively.
Finally, the filament containing L1689B has an integrated line mass of 19.1 \msun\,pc$^{-1}$ and a temperature of 15.5\,K, and so a critical line mass ratio of 0.90.  If we instead adopt a temperature of 12\,K for the filaments, consistent with the temperature which we assume in the clumps/cores, these ratios increase by a factor $\sim 2.2$.
\citet{arzoumanian2019} consider filaments with $0.5 < (M/L)/(M/L)_{crit}< 2$ to be marginally gravitationally unstable.
Thus, even in the absence of magnetic or turbulent support, the filaments of L1689 are only marginally unstable.
{Despite the demonstrable gravitational instability of L1689N, the marginal stability of SMM-16, and signs of infall in L1689B \citep[e.g.][]{lee2001}, all three clumps appear to have formed within, or at the junction of, filaments which are not themselves significantly unstable, perhaps suggesting that the fragmentation of the filaments is driven by turbulent processes (e.g. \citealt{clarke2016}).  Gravitationally bound cores have previously been found in filaments with $(M/L)<(M/L)_{crit}$ in the California molecular cloud \citep{zhang2020}, indicating that such a scenario is not unique to L1689.}

\subsubsection{Magnetic field alignment within clumps and cores}

In all three of the clumps/cores in L1689 which we observe, the {mean} plane-of-sky magnetic field direction observed with POL-2 is {significantly offset from} the plane-of-sky major axis of the filament in which the clump/core is embedded (see Figure~\ref{fig:angles}).

A number of recent theoretical and numerical studies have predicted that magnetic fields should turn to become perpendicular to magnetically super-critical filaments \citep[see][for a recent review]{hennebelle2019}.  \citet{seifried2020} find that magnetic fields which are initially parallel to filamentary structures at low densities become perpendicular at high densities, for initial magnetic field strengths $> 5$\,$\mu$G.  They identify this transition as occurring at $n\sim 10^{2}-10^{3}$\,cm$^{-3}$ (or $N\sim 10^{21}-10^{21.5}$\,cm$^{-2}$) and at $\lambda\sim 1$, with the mass distribution being magnetically sub-critical on large scales, and super-critical in the dense material.  This change of orientation is taken to be indicative of compressive motions, resulting from either converging flows or gravitational collapse, and of an initially dynamically important magnetic field which suppresses turbulent motions and promotes accretion of material and gravitational collapse \citep{soler2017,seifried2020}.  This picture is broadly consistent with what we see in L1689: a cloud-parallel magnetic field threading a magnetically sub-critical mass distribution at low column densities \citep{planck2016a}, which becomes {approximately perpendicular} to the high-column-density filaments in regions of potential gravitational instability.  However, it should be noted that while models predict super-critical mass-to-flux ratios in filaments with perpendicular fields, we infer trans-critical mass-to-flux ratios in all our clumps, despite their having densities $\gtrsim 10^{5}$\,cm$^{-3}$.

The behavior which we see in L1689 is consistent with other recent BISTRO survey observations of nearby star-forming regions, including Orion A \citep{wardthompson2017, pattle2017a}, IC5146 \citep{wang2019} and NGC 1333 \citep{doi2020}.  Magnetic fields have also been observed to be perpendicular to filaments in POL-2 observations of more distant massive infrared dark clouds (IRDCs), including G34.43+0.24 (\citealt{soam2019}; see also \citealt{tang2019}) and G035.39-00.33 \citep{liu2019}.  However, the proximity of L1689 allows us to examine the transition from cloud-parallel to filament-perpendicular fields at higher physical resolution than is possible elsewhere.

\subsection{Identifying the transition from sub- to super-critical dynamics}

The fact that filament directions in Ophiuchus are preferentially either parallel or perpendicular to the cloud-scale magnetic field direction, along with their lack of significant gravitational instability, suggests that the magnetic field may be dynamically important in filament formation.  However, the magnetic field direction in the dense clumps and cores which we observe appears 
{not to be consistently set by} cloud-scale magnetic field direction, suggesting that a transition from magnetically sub-critical to super-critical dynamics {may occur at size scales between those resolved by $Planck$ and those observable with POL-2.}

Although the direction of magnetic fields in our regions appears not to be determined by the cloud-scale field direction, the relative orientation of the cloud- and clump-scale fields may still influence their evolution.  The apparent similarity between cloud- and clump-scale magnetic field directions in L1689N and L1689B (if not a projection effect) suggests that material can accrete onto these regions along magnetic field lines more efficiently than is the case in SMM-16.

A key unanswered question is whether the magnetic field direction in the filaments in which the clumps/cores are embedded is perpendicular to the filament, filament-parallel, or inherited from the cloud-scale magnetic field direction.  At present, the best tool available for examining magnetic field behavior in such intermediate-scale low-surface-brightness structures and identifying the column density at which the transition from {super-critical to sub-critical dynamics} occurs is high-resolution NIR extinction polarimetry \citep[e.g.][]{kwon2015}.

\section{Summary}
\label{sec:summary}

In this paper we have presented JCMT POL-2 observations of the L1689 molecular cloud in Ophiuchus, specifically of the star-forming clump L1689N, the starless clump SMM-16, and the starless core L1689B.  Our key results are:
\begin{enumerate}

    \item We noted that revised distance estimates to the L1688 and L1689 molecular clouds suggest that the two clouds are located at approximately equal distances from the Sco OB2 association, feedback from which is thought to strongly influence the evolution of both clouds.
    
    \item L1689N has a linear magnetic field running approximately NE/SW, except on the position of the IRAS 16293-2422 protostellar system.  The NE/SW field shows some signs of {deviation from linearity in the periphery of the clump}, but is on average {approximately} perpendicular to major axis of the clump, and to the {local direction of the L1689/L1712 filamentary streamer, as} identified in $Herschel$ observations.  L1689N shows good agreement between magnetic field directions observed with $Planck$ and with POL-2.  SMM-16 has {an approximately linear} magnetic field running E/W in its center, which curves towards running N/S in its periphery.  The central E/W field is approximately perpendicular both to the field observed by $Planck$, and to the filament in which SMM-16 is embedded.  L1689B has a linear magnetic field running N/S, approximately parallel to the magnetic field direction observed with $Planck$, and perpendicular to the major axis of the {L1689/L1712 filamentary streamer,} in which it is embedded.
    
    \item L1689N has a plane-of-sky magnetic field strength $B_{{\rm pos}} = 366\pm209\,\mu$G, a trans-critical mass to flux ratio $\lambda = 0.79\pm 0.45$ ($\lambda = 1.52\pm 0.86$ including the mass of the IRAS 16293-2433 system), and sub-Alfv\'enic non-thermal motions with an Alfv\'en Mach number $\mathcal{M}_{A} = 0.23\pm 0.05$.  SMM-16 has a field strength $B_{{\rm pos}} = 284\pm 127\,\mu$G, a magnetically trans-critical mass-to-flux ratio $\lambda = 0.81\pm 0.34$, and sub-Alfv\'{e}nic non-thermal motions with $\mathcal{M}_{A} = 0.34\pm 0.06$. L1689B has a field strength $B_{{\rm pos}} = 72\pm60\,\mu$G, a trans-critical mass-to-flux ratio $\lambda = 0.95\pm 0.81$, and sub-Alfv\'{e}nic non-thermal motions with $\mathcal{M}_{A} = 0.47\pm 0.26$.  The uncertainties on $B_{\rm pos}$ and $\lambda$ are dominated by systematic uncertainty on dust opacity; their statistical uncertainties are $15$\%, $11$\% and $46$\% on L1689N, SMM-16 and L1689B respectively.
    
    \item We found that L1689N is sufficiently massive to be unstable to gravitational fragmentation and collapse only when the mass of the IRAS 16293-24322 protostellar system is accounted for, and is otherwise energetically in approximate equipartition.  In SMM-16 and L1689B the gravitational, magnetic and kinetic energies are comparable to one another.
    
    \item We found that dust grains in all three regions retain some degree of alignment with respect to the magnetic field at high column densities, suggesting that grain growth has occurred.  Grains appear to be similarly well-aligned in each region despite their differing star formation histories.
     
    \item In all three regions, the plane-of-sky magnetic field direction is, {in the clumps' centres, approximately perpendicular} to the plane-of-sky major axis of the filament in which the clump is embedded, appearing to be set by the direction of the filament rather than by the large-scale magnetic field direction, in keeping with predictions from recent numerical modelling.  However, the filaments in which the clumps are embedded are not themselves significantly gravitationally unstable, and appear to have formed in a magnetically sub-critical environment.  This suggests that a transition from magnetically sub-critical to super-critical gas dynamics may occur {on size scales between those resolved by $Planck$ and} those that we observe with POL-2.
    
    \item We propose that star formation in L1689N is, on large scales, magnetically-regulated, with consistent magnetic field morphology from cloud to clump scales, {and a clump major axis perpendicular to the magnetic field direction}.  The magnetic field configuration may allow unrestricted flow of material onto L1689N along magnetic field lines, potentially provoking gravitational collapse.  However, the fragmentation of the clump perpendicular to the large-scale field direction and the misalignment of the magnetic field in the IRAS 16293-2422 protostar with that of the larger clump suggest that the region is not magnetically regulated on all size scales.  Conversely, in SMM-16, the misalignment between the magnetic field in the clump and that in the surrounding cloud may prevent the clump from efficiently acquiring additional mass from its surroundings, potentially inhibiting gravitational collapse.

\end{enumerate}

\acknowledgments

K.P. and S.P.L. acknowledge support from the Ministry of Science and Technology (Taiwan) under grant No. 106-2119-M-007-021-MY3.  J.D.F. and D.J. are supported by the National Research Council of Canada and by a Natural Sciences and Engineering Research Council of Canada (NSERC) Discovery Grant.  T.H. acknowledges the support from the National Research Foundation of Korea (NRF) grants funded by the Korea government (MSIT) through the Mid-career Research Program (2019R1A2C1087045).  D.A. acknowledges support by FCT/MCTES through national funds (PIDDAC) by the grants UID/FIS/04434/2019 \& UIDB/04434/2020.  Y.D. was supported by Grants-in-Aid for Scientific Research (18H01250) from the Japan Society for the Promotion of Science (JSPS).  C.L.H.H. acknowledges the support of the NAOJ Fellowship as well as JSPS KAKENHI grants 18K13586 and 20K14527.  J.K. is supported by JSPS KAKENHI grant No. 19K14775.  W.K. was supported by the New Faculty Startup Fund from Seoul National University and by the Basic Science Research Program through the NRF (NRF-2016R1C1B2013642).  C.W.L. is supported by the Basic Science Research Program through the NRF funded by the Ministry of Education, Science and Technology (NRF-2019R1A2C1010851).  T.L. acknowledges support from the international partnership program of the Chinese Academy of Sciences through grant No.114231KYSB20200009 and the support from the National Natural Science Foundation of China (NSFC) through grant NSFC No.12073061.  M. Tamura is supported by JSPS KAKENHI grant Nos. 18H05442, 15H02063, and 22000005.  This research is partially supported by Grants-in-Aid for Scientific Researches from the Japan Society for Promotion of Science (KAKENHI 19H0193810). 
The James Clerk Maxwell Telescope is operated by the East Asian Observatory on behalf of The National Astronomical Observatory of Japan; Academia Sinica Institute of Astronomy and Astrophysics; the Korea Astronomy and Space Science Institute; Center for Astronomical Mega-Science (as well as the National Key R\&D Program of China with No. 2017YFA0402700). Additional funding support is provided by the Science and Technology Facilities Council of the United Kingdom and participating universities in the United Kingdom, Canada and Ireland.
The authors wish to recognize and acknowledge the very significant cultural role and reverence that the summit of Maunakea has always had within the indigenous Hawaiian community. We are most fortunate to have the opportunity to conduct observations from this mountain.

\vspace{5mm}
\facilities{James Clerk Maxwell Telescope (JCMT)}
\software{Starlink \citep{currie2014}, Astropy \citep{astropy, astropy2018}, Simbad \citep{simbad}}

\clearpage

\appendix

\section{Polarization properties}
\label{sec:appendix_pol}

The formulae used to determined derived polarization proerties are as follows: polarized intensity is given by
\begin{equation}
    P = \sqrt{Q^{2} + U^{2}},
    \label{eq:pi}
\end{equation}
and debiased polarized intensity,
\begin{equation}
    P_{db} = \sqrt{Q^{2} + U^{2} - \frac{1}{2}(V_{Q} + V_{U})},
\end{equation}
where $V_{Q}$ is the variance of $Q$ and $V_{U}$ is the variance of U.  We also take uncertainties on $Q$, $U$ and $I$ to be $\delta Q = \sqrt{V_{Q}}$, $\delta U = \sqrt{V_{U}}$ and $\delta I = \sqrt{V_{I}}$, respectively.  Polarization fraction is then given by
\begin{equation}
    p = \frac{P}{I},
\end{equation}
and debiased polarization fraction by
\begin{equation}
    p_{db} = \frac{P_{db}}{I}.
\end{equation}
Uncertainties on polarization fraction (both debiased and non-debiased) are given by
\begin{equation}
    \delta p = \sqrt{\frac{Q^{2}\delta Q^{2}+U^{2}\delta U^{2}}{I^{2}(Q^{2}+U^{2})} + \frac{\delta I^{2}(Q^{2}+U^{2})}{I^{4}}}.
\end{equation}
Polarization angle is given by
\begin{equation}
    \theta_{p} = \frac{1}{2}\arctan\left(\frac{U}{Q}\right),
\end{equation}
and uncertainty on polarization angle by
\begin{equation}
    \delta\theta_{p} = \frac{1}{2}\frac{\sqrt{Q^{2}\delta U^{2} + U^{2}\delta Q^{2}}}{Q^{2}+U^{2}}.
\end{equation}

\section{Davis-Chandrasekhar-Fermi Analysis}
\label{sec:appendix_dcf}

In this appendix we describe how we arrived at the values of gas density, magnetic field angle dispersion and non-thermal velocity dispersion which we used in the Davis-Chandrasekhar-Fermi (DCF) analysis described in Section~\ref{sec:mag_fields}.

\subsection{Gas density}

We estimated the size of each of the clumps by fitting a 2D Gaussian distribution to the emission associated with the clump.  In the case of L1689N, we masked emission in a
the 40$^{\prime\prime}$-diameter region surrounding the IRAS 16293-2422 protostar, as we aim to investigate the magnetic field strength in the larger clump, rather than the behavior of IRAS 16293-2422 itself, the magnetic field of which appears to be behaving quite differently to that in the clump in which it is embedded.  The best-fit Gaussian distributions are listed in Table~\ref{tab:gaussians}.
The 1-FWHM contours of the fitted ellipses are shown in Figure~\ref{fig:ellipses}.

We calculated gas mass $M(F_{\nu}, D, \kappa_{\nu},T)$ using the \citet{hildebrand1983} relation (equation~\ref{eq:mass_main}).
We determined $F_{\nu}$ by summing all emission within the aperture defined by the 1-FWHM-diameter ellipse for each clump, as defined in Table~\ref{tab:gaussians}.
We filled the masked pixels covering IRAS 16293-2422 with the mean flux density in the pixels bordering the masked region (0.0793 Jy/pixel).
We took $D = 144.2$\,pc \citep{ortizleon2018}, and $\kappa_{\nu}=0.0125$\,cm$^{2}$g$^{-1}$ \citep[e.g.][]{johnstone2017}.  We estimated gas temperatures in L1689N and SMM-16 by averaging the dust temperatures determined from spectral energy distribution (SED) fitting of $Herschel$ and SCUBA-2 observations in each region by \citet{pattle2015}, assuming that the gas and dust are well-coupled.  In L1689N we averaged the temperatures of their sources SMM 19, 22, 23, 24, 25 and 26 (excluding SMM 20, their identifier for IRAS 16293-2422) to get a mean temperature of $11.7\pm1.4$\,K.  In SMM-16 we averaged the temperatures of their sources SMM 16a, 16b and 16c to get a mean temperature of $12.2 \pm 0.4$\,K.  Thus for simplicity we took $T=12$\,K in both clumps, with uncertainties as given above.  We also took $T=12$\,K for L1689B, following \citet{redman2002}.

The uncertainty on our mass estimates is dominated by the systematic uncertainty on dust opacity.  We conservatively took our value of $\kappa_{\nu}$ to be accurate to $\sim$50\% \citep{roy2014}.  We note that this results in a significantly larger mass uncertainty than is usual; however, we adopt this value in order to demonstrate the plausible range of magnetic field values associated with our measurements.  We further took the uncertainty on $F_\nu$ to be dominated by the SCUBA-2 850$\mu$m calibration uncertainty of 10\% \citep{dempsey2013}, and assumed a representative uncertainty of $\pm 1.5$\,K on $T$.  We defined a representative source size $R$ as described in equation~\ref{eq:R}
and thus calculated average column and volume densities using equations~\ref{eq:coldens_main} nad \ref{eq:density_main} respectively.  The flux densities, masses and column and volume densities determined for each clump are listed in Table~\ref{tab:properties}.

\subsection{Angle dispersion}

\begin{figure}
    \centering
    \includegraphics[width=0.47\textwidth]{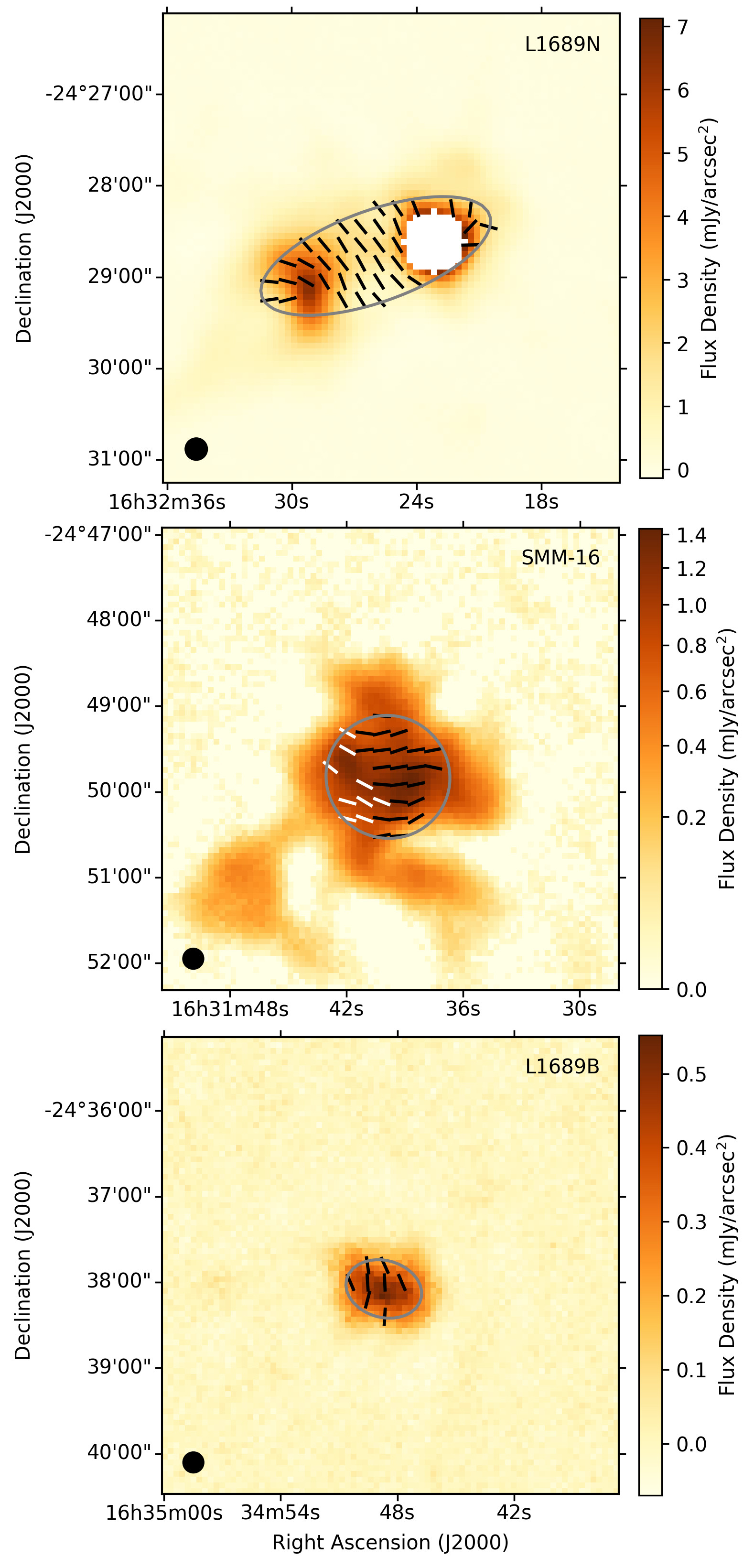}
    \caption{Areas over which magnetic field strengths were estimated, enclosed by 1-FWHM contours of ellipses fitted to Stokes I emission, in L1689N (top), SMM-16 (middle) and L1689B (bottom).  Masked pixels in L1689N cover the IRAS 16293-2422 protostellar system.  White half-vectors in SMM-16 indicate the area of ordered deviation identified as the `East' Gaussian component.}
    \label{fig:ellipses}
\end{figure}

\begin{figure}
    \centering
    \includegraphics[width=0.47\textwidth]{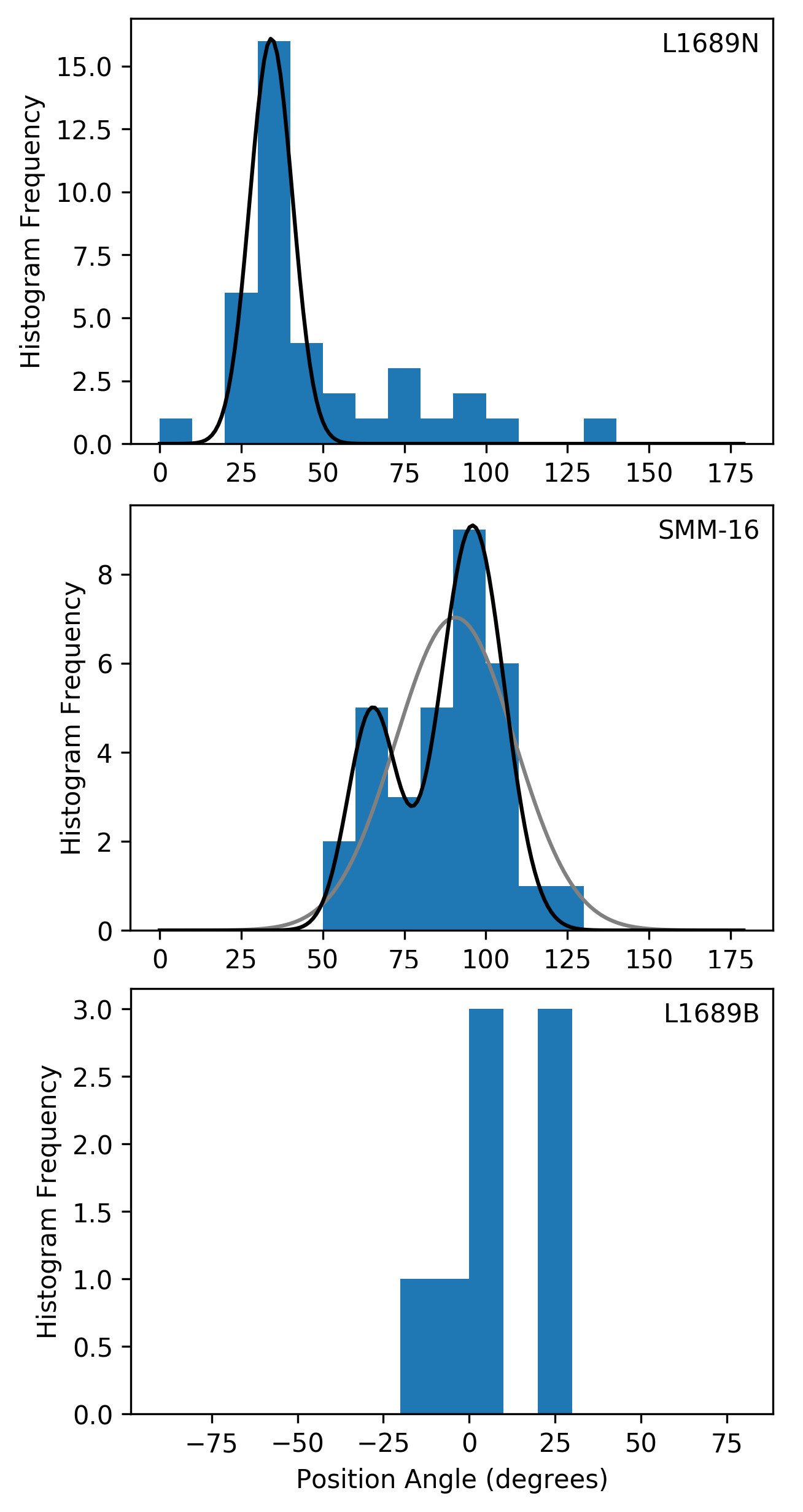}
    \caption{Histograms of magnetic field angle in L1689N (top), SMM-16 (middle) and L1689B (bottom), within areas enclosed by fitted ellipses.  Position angles are in degrees E of N, in the range $0 \leq \theta < 180^{\circ}$ for L1689N and SMM-16, and in the range $-90 \leq \theta < 90^{\circ}$ for L1689B.  In L1689N, the best-fit Gaussian model is shown in black.  In SMM-16, the best-fit single-Gaussian model is shown in grey, and the best-fit double Gaussian model is shown in black.  L1689B does not have enough half-vectors to accurately fit a Gaussian model.}
    \label{fig:histograms}
\end{figure}

The dispersion in angle $\sigma_{\theta}$ in the DCF relation measures intrinsic dispersion caused by Alfv\'enic distortion of the magnetic field by non-thermal motions.  However, the observed distribution of magnetic field angles also contains contributions from large-scale field structure and from measurement uncertainties.  Various methods have been proposed to account for ordered variation in magnetic field (see \citealt{pattle2019} for a recent review). Here, we choose to measure the dispersion in magnetic field position angle by fitting a Gaussian distribution to the distributions of position angles.  This method is suitable when a clearly identifiable and dominant linear magnetic field component exists -- here, the north-east/south-west component in L1689N, the east/west component in SMM-16, {and the north/south component in L1689B.}

We selected half-vectors where $p_{db}/\delta p > 3$, $I/\delta I > 5$ and $\delta \theta < 10^{\circ}$.  For self-consistency, we restricted our sample to those half-vectors contained within 1 FWHM of the fitted Gaussian over which we estimated volume density.  In L1689N, we excluded half-vectors coincident with the masked region around the position of IRAS 16293-2422.  The half-vectors included in our CF analysis are shown in Figure~\ref{fig:ellipses}.

For L1689N and SMM-16, we fitted the histogram of position angles with a Gaussian function, as shown in Figure~\ref{fig:histograms}.  The standard deviation of the fitted Gaussian was then taken to be the dispersion in angle $\sigma_{\theta}$.  We found that L1689N was well-fitted by a single Gaussian, with $\sigma_{\theta} = 6.6^{\circ}\pm 0.6^{\circ}$ and a mean value of $\langle\theta\rangle = 34.2^{\circ} \pm 0.8^{\circ}$.  As SMM-16 shows ordered variation on the eastern side of the {clump}, we fitted both single- and double-Gaussian models.  For the single-Gaussian model, we found $\sigma_{\theta}=18.2^{\circ}\pm 6.7^{\circ}$ ($\langle\theta\rangle = 90.6^{\circ} \pm 2.6^{\circ}$), whereas for the central component of the double-Gaussian model, we found $\sigma_{\theta} = 9.7^{\circ}\pm 0.1^{\circ}$ ($\langle\theta\rangle = 95.9^{\circ} \pm 0.4^{\circ}$), and for the eastern component, $\sigma_{\theta} = 7.5^{\circ}\pm 0.5^{\circ}$ ($\langle\theta\rangle = 65.2^{\circ} \pm 0.7^{\circ}$).
The half-vectors covered by the two Gaussian components are marked on Figure~\ref{fig:ellipses}, where black half-vectors indicate the central component.  As the two Gaussian components are spatially distinct, we consider the two-Gaussian fit to produce dispersion values more representative of the underlying dispersion in angle, and so take $\sigma_{\theta} = 9.7^{\circ}\pm 0.1^{\circ}$ for SMM-16.
L1689B does not have enough well-characterized half-vectors to accurately fit a Gaussian model, and so we instead list the mean and standard deviation of the position angles of the half-vectors shown in Figure~\ref{fig:ellipses} in Table~\ref{tab:properties}.  We took the uncertainty on the measured standard deviation to be $\sigma_{\theta}/\sqrt{N}$, where $N$ is the number of half-vectors included, and so find $\sigma_{\theta} = 13.3^{\circ} \pm 4.7^{\circ}$.

The effect of measurement uncertainty on measured angular dispersion is small while measurement uncertainties are less than or similar to the intrinsic dispersion \citep{pattle2017a}.  We further mitigated the effect of measurement uncertainties by choosing histogram bins of 10$^{\circ}$, larger than the maximum allowed uncertainty in angle.  In L1689N, the mean measurement uncertainty was 3.4$^{\circ}$, the median, 3.1$^{\circ}$, and the maximum, 8.6$^{\circ}$, suggesting that our measured dispersion $\sigma_{\theta} = 6.6^{\circ}\pm 0.6^{\circ}$ does not need to be corrected for the effects of measurement uncertainty.  Similarly, in SMM-16, the mean measurement uncertainty was 5.8$^{\circ}$, the median, 5.5$^{\circ}$, and the maximum, 9.7$^{\circ}$, again suggesting that our measured dispersion, $\sigma_{\theta} = 9.7^{\circ}\pm 0.1^{\circ}$, does not require correction.

\subsection{Gas velocity dispersion}

We took gas velocity dispersions for {L1689N and SMM-16} from N$_{2}$H$^{+}$ $J=1\to 0$ measurements made by \citet{pan2017} using the Purple Mountain Observatory Delingha 13.7 m telescope, with a beam size of $\sim 60^{\prime\prime}$ at 3.22\,mm.  N$_{2}$H$^{+}$ is a dense gas tracer with a critical density of $10^{5}-10^{6}$\,cm$^{-3}$ \citep{pan2017}, consistent with the mean volume densities in our clumps.  We choose these measurements to ensure that both clumps are measured self-consistently, and because the PMO Delingha beam size is comparable to the size of the clumps, suggesting that the linewidths will be representative of the dense gas in each clump.  \citet{pan2017} list FWHM linewidths of $\Delta v = 0.62\pm 0.04$\,km\,s$^{-1}$ and $\Delta v = 0.59\pm 0.04$\,km\,s$^{-1}$ in L1689N and SMM-16 respectively (L1689NW and L1689W in their nomenclature).

We took the gas velocity dispersion of L1689B from N$_{2}$H$^{+}$ $J=1\to 0$ measurements made by \citet{lee2001} using the FCRAO 14m telescope, with a beam size of $\sim 52^{\prime\prime}$.  \citet{lee2001} list a FWHM linewidth of $\Delta v = 0.32\pm 0.03$\,km\,s$^{-1}$ for the source, averaged over 8 pointings.

We estimated the non-thermal component of these linewidths using the relation
\begin{equation}
    \Delta v_{\textsc{nt}} = \sqrt{(\Delta v)^{2} - 8\,\ln2\,\frac{k_{\textsc{b}}T}{29\,m_{\textsc{H}}}},
\end{equation}
continuing to take $T=12$\,K in each clump, and noting that N$_{2}$H$^{+}$ has a mass of 29\,m$_{\textsc{h}}$.  The correction for the thermal linewidth component is small.  We assume that the non-thermal velocity component represents random turbulent motion, rather than infall motions or other systemic velocity shifts within the clump.  The $\Delta v_{\textsc{nt}}$ values for each {clump} are listed in Table~\ref{tab:properties}.  

\section{Grain alignment in L1689} \label{sec:appendix_grains}

Observations of polarized dust emission typically show a power-law dependence, $p\propto I^{-\alpha}$, where $0\leq \alpha \leq 1$.  A steeper index (higher $\alpha$) indicates poorer grain alignment: $\alpha = 0$ indicates that grains are equally well aligned at all depths, while $\alpha = 1$ indicates either a total absence of aligned grains, or that all observed polarized emission is produced in a thin layer at the surface of the cloud \citep{pattle2019a}.

We applied the method described by \citet{pattle2019a} to {our observations of L1689}.  We assume that the underlying relationship between $p$ and $I$ can be parameterised as
\begin{equation}
    p = p_{\sigma_{QU}}\left(\frac{I}{\sigma_{QU}}\right)^{-\alpha}
    \label{eq:polfrac}
\end{equation}
where $p_{\sigma_{QU}}$ is the polarization fraction at the RMS noise level of the data $\sigma_{QU}$, and $\alpha$ is a power-law index in the range $0 \leq \alpha \leq 1$.  We take $\sigma_{QU}$ to be a directly measurable property of the data set, and aim to infer $p_{\sigma_{QU}}$ and $\alpha$.

In order to determine $\alpha$ and $p_{\sigma_{QU}}$ we  fitted the relationship between $I$ and \emph{non-debiased} observed polarization fraction $p^{\prime}$ with the mean of the Rice distribution, such that
\begin{equation}
    p^{\prime}(I) = \sqrt{\frac{\pi}{2}}\sigma_{p}(I)\mathcal{L}_{\frac{1}{2}}\left(-\frac{p(I)^{2}}{2\sigma_{p}(I)^{2}}\right)
    \label{eq:ricemean}
\end{equation}
where $\sigma_{p}$ is uncertainty on $p$ and $\mathcal{L}_{\frac{1}{2}}$ is a Laguerre polynomial of order $\frac{1}{2}$.

We took $p(I)$ to be as given in equation~\ref{eq:polfrac} and $\sigma_{p}(I)\approx \sigma_{QU}/I$ (see \citealt{pattle2019a} for a discussion of this approximation), and so equation~\ref{eq:ricemean} becomes
\begin{equation}
    p^{\prime}(I) = \sqrt{\frac{\pi}{2}}\left(\frac{I}{\sigma_{QU}}\right)^{-1}\mathcal{L}_{\frac{1}{2}}\left(-\frac{p_{\sigma_{QU}}^{2}}{2}\left(\frac{I}{\sigma_{QU}}\right)^{2(1-\alpha)}\right).
    \label{eq:rmfit}
\end{equation}

We restricted each data set to the central 3-arcminute diameter region over which exposure time is approximately constant and so RMS noise is approximately flat \citep{friberg2016}.  We then estimated an RMS noise value in our Stokes $Q$ and $U$ data,
\begin{equation}
    \sigma_{QU} = \frac{1}{2N}\sum_{i=1}^{N}\left(\sqrt{V_{Q,i}} + \sqrt{V_{U,i}}\right),
\end{equation}
where $N$ is the number of pixels in the central 3-arcminute-diameter region of the data set and $V_{Q}$ and $V_{U}$ are the variance values associated with each pixel, as determined in the data reduction process.  Our measured values of $\sigma_{QU}$ are listed in Table~\ref{tab:polfrac}.

We then fitted each data set using equation~\ref{eq:rmfit}.  The results of this fitting are listed in Table~\ref{tab:polfrac} and shown in Figure~\ref{fig:polfrac}.  We find that the fitting results for all regions agree with one another within error.  In L1689N we find $\alpha=0.55\pm0.03$ while in SMM-16 we find $\alpha=0.59\pm0.06$, {and in L1689B, $\alpha = 0.56\pm 0.09$}.  This suggests that in both regions, grains become less well-aligned with the magnetic field as density increases, but that some degree of alignment persists at the highest densities which we observe.  
 
In both {IRAS 16293 and SMM-16} the reduced-$\chi^{2}$ values of the best-fitting models are significantly smaller than those of the null-hypothesis behavior.  The deviation of the data from the null hypothesis behavior can also be clearly seen in Figure~\ref{fig:polfrac}.  However, in both regions, and particularly in L1689N, the reduced chi-squared values of the best-fit models are greater than unity.  This suggests that there is more variation in the $p-I$ relationship in L1689N than can be explained with this simple model alone.
 
In L1689B, which is significantly less bright than either L1689N or SMM-16, the reduced-$\chi^{2}$ value of the best-fitting model is somewhat smaller than that of the null hypothesis, but both are similar to unity.  While Figure~\ref{fig:polfrac} shows a clear deviation from the null hypothesis behavior in L1689N and SMM-16, such a deviation is less apparent, although still somewhat visible, in L1689B.  The level of grain alignment in L1689B is thus less well-characterised than in either of the other regions, but the region shows behavior consistent with some degree of grain alignment being retained to high densities.

 \begin{table*}
     \centering
     \begin{tabular}{c c c c c c c c}
        \hline
         & & & Null & \multicolumn{4}{c}{Ricean-mean model} \\ \cline{5-8}
        Region & $\sigma_{QU}$ (mJy/beam) & $N$ & $\chi^{2}/N$ & $\alpha$ & $p_{\sigma_{QU}}$ & $p_{100\,{\rm mJy\,beam}^{-1}}$ & $\chi^{2}/(N-2)$ \\
        \hline
         L1689N  & $0.83\pm0.06$ & 160 & 84.9 & $0.55\pm0.03$ & $0.48\pm0.11$ & $0.035\pm0.013$ & 19.9 \\
         SMM-16 & $0.81\pm 0.06$ & 160 & 6.0 & $0.59\pm0.06$ & $0.41\pm0.13$ & $0.024\pm0.015$ & 1.9 \\
         L1689B & $0.75\pm 0.05$ & 94 & 2.1 & $0.56\pm 0.09$ & $0.43 \pm 0.17$ & $0.028\pm 0.024$ & 1.0 \\
        \hline
     \end{tabular}
     \caption{Results of fitting Ricean-mean model to the $p-I$ relations of L1689N, SMM-16 and L1689B.}
     \label{tab:polfrac}
 \end{table*}


\begin{thebibliography}{}
\expandafter\ifx\csname natexlab\endcsname\relax\def\natexlab#1{#1}\fi
\providecommand{\url}[1]{\href{#1}{#1}}

\bibitem[{{Alves} {et~al.}(2014){Alves}, {Frau}, {Girart}, {Franco}, {Santos},
  \& {Wiesemeyer}}]{alves2014}
{Alves}, F.~O., {Frau}, P., {Girart}, J.~M., {et~al.} 2014, \aap, 569, L1

\bibitem[{{Alves} {et~al.}(2015){Alves}, {Frau}, {Girart}, {Franco}, {Santos},
  \& {Wiesemeyer}}]{alves2015}
---. 2015, \aap, 574, C4

\bibitem[{{Andersson} {et~al.}(2015){Andersson}, {Lazarian}, \&
  {Vaillancourt}}]{andersson2015}
{Andersson}, B.-G., {Lazarian}, A., \& {Vaillancourt}, J.~E. 2015, \araa, 53,
  501

\bibitem[{{Andr{\'e}} {et~al.}(2014){Andr{\'e}}, {Di Francesco},
  {Ward-Thompson}, {Inutsuka}, {Pudritz}, \& {Pineda}}]{andre2014}
{Andr{\'e}}, P., {Di Francesco}, J., {Ward-Thompson}, D., {et~al.} 2014,
  Protostars and Planets VI, 27

\bibitem[{{Andr{\'e}} {et~al.}(2010){Andr{\'e}}, {Men'shchikov}, {Bontemps},
  {K{\"o}nyves}, {Motte}, {Schneider}, {Didelon}, {Minier}, {Saraceno},
  {Ward-Thompson}, {Di Francesco}, {White}, {Molinari}, {Testi}, {Abergel},
  {Griffin}, {Henning}, {Royer}, {Mer{\'{\i}}n}, {Vavrek}, {Attard},
  {Arzoumanian}, {Wilson}, {Ade}, {Aussel}, {Baluteau}, {Benedettini},
  {Bernard}, {Blommaert}, {Cambr{\'e}sy}, {Cox}, {di Giorgio}, {Hargrave},
  {Hennemann}, {Huang}, {Kirk}, {Krause}, {Launhardt}, {Leeks}, {Le Pennec},
  {Li}, {Martin}, {Maury}, {Olofsson}, {Omont}, {Peretto}, {Pezzuto}, {Prusti},
  {Roussel}, {Russeil}, {Sauvage}, {Sibthorpe}, {Sicilia-Aguilar}, {Spinoglio},
  {Waelkens}, {Woodcraft}, \& {Zavagno}}]{andre2010}
{Andr{\'e}}, P., {Men'shchikov}, A., {Bontemps}, S., {et~al.} 2010, {\aap},
  518, {L102}

\bibitem[{{Arzoumanian} {et~al.}(2019){Arzoumanian}, {Andr{\'e}},
  {K{\"o}nyves}, {Palmeirim}, {Roy}, {Schneider}, {Benedettini}, {Didelon}, {Di
  Francesco}, {Kirk}, \& {Ladjelate}}]{arzoumanian2019}
{Arzoumanian}, D., {Andr{\'e}}, P., {K{\"o}nyves}, V., {et~al.} 2019, \aap,
  621, A42

\bibitem[{{Astropy Collaboration} {et~al.}(2013){Astropy Collaboration},
  {Robitaille}, {Tollerud}, {Greenfield}, {Droettboom}, {Bray}, {Aldcroft},
  {Davis}, {Ginsburg}, {Price-Whelan}, {Kerzendorf}, {Conley}, {Crighton},
  {Barbary}, {Muna}, {Ferguson}, {Grollier}, {Parikh}, {Nair}, {Unther},
  {Deil}, {Woillez}, {Conseil}, {Kramer}, {Turner}, {Singer}, {Fox}, {Weaver},
  {Zabalza}, {Edwards}, {Azalee Bostroem}, {Burke}, {Casey}, {Crawford},
  {Dencheva}, {Ely}, {Jenness}, {Labrie}, {Lian Lim}, {Pierfederici},
  {Pontzen}, {Ptak}, {Refsdal}, {Servillat}, \& {Streicher}}]{astropy}
{Astropy Collaboration}, {Robitaille}, T.~P., {Tollerud}, E.~J., {et~al.} 2013,
  \aap, 558, A33

\bibitem[{{Astropy Collaboration} {et~al.}(2018){Astropy Collaboration},
  {Price-Whelan}, {Sip{\H o}cz}, {G{\"u}nther}, {Lim}, {Crawford}, {Conseil},
  {Shupe}, {Craig}, {Dencheva}, {Ginsburg}, {VanderPlas}, {Bradley},
  {P{\'e}rez-Su{\'a}rez}, {de Val-Borro}, {Aldcroft}, {Cruz}, {Robitaille},
  {Tollerud}, {Ardelean}, {Babej}, {Bach}, {Bachetti}, {Bakanov}, {Bamford},
  {Barentsen}, {Barmby}, {Baumbach}, {Berry}, {Biscani}, {Boquien}, {Bostroem},
  {Bouma}, {Brammer}, {Bray}, {Breytenbach}, {Buddelmeijer}, {Burke},
  {Calderone}, {Cano Rodr{\'{\i}}guez}, {Cara}, {Cardoso}, {Cheedella},
  {Copin}, {Corrales}, {Crichton}, {D'Avella}, {Deil}, {Depagne}, {Dietrich},
  {Donath}, {Droettboom}, {Earl}, {Erben}, {Fabbro}, {Ferreira}, {Finethy},
  {Fox}, {Garrison}, {Gibbons}, {Goldstein}, {Gommers}, {Greco}, {Greenfield},
  {Groener}, {Grollier}, {Hagen}, {Hirst}, {Homeier}, {Horton}, {Hosseinzadeh},
  {Hu}, {Hunkeler}, {Ivezi{\'c}}, {Jain}, {Jenness}, {Kanarek}, {Kendrew},
  {Kern}, {Kerzendorf}, {Khvalko}, {King}, {Kirkby}, {Kulkarni}, {Kumar},
  {Lee}, {Lenz}, {Littlefair}, {Ma}, {Macleod}, {Mastropietro}, {McCully},
  {Montagnac}, {Morris}, {Mueller}, {Mumford}, {Muna}, {Murphy}, {Nelson},
  {Nguyen}, {Ninan}, {N{\"o}the}, {Ogaz}, {Oh}, {Parejko}, {Parley}, {Pascual},
  {Patil}, {Patil}, {Plunkett}, {Prochaska}, {Rastogi}, {Reddy Janga},
  {Sabater}, {Sakurikar}, {Seifert}, {Sherbert}, {Sherwood-Taylor}, {Shih},
  {Sick}, {Silbiger}, {Singanamalla}, {Singer}, {Sladen}, {Sooley},
  {Sornarajah}, {Streicher}, {Teuben}, {Thomas}, {Tremblay}, {Turner},
  {Terr{\'o}n}, {van Kerkwijk}, {de la Vega}, {Watkins}, {Weaver}, {Whitmore},
  {Woillez}, {Zabalza}, \& {Astropy Contributors}}]{astropy2018}
{Astropy Collaboration}, {Price-Whelan}, A.~M., {Sip{\H o}cz}, B.~M., {et~al.}
  2018, \aj, 156, 123

\bibitem[{{Bacmann} {et~al.}(2016){Bacmann}, {Garc{\'\i}a-Garc{\'\i}a}, \&
  {Faure}}]{bacmann2016}
{Bacmann}, A., {Garc{\'\i}a-Garc{\'\i}a}, E., \& {Faure}, A. 2016, \aap, 588,
  L8

\bibitem[{{Bethell} {et~al.}(2007){Bethell}, {Chepurnov}, {Lazarian}, \&
  {Kim}}]{bethell2007}
{Bethell}, T.~J., {Chepurnov}, A., {Lazarian}, A., \& {Kim}, J. 2007, \apj,
  663, 1055

\bibitem[{{Caselli} {et~al.}(2002){Caselli}, {Benson}, {Myers}, \&
  {Tafalla}}]{caselli2002a}
{Caselli}, P., {Benson}, P.~J., {Myers}, P.~C., \& {Tafalla}, M. 2002, \apj,
  572, 238

\bibitem[{{Chandler} {et~al.}(2005){Chandler}, {Brogan}, {Shirley}, \&
  {Loinard}}]{chandler2005}
{Chandler}, C.~J., {Brogan}, C.~L., {Shirley}, Y.~L., \& {Loinard}, L. 2005,
  \apj, 632, 371

\bibitem[{{Chandrasekhar} \& {Fermi}(1953)}]{chandrasekhar1953}
{Chandrasekhar}, S., \& {Fermi}, E. 1953, \apj, 118, 113

\bibitem[{{Chapin} {et~al.}(2013){Chapin}, {Berry}, {Gibb}, {Jenness}, {Scott},
  {Tilanus}, {Economou}, \& {Holland}}]{chapin2013}
{Chapin}, E.~L., {Berry}, D.~S., {Gibb}, A.~G., {et~al.} 2013, \mnras, 430,
  2545

\bibitem[{{Chitsazzadeh} {et~al.}(2014){Chitsazzadeh}, {Di Francesco},
  {Schnee}, {Friesen}, {Shimajiri}, {Langston}, {Sadavoy}, {Bourke}, {Keto},
  {Pineda}, {Takakuwa}, \& {Tatematsu}}]{chitsazzadeh2014}
{Chitsazzadeh}, S., {Di Francesco}, J., {Schnee}, S., {et~al.} 2014, \apj, 790,
  129

\bibitem[{{Clarke} {et~al.}(2016){Clarke}, {Whitworth}, \&
  {Hubber}}]{clarke2016}
{Clarke}, S.~D., {Whitworth}, A.~P., \& {Hubber}, D.~A. 2016, \mnras, 458, 319

\bibitem[{{Coud{\'e}} {et~al.}(2019){Coud{\'e}}, {Bastien}, {Houde}, {Sadavoy},
  {Friesen}, {Di Francesco}, {Johnstone}, {Mairs}, {Hasegawa}, {Kwon}, {Lai},
  {Qiu}, {Ward-Thompson}, {Berry}, {Chen}, {Fiege}, {Franzmann}, {Hatchell},
  {Lacaille}, {Matthews}, {Moriarty-Schieven}, {Pon}, {Andr{\'e}},
  {Arzoumanian}, {Aso}, {Byun}, {Chakali}, {Chen}, {Chen}, {Ching}, {Cho},
  {Choi}, {Chrysostomou}, {Chung}, {Doi}, {Drabek-Maunder}, {Dowell}, {Eyres},
  {Falle}, {Friberg}, {Fuller}, {Furuya}, {Gledhill}, {Graves}, {Greaves},
  {Griffin}, {Gu}, {Hayashi}, {Hoang}, {Holland }, {Inoue}, {Inutsuka},
  {Iwasaki}, {Jeong}, {Kanamori}, {Kataoka}, {Kang}, {Kang}, {Kang},
  {Kawabata}, {Kemper}, {Kim}, {Kim}, {Kim}, {Kim}, {Kim}, {Kim}, {Kirk},
  {Kobayashi}, {Koch}, {Kwon}, {Lee}, {Lee}, {Lee}, {Li}, {Li}, {Li}, {Liu},
  {Liu}, {Liu}, {Liu}, {van Loo}, {Lyo}, {Matsumura}, {Nagata}, {Nakamura},
  {Nakanishi}, {Ohashi}, {Onaka}, {Parsons}, {Pattle}, {Peretto}, {Pyo},
  {Qian}, {Rao}, {Rawlings}, {Retter}, {Richer}, {Rigby}, {Robitaille},
  {Saito}, {Savini}, {Scaife}, {Seta}, {Shinnaga}, {Soam}, {Tamura}, {Tang},
  {Tomisaka}, {Tsukamoto}, {Wang}, {Wang}, {Whitworth}, {Yen}, {Yoo}, {Yuan},
  {Zenko}, {Zhang}, {Zhang}, {Zhou}, \& {Zhu}}]{coude2019}
{Coud{\'e}}, S., {Bastien}, P., {Houde}, M., {et~al.} 2019, arXiv e-prints,
  arXiv:1904.07221

\bibitem[{{Crapsi} {et~al.}(2005){Crapsi}, {Caselli}, {Walmsley}, {Myers},
  {Tafalla}, {Lee}, \& {Bourke}}]{crapsi2005}
{Crapsi}, A., {Caselli}, P., {Walmsley}, C.~M., {et~al.} 2005, \apj, 619, 379

\bibitem[{{Crutcher} {et~al.}(2004){Crutcher}, {Nutter}, {Ward-Thompson}, \&
  {Kirk}}]{crutcher2004}
{Crutcher}, R.~M., {Nutter}, D.~J., {Ward-Thompson}, D., \& {Kirk}, J.~M. 2004,
  \apj, 600, 279

\bibitem[{{Crutcher} {et~al.}(2010){Crutcher}, {Wandelt}, {Heiles},
  {Falgarone}, \& {Troland}}]{crutcher2010}
{Crutcher}, R.~M., {Wandelt}, B., {Heiles}, C., {Falgarone}, E., \& {Troland},
  T.~H. 2010, \apj, 725, 466

\bibitem[{{Currie} {et~al.}(2014){Currie}, {Berry}, {Jenness}, {Gibb}, {Bell},
  \& {Draper}}]{currie2014}
{Currie}, M.~J., {Berry}, D.~S., {Jenness}, T., {et~al.} 2014, in Astronomical
  Society of the Pacific Conference Series, Vol. 485, Astronomical Data
  Analysis Software and Systems XXIII, ed. N.~{Manset} \& P.~{Forshay}, 391

\bibitem[{{Davis}(1951)}]{davis1951a}
{Davis}, L. 1951, Physical Review, 81, 890

\bibitem[{{Davis} \& {Greenstein}(1951)}]{davis1951}
{Davis}, Jr., L., \& {Greenstein}, J.~L. 1951, \apj, 114, 206

\bibitem[{{de Zeeuw} {et~al.}(1999){de Zeeuw}, {Hoogerwerf}, {de Bruijne},
  {Brown}, \& {Blaauw}}]{dezeeuw1999}
{de Zeeuw}, P.~T., {Hoogerwerf}, R., {de Bruijne}, J.~H.~J., {Brown}, A.~G.~A.,
  \& {Blaauw}, A. 1999, \aj, 117, 354

\bibitem[{{Dempsey} {et~al.}(2013){Dempsey}, {Friberg}, {Jenness}, {Tilanus},
  {Thomas}, {Holland}, {Bintley}, {Berry}, {Chapin}, {Chrysostomou}, {Davis},
  {Gibb}, {Parsons}, \& {Robson}}]{dempsey2013}
{Dempsey}, J.~T., {Friberg}, P., {Jenness}, T., {et~al.} 2013, \mnras, 430,
  2534

\bibitem[{{Doi} {et~al.}(2020){Doi}, {Hasegawa}, {Furuya}, {Coud'e}, {Hull},
  {Arzoumanian}, {Bastien}, {Chen}, {Francesco}, {Friesen}, {Houde},
  {Inutsuka}, {Mairs}, {Matsumura}, {Onaka}, {Sadavoy}, {Shimajiri}, {Tahani},
  {Tomisaka}, {Eswaraiah}, {Koch}, {Pattle}, {Lee}, {Tamura}, {Berry}, {Ching},
  {Jihye Hwang}, {Kwon}, {Soam}, {Jia-Wei Wang}, {Lai}, {Qiu}, {Ward-Thompson},
  {Byun}, {Chen}, {Chen}, {Chen}, {Cho}, {Choi}, {Choi}, {Chrysostomou},
  {Chung}, {Diep}, {Duan}, {Fanciullo}, {Fiege}, {Franzmann}, {Friberg},
  {Fuller}, {Gledhill}, {Graves}, {Greaves}, {Griffin}, {Qilao Gu}, {Han},
  {Hatchell}, {Hayashi}, {Hoang}, {Inoue}, {Iwasaki}, {Jeong}, {Johnstone},
  {Kanamori}, {Kang}, {Kang}, {Kang}, {Kataoka}, {Kawabata}, {Kemper}, {Kim},
  {Kim}, {Kim}, {Kim}, {Kim}, {Shinyoung Kim}, {Kirk}, {Kobayashi}, {Konyves},
  {Takayoshi Kusune}, {Kwon}, {Lacaille}, {Law}, {Lee}, {Hyeseung Lee}, {Lee},
  {Lee}, {Lee}, {Li}, {Li}, {Li}, {Liu}, {Junhao Liu}, {Liu}, {Liu}, {Looze},
  {Lyo}, {Matthews}, {Moriarty-Schieven}, {Nagata}, {Nakamura}, {Nakanishi},
  {Ohashi}, {Park}, {Parsons}, {Peretto}, {Pyo}, {Qian}, {Rao}, {Rawlings},
  {Retter}, {Richer}, {Rigby}, {Saito}, {Savini}, {Scaife}, {Seta}, {Shinnaga},
  {Tang}, {Tsukamoto}, {Viti}, {Wang}, {Whitworth}, {Yen}, {Yoo}, {Yuan},
  {Yun}, {Zenko}, {Zhang}, {Zhang}, {Zhang}, {Zhou}, {Zhu}, {Andr'e}, {Dowell},
  {Eyres}, {Falle}, {Loo}, \& {Robitaille}}]{doi2020}
{Doi}, Y., {Hasegawa}, T., {Furuya}, R.~S., {et~al.} 2020, arXiv e-prints,
  arXiv:2007.00176

\bibitem[{{Dolginov} \& {Mitrofanov}(1976)}]{dolginov1976}
{Dolginov}, A.~Z., \& {Mitrofanov}, I.~G. 1976, \apss, 43, 291

\bibitem[{{Draine}(2011)}]{draine2011}
{Draine}, B.~T. 2011, {Physics of the Interstellar and Intergalactic Medium}

\bibitem[{{Draine} \& {Weingartner}(1996)}]{draine1996}
{Draine}, B.~T., \& {Weingartner}, J.~C. 1996, \apj, 470, 551

\bibitem[{{Draine} {et~al.}(2007){Draine}, {Dale}, {Bendo}, {Gordon}, {Smith},
  {Armus}, {Engelbracht}, {Helou}, {Kennicutt}, {Li}, {Roussel}, {Walter},
  {Calzetti}, {Moustakas}, {Murphy}, {Rieke}, {Bot}, {Hollenbach}, {Sheth}, \&
  {Teplitz}}]{draine2007}
{Draine}, B.~T., {Dale}, D.~A., {Bendo}, G., {et~al.} 2007, \apj, 663, 866

\bibitem[{{Enoch} {et~al.}(2009){Enoch}, {Evans}, {Sargent}, \&
  {Glenn}}]{enoch2009}
{Enoch}, M.~L., {Evans}, II, N.~J., {Sargent}, A.~I., \& {Glenn}, J. 2009,
  \apj, 692, 973

\bibitem[{{Fiedler} \& {Mouschovias}(1993)}]{fiedler1993}
{Fiedler}, R.~A., \& {Mouschovias}, T.~C. 1993, \apj, 415, 680

\bibitem[{Friberg {et~al.}(2016)Friberg, Bastien, Berry, Savini, Graves, \&
  Pattle}]{friberg2016}
Friberg, P., Bastien, P., Berry, D., {et~al.} 2016, Proc. SPIE, 9914, 991403

\bibitem[{{Friberg} {et~al.}(2018){Friberg}, {Berry}, {Savini}, {Bintley},
  {Dempsey}, {Graves}, \& {Parsons}}]{friberg2018}
{Friberg}, P., {Berry}, D., {Savini}, G., {et~al.} 2018, in Society of
  Photo-Optical Instrumentation Engineers (SPIE) Conference Series, Vol. 10708,
  Millimeter, Submillimeter, and Far-Infrared Detectors and Instrumentation for
  Astronomy IX, 107083M

\bibitem[{{Girart} {et~al.}(2006){Girart}, {Rao}, \& {Marrone}}]{girart2006}
{Girart}, J.~M., {Rao}, R., \& {Marrone}, D.~P. 2006, Science, 313, 812

\bibitem[{{G{\'o}mez} {et~al.}(2018){G{\'o}mez}, {V{\'a}zquez-Semadeni}, \&
  {Zamora-Avil{\'e}s}}]{gomez2018}
{G{\'o}mez}, G.~C., {V{\'a}zquez-Semadeni}, E., \& {Zamora-Avil{\'e}s}, M.
  2018, \mnras, 480, 2939

\bibitem[{{Guillet} {et~al.}(2020){Guillet}, {Girart}, {Maury}, \&
  {Alves}}]{guillet2020}
{Guillet}, V., {Girart}, J.~M., {Maury}, A.~J., \& {Alves}, F.~O. 2020, \aap,
  634, L15

\bibitem[{{Heiles} \& {Troland}(2005)}]{heiles2005}
{Heiles}, C., \& {Troland}, T.~H. 2005, \apj, 624, 773

\bibitem[{{Heitsch} {et~al.}(2001){Heitsch}, {Zweibel}, {Mac Low}, {Li}, \&
  {Norman}}]{heitsch2001}
{Heitsch}, F., {Zweibel}, E.~G., {Mac Low}, M.-M., {Li}, P., \& {Norman}, M.~L.
  2001, \apj, 561, 800

\bibitem[{{Hennebelle} \& {Inutsuka}(2019)}]{hennebelle2019}
{Hennebelle}, P., \& {Inutsuka}, S.-i. 2019, Frontiers in Astronomy and Space
  Sciences, 6, 5

\bibitem[{Hildebrand(1983)}]{hildebrand1983}
Hildebrand, R.~H. 1983, {Q. Jl R. astr. Soc.}, 24, 267

\bibitem[{{Hoang} \& {Lazarian}(2008)}]{hoang2008}
{Hoang}, T., \& {Lazarian}, A. 2008, \mnras, 388, 117

\bibitem[{{Hull} \& {Zhang}(2019)}]{hull2019}
{Hull}, C. L.~H., \& {Zhang}, Q. 2019, Frontiers in Astronomy and Space
  Sciences, 6, 3

\bibitem[{{Hull} {et~al.}(2013){Hull}, {Plambeck}, {Bolatto}, {Bower},
  {Carpenter}, {Crutcher}, {Fiege}, {Franzmann}, {Hakobian}, {Heiles}, {Houde},
  {Hughes}, {Jameson}, {Kwon}, {Lamb}, {Looney}, {Matthews}, {Mundy}, {Pillai},
  {Pound}, {Stephens}, {Tobin}, {Vaillancourt}, {Volgenau}, \&
  {Wright}}]{hull2013}
{Hull}, C.~L.~H., {Plambeck}, R.~L., {Bolatto}, A.~D., {et~al.} 2013, \apj,
  768, 159

\bibitem[{{Jacobsen} {et~al.}(2018){Jacobsen}, {J{\o}rgensen}, {van der Wiel},
  {Calcutt}, {Bourke}, {Brinch}, {Coutens}, {Drozdovskaya}, {Kristensen},
  {M{\"u}ller}, \& {Wampfler}}]{jacobsen2018}
{Jacobsen}, S.~K., {J{\o}rgensen}, J.~K., {van der Wiel}, M.~H.~D., {et~al.}
  2018, \aap, 612, A72

\bibitem[{{Jessop} \& {Ward-Thompson}(2000)}]{jessop2000}
{Jessop}, N.~E., \& {Ward-Thompson}, D. 2000, \mnras, 311, 63

\bibitem[{{Johnstone} {et~al.}(2017){Johnstone}, {Ciccone}, {Kirk}, {Mairs},
  {Buckle}, {Berry}, {Broekhoven-Fiene}, {Currie}, {Hatchell}, {Jenness},
  {Mottram}, {Pattle}, {Tisi}, {Di Francesco}, {Hogerheijde}, {Ward-Thompson},
  {Bastien}, {Bresnahan}, {Butner}, {Chen}, {Chrysostomou}, {Coud{\'e}},
  {Davis}, {Drabek-Maunder}, {Duarte-Cabral}, {Fich}, {Fiege}, {Friberg},
  {Friesen}, {Fuller}, {Graves}, {Greaves}, {Gregson}, {Holland}, {Joncas},
  {Kirk}, {Knee}, {Marsh}, {Matthews}, {Moriarty-Schieven}, {Mowat}, {Nutter},
  {Pineda}, {Salji}, {Rawlings}, {Richer}, {Robertson}, {Rosolowsky}, {Rumble},
  {Sadavoy}, {Thomas}, {Tothill}, {Viti}, {White}, {Wouterloot}, {Yates}, \&
  {Zhu}}]{johnstone2017}
{Johnstone}, D., {Ciccone}, S., {Kirk}, H., {et~al.} 2017, \apj, 836, 132

\bibitem[{{Jones} {et~al.}(2015){Jones}, {Bagley}, {Krejny}, {Andersson}, \&
  {Bastien}}]{jones2015}
{Jones}, T.~J., {Bagley}, M., {Krejny}, M., {Andersson}, B.-G., \& {Bastien},
  P. 2015, \aj, 149, 31

\bibitem[{{J{\o}rgensen} {et~al.}(2016){J{\o}rgensen}, {van der Wiel},
  {Coutens}, {Lykke}, {M{\"u}ller}, {van Dishoeck}, {Calcutt}, {Bjerkeli},
  {Bourke}, {Drozdovskaya}, {Favre}, {Fayolle}, {Garrod}, {Jacobsen},
  {{\"O}berg}, {Persson}, \& {Wampfler}}]{jorgensen2016}
{J{\o}rgensen}, J.~K., {van der Wiel}, M.~H.~D., {Coutens}, A., {et~al.} 2016,
  \aap, 595, A117

\bibitem[{{Juvela} {et~al.}(2015){Juvela}, {Ristorcelli}, {Marshall},
  {Montillaud}, {Pelkonen}, {Ysard}, {McGehee}, {Paladini}, {Pagani},
  {Malinen}, {Rivera-Ingraham}, {Lef{\`e}vre}, {T{\'o}th}, {Montier},
  {Bernard}, \& {Martin}}]{juvela2015}
{Juvela}, M., {Ristorcelli}, I., {Marshall}, D.~J., {et~al.} 2015, \aap, 584,
  A93

\bibitem[{{Kataoka} {et~al.}(2015){Kataoka}, {Muto}, {Momose}, {Tsukagoshi},
  {Fukagawa}, {Shibai}, {Hanawa}, {Murakawa}, \& {Dullemond}}]{kataoka2015}
{Kataoka}, A., {Muto}, T., {Momose}, M., {et~al.} 2015, \apj, 809, 78

\bibitem[{{Kim} {et~al.}(2020){Kim}, {Lee}, {Gopinathan}, {Tafalla}, {Sohn},
  {Kim}, {Kim}, {Soam}, \& {Myers}}]{kim2020}
{Kim}, S., {Lee}, C.~W., {Gopinathan}, M., {et~al.} 2020, \apj, 891, 169

\bibitem[{{Kirchschlager} {et~al.}(2019){Kirchschlager}, {Bertrang}, \&
  {Flock}}]{kirchschlager2019}
{Kirchschlager}, F., {Bertrang}, G. H.~M., \& {Flock}, M. 2019, \mnras, 488,
  1211

\bibitem[{{Kirk} {et~al.}(2017){Kirk}, {Dunham}, {Di Francesco}, {Johnstone},
  {Offner}, {Sadavoy}, {Tobin}, {Arce}, {Bourke}, {Mairs}, {Myers}, {Pineda},
  {Schnee}, \& {Shirley}}]{kirk2017}
{Kirk}, H., {Dunham}, M.~M., {Di Francesco}, J., {et~al.} 2017, \apj, 838, 114

\bibitem[{Kirk {et~al.}(2007)Kirk, Ward-Thompson, \& Andr\'{e}}]{kirk2007}
Kirk, J.~M., Ward-Thompson, D., \& Andr\'{e}, P. 2007, \mnras, 375, 843

\bibitem[{{Kirk} {et~al.}(2006){Kirk}, {Ward-Thompson}, \&
  {Crutcher}}]{kirk2006}
{Kirk}, J.~M., {Ward-Thompson}, D., \& {Crutcher}, R.~M. 2006, \mnras, 369,
  1445

\bibitem[{{Ko} {et~al.}(2020){Ko}, {Liu}, {Lai}, {Ching}, {Rao}, \&
  {Girart}}]{ko2020}
{Ko}, C.-L., {Liu}, H.~B., {Lai}, S.-P., {et~al.} 2020, \apj, 889, 172

\bibitem[{{K{\"o}nyves} {et~al.}(2010){K{\"o}nyves}, {Andr{\'e}},
  {Men'shchikov}, {Schneider}, {Arzoumanian}, {Bontemps}, {Attard}, {Motte},
  {Didelon}, {Maury}, {Abergel}, {Ali}, {Baluteau}, {Bernard}, {Cambr{\'e}sy},
  {Cox}, {Di Francesco}, {di Giorgio}, {Griffin}, {Hargrave}, {Huang}, {Kirk},
  {Li}, {Martin}, {Minier}, {Molinari}, {Olofsson}, {Pezzuto}, {Russeil},
  {Roussel}, {Saraceno}, {Sauvage}, {Sibthorpe}, {Spinoglio}, {Testi},
  {Ward-Thompson}, {White}, {Wilson}, {Woodcraft}, \& {Zavagno}}]{konyves2010}
{K{\"o}nyves}, V., {Andr{\'e}}, P., {Men'shchikov}, A., {et~al.} 2010, \aap,
  518, L106

\bibitem[{{Krumholz} \& {Federrath}(2019)}]{krumholz2019}
{Krumholz}, M.~R., \& {Federrath}, C. 2019, Frontiers in Astronomy and Space
  Sciences, 6, 7

\bibitem[{{Kwon} {et~al.}(2015){Kwon}, {Tamura}, {Hough}, {Nakajima},
  {Nishiyama}, {Kusakabe}, {Nagata}, \& {Kandori}}]{kwon2015}
{Kwon}, J., {Tamura}, M., {Hough}, J.~H., {et~al.} 2015, \apjs, 220, 17

\bibitem[{{Kwon} {et~al.}(2018){Kwon}, {Doi}, {Tamura}, {Matsumura}, {Pattle},
  {Berry}, {Sadavoy}, {Matthews}, {Ward-Thompson}, {Hasegawa}, {Furuya}, {Pon},
  {Di Francesco}, {Arzoumanian}, {Hayashi}, {Kawabata}, {Onaka}, {Choi},
  {Kang}, {Hoang}, {Lee}, {Lee}, {Liu}, {Liu}, {Inutsuka}, {Eswaraiah},
  {Bastien}, {Kwon}, {Lai}, {Qiu}, {Coud{\'e}}, {Franzmann}, {Friberg},
  {Graves}, {Greaves}, {Houde}, {Johnstone}, {Kirk}, {Koch}, {Li}, {Parsons},
  {Rao}, {Rawlings}, {Shinnaga}, {van Loo}, {Aso}, {Byun}, {Chen}, {Chen},
  {Chen}, {Ching}, {Cho}, {Chrysostomou}, {Chung}, {Drabek-Maunder}, {Eyres},
  {Fiege}, {Friesen}, {Fuller}, {Gledhill}, {Griffin}, {Gu}, {Hatchell},
  {Holland}, {Inoue}, {Iwasaki}, {Jeong}, {Kang}, {Kang}, {Kemper}, {Kim},
  {Kim}, {Kim}, {Kim}, {Kim}, {Kim}, {Lacaille}, {Lee}, {Li}, {Li}, {Liu},
  {Liu}, {Lyo}, {Mairs}, {Moriarty-Schieven}, {Nakamura}, {Nakanishi},
  {Ohashi}, {Peretto}, {Pyo}, {Qian}, {Retter}, {Richer}, {Rigby},
  {Robitaille}, {Savini}, {Scaife}, {Soam}, {Tang}, {Tomisaka}, {Wang}, {Wang},
  {Whitworth}, {Yen}, {Yoo}, {Yuan}, {Zhang}, {Zhang}, {Zhou}, {Zhu},
  {Andr{\'e}}, {Dowell}, {Falle}, {Tsukamoto}, {Nakagawa}, {Kanamori},
  {Kataoka}, {Kobayashi}, {Nagata}, {Saito}, {Seta}, \& {Zenko}}]{kwon2018}
{Kwon}, J., {Doi}, Y., {Tamura}, M., {et~al.} 2018, \apj, 859, 4

\bibitem[{{Ladjelate} {et~al.}(2020){Ladjelate}, {Andr{\'e}}, {K{\"o}nyves},
  {Ward-Thompson}, {Men'shchikov}, {Bracco}, {Palmeirim}, {Roy}, {Shimajiri},
  {Kirk}, {Arzoumanian}, {Benedettini}, {Di Francesco}, {Fiorellino},
  {Schneider}, \& {Pezzuto}}]{ladjelate2020}
{Ladjelate}, B., {Andr{\'e}}, P., {K{\"o}nyves}, V., {et~al.} 2020, arXiv
  e-prints, arXiv:2001.11036

\bibitem[{{Lazarian} \& {Hoang}(2007)}]{lazarian2007}
{Lazarian}, A., \& {Hoang}, T. 2007, \mnras, 378, 910

\bibitem[{{Lee} \& {Myers}(2011)}]{lee2011}
{Lee}, C.~W., \& {Myers}, P.~C. 2011, \apj, 734, 60

\bibitem[{{Lee} {et~al.}(1999){Lee}, {Myers}, \& {Tafalla}}]{lee1999a}
{Lee}, C.~W., {Myers}, P.~C., \& {Tafalla}, M. 1999, \apj, 526, 788

\bibitem[{{Lee} {et~al.}(2001){Lee}, {Myers}, \& {Tafalla}}]{lee2001}
---. 2001, \apjs, 136, 703

\bibitem[{{Lee} {et~al.}(2020){Lee}, {Hoang}, {Le}, \& {Cho}}]{lee2020}
{Lee}, H., {Hoang}, T., {Le}, N., \& {Cho}, J. 2020, \apj, 896, 44

\bibitem[{{Lee} {et~al.}(2003){Lee}, {Evans}, {Shirley}, \&
  {Tatematsu}}]{lee2003}
{Lee}, J.-E., {Evans}, Neal~J., I., {Shirley}, Y.~L., \& {Tatematsu}, K. 2003,
  \apj, 583, 789

\bibitem[{{Li} {et~al.}(2013){Li}, {Fang}, {Henning}, \&
  {Kainulainen}}]{li2013}
{Li}, H.-b., {Fang}, M., {Henning}, T., \& {Kainulainen}, J. 2013, \mnras, 436,
  3707

\bibitem[{{Li} {et~al.}(2017){Li}, {Jiang}, {Fan}, {Gu}, \& {Zhang}}]{li2017}
{Li}, H.-B., {Jiang}, H., {Fan}, X., {Gu}, Q., \& {Zhang}, Y. 2017, Nature
  Astronomy, 1, 0158

\bibitem[{{Liu} {et~al.}(2019){Liu}, {Qiu}, {Berry}, {Di Francesco}, {Bastien},
  {Koch}, {Furuya}, {Kim}, {Coud{\'e}}, {Lee}, {Soam}, {Eswaraiah}, {Li},
  {Hwang}, {Lyo}, {Pattle}, {Hasegawa}, {Kwon}, {Lai}, {Ward-Thompson},
  {Ching}, {Chen}, {Gu}, {Li}, {Li}, {Liu}, {Qian}, {Wang}, {Yuan}, {Zhang},
  {Zhang}, {Zhang}, {Zhou}, {Zhu}, {Andr{\'e}}, {Arzoumanian}, {Aso}, {Byun},
  {Chen}, {Chen}, {Chen}, {Cho}, {Choi}, {Chrysostomou}, {Chung}, {Doi},
  {Drabek-Maunder}, {Dowell}, {Eyres}, {Falle}, {Fanciullo}, {Fiege},
  {Franzmann}, {Friberg}, {Friesen}, {Fuller}, {Gledhill}, {Graves}, {Greaves},
  {Griffin}, {Han}, {Hatchell}, {Hayashi}, {Hoang}, {Holland}, {Houde},
  {Inoue}, {Inutsuka}, {Iwasaki}, {Jeong}, {Johnstone}, {Kanamori}, {Kang},
  {Kang}, {Kang}, {Kataoka}, {Kawabata}, {Kemper}, {Kim}, {Kim}, {Kim}, {Kim},
  {Kim}, {Kirk}, {Kobayashi}, {Kusune}, {Kwon}, {Lacaille}, {Lee}, {Lee},
  {Lee}, {Lee}, {Liu}, {Liu}, {van Loo}, {Mairs}, {Matsumura}, {Matthews},
  {Moriarty-Schieven}, {Nagata}, {Nakamura}, {Nakanishi}, {Ohashi}, {Onaka},
  {Parker}, {Parsons}, {Pascale}, {Peretto}, {Pon}, {Pyo}, {Rao}, {Rawlings},
  {Retter}, {Richer}, {Rigby}, {Robitaille}, {Sadavoy}, {Saito}, {Savini},
  {Scaife}, {Seta}, {Shinnaga}, {Tamura}, {Tang}, {Tomisaka}, {Tsukamoto},
  {Wang}, {Whitworth}, {Yen}, {Yoo}, \& {Zenko}}]{liu2019}
{Liu}, J., {Qiu}, K., {Berry}, D., {et~al.} 2019, arXiv e-prints,
  arXiv:1902.07734

\bibitem[{Loren(1989)}]{loren1989a}
Loren, R.~B. 1989, \apj, 338, 902

\bibitem[{{Loren} {et~al.}(1990){Loren}, {Wootten}, \& {Wilking}}]{loren1990}
{Loren}, R.~B., {Wootten}, A., \& {Wilking}, B.~A. 1990, \apj, 365, 269

\bibitem[{{Lynds}(1962)}]{lynds1962}
{Lynds}, B.~T. 1962, \apjs, 7, 1

\bibitem[{{Mairs} {et~al.}(2015){Mairs}, {Johnstone}, {Kirk}, {Graves},
  {Buckle}, {Beaulieu}, {Berry}, {Broekhoven-Fiene}, {Currie}, {Fich},
  {Hatchell}, {Jenness}, {Mottram}, {Nutter}, {Pattle}, {Pineda}, {Salji},
  {Francesco}, {Hogerheijde}, {Ward-Thompson}, \& {JCMT Gould Belt survey
  Team}}]{mairs2015}
{Mairs}, S., {Johnstone}, D., {Kirk}, H., {et~al.} 2015, \mnras, 454, 2557

\bibitem[{{Mairs} {et~al.}(2017){Mairs}, {Lane}, {Johnstone}, {Kirk},
  {Lacaille}, {Bower}, {Bell}, {Graves}, {Chapman}, \& {JCMT Transient
  Team}}]{mairs2017}
{Mairs}, S., {Lane}, J., {Johnstone}, D., {et~al.} 2017, \apj, 843, 55

\bibitem[{{Mamajek}(2008)}]{mamajek2008}
{Mamajek}, E.~E. 2008, Astronomische Nachrichten, 329, 10

\bibitem[{{Mathis} {et~al.}(1983){Mathis}, {Mezger}, \& {Panagia}}]{mathis1983}
{Mathis}, J.~S., {Mezger}, P.~G., \& {Panagia}, N. 1983, \aap, 500, 259

\bibitem[{{Mathis} {et~al.}(1977){Mathis}, {Rumpl}, \&
  {Nordsieck}}]{mathis1977}
{Mathis}, J.~S., {Rumpl}, W., \& {Nordsieck}, K.~H. 1977, \apj, 217, 425

\bibitem[{{Mizuno} {et~al.}(1990){Mizuno}, {Fukui}, {Iwata}, {Nozawa}, \&
  {Takano}}]{mizuno1990}
{Mizuno}, A., {Fukui}, Y., {Iwata}, T., {Nozawa}, S., \& {Takano}, T. 1990,
  \apj, 356, 184

\bibitem[{{Mouschovias} \& {Spitzer}(1976)}]{mouschovias1976}
{Mouschovias}, T.~C., \& {Spitzer}, L., J. 1976, \apj, 210, 326

\bibitem[{{Mundy} {et~al.}(1992){Mundy}, {Wootten}, {Wilking}, {Blake}, \&
  {Sargent}}]{mundy1992}
{Mundy}, L.~G., {Wootten}, A., {Wilking}, B.~A., {Blake}, G.~A., \& {Sargent},
  A.~I. 1992, \apj, 385, 306

\bibitem[{{Nutter} {et~al.}(2006){Nutter}, {Ward-Thompson}, \&
  {Andr{\'e}}}]{nutter2006}
{Nutter}, D., {Ward-Thompson}, D., \& {Andr{\'e}}, P. 2006, \mnras, 368, 1833

\bibitem[{{Ortiz-Le{\'o}n} {et~al.}(2018){Ortiz-Le{\'o}n}, {Loinard}, {Dzib},
  {Kounkel}, {Galli}, {Tobin}, {Evans}, {Hartmann}, {Rodr{\'{\i}}guez},
  {Brice{\~n}o}, {Torres}, \& {Mioduszewski}}]{ortizleon2018}
{Ortiz-Le{\'o}n}, G.~N., {Loinard}, L., {Dzib}, S.~A., {et~al.} 2018, \apjl,
  869, L33

\bibitem[{{Ostriker} {et~al.}(2001){Ostriker}, {Stone}, \&
  {Gammie}}]{ostriker2001}
{Ostriker}, E.~C., {Stone}, J.~M., \& {Gammie}, C.~F. 2001, \apj, 546, 980

\bibitem[{{Ostriker}(1964)}]{ostriker1964}
{Ostriker}, J. 1964, \apj, 140, 1056

\bibitem[{{Palmeirim} {et~al.}(2013){Palmeirim}, {Andr{\'e}}, {Kirk},
  {Ward-Thompson}, {Arzoumanian}, {K{\"o}nyves}, {Didelon}, {Schneider},
  {Benedettini}, {Bontemps}, {Di Francesco}, {Elia}, {Griffin}, {Hennemann},
  {Hill}, {Martin}, {Men'shchikov}, {Molinari}, {Motte}, {Nguyen Luong},
  {Nutter}, {Peretto}, {Pezzuto}, {Roy}, {Rygl}, {Spinoglio}, \&
  {White}}]{palmeirim2013}
{Palmeirim}, P., {Andr{\'e}}, P., {Kirk}, J., {et~al.} 2013, \aap, 550, A38

\bibitem[{{Pan} {et~al.}(2017){Pan}, {Li}, {Chang}, {Qian}, {Bergin}, \&
  {Wang}}]{pan2017}
{Pan}, Z., {Li}, D., {Chang}, Q., {et~al.} 2017, \apj, 836, 194

\bibitem[{{Pattle} \& {Fissel}(2019)}]{pattle2019}
{Pattle}, K., \& {Fissel}, L. 2019, Frontiers in Astronomy and Space Sciences,
  6, 15

\bibitem[{{Pattle} {et~al.}(2015){Pattle}, {Ward-Thompson}, {Kirk}, {White},
  {Drabek-Maunder}, {Buckle}, {Beaulieu}, {Berry}, {Broekhoven-Fiene},
  {Currie}, {Fich}, {Hatchell}, {Kirk}, {Jenness}, {Johnstone}, {Mottram},
  {Nutter}, {Pineda}, {Quinn}, {Salji}, {Tisi}, {Walker-Smith}, {Francesco},
  {Hogerheijde}, {Andr{\'e}}, {Bastien}, {Bresnahan}, {Butner}, {Chen},
  {Chrysostomou}, {Coude}, {Davis}, {Duarte-Cabral}, {Fiege}, {Friberg},
  {Friesen}, {Fuller}, {Graves}, {Greaves}, {Gregson}, {Griffin}, {Holland},
  {Joncas}, {Knee}, {K{\"o}nyves}, {Mairs}, {Marsh}, {Matthews},
  {Moriarty-Schieven}, {Rawlings}, {Richer}, {Robertson}, {Rosolowsky},
  {Rumble}, {Sadavoy}, {Spinoglio}, {Thomas}, {Tothill}, {Viti}, {Wouterloot},
  {Yates}, \& {Zhu}}]{pattle2015}
{Pattle}, K., {Ward-Thompson}, D., {Kirk}, J.~M., {et~al.} 2015, \mnras, 450,
  1094

\bibitem[{{Pattle} {et~al.}(2017){Pattle}, {Ward-Thompson}, {Berry},
  {Hatchell}, {Chen}, {Pon}, {Koch}, {Kwon}, {Kim}, {Bastien}, {Cho},
  {Coud{\'e}}, {Di Francesco}, {Fuller}, {Furuya}, {Graves}, {Johnstone},
  {Kirk}, {Kwon}, {Lee}, {Matthews}, {Mottram}, {Parsons}, {Sadavoy},
  {Shinnaga}, {Soam}, {Hasegawa}, {Lai}, {Qiu}, \& {Friberg}}]{pattle2017a}
{Pattle}, K., {Ward-Thompson}, D., {Berry}, D., {et~al.} 2017, \apj, 846, 122

\bibitem[{{Pattle} {et~al.}(2019){Pattle}, {Lai}, {Hasegawa}, {Wang}, {Furuya},
  {Ward-Thompson}, {Bastien}, {Coud{\'e}}, {Eswaraiah}, {Fanciullo}, {di
  Francesco}, {Hoang}, {Kim}, {Kwon}, {Lee}, {Liu}, {Liu}, {Matsumura},
  {Onaka}, {Sadavoy}, \& {Soam}}]{pattle2019a}
{Pattle}, K., {Lai}, S.-P., {Hasegawa}, T., {et~al.} 2019, \apj, 880, 27

\bibitem[{{Pineda} {et~al.}(2012){Pineda}, {Maury}, {Fuller}, {Testi},
  {Garc{\'\i}a-Appadoo}, {Peck}, {Villard}, {Corder}, {van Kempen}, {Turner},
  {Tachihara}, \& {Dent}}]{pineda2012}
{Pineda}, J.~E., {Maury}, A.~J., {Fuller}, G.~A., {et~al.} 2012, \aap, 544, L7

\bibitem[{{Planck Collaboration} {et~al.}(2015){Planck Collaboration}, {Ade},
  {Aghanim}, {Alina}, {Alves}, {Armitage-Caplan}, {Arnaud}, {Arzoumanian},
  {Ashdown}, {Atrio-Barandela}, \& et~al.}]{planck2015}
{Planck Collaboration}, {Ade}, P.~A.~R., {Aghanim}, N., {et~al.} 2015, \aap,
  576, A104

\bibitem[{{Planck Collaboration} {et~al.}(2016){Planck Collaboration}, {Ade},
  {Aghanim}, {Alves}, {Arnaud}, {Arzoumanian}, {Ashdown}, {Aumont},
  {Baccigalupi}, {Banday}, {Barreiro}, {Bartolo}, {Battaner}, {Benabed},
  {Beno{\^i}t}, {Benoit-L{\'e}vy}, {Bernard}, {Bersanelli}, {Bielewicz},
  {Bock}, {Bonavera}, {Bond}, {Borrill}, {Bouchet}, {Boulanger}, {Bracco},
  {Burigana}, {Calabrese}, {Cardoso}, {Catalano}, {Chiang}, {Christensen},
  {Colombo}, {Combet}, {Couchot}, {Crill}, {Curto}, {Cuttaia}, {Danese},
  {Davies}, {Davis}, {de Bernardis}, {de Rosa}, {de Zotti}, {Delabrouille},
  {Dickinson}, {Diego}, {Dole}, {Donzelli}, {Dor{\'e}}, {Douspis}, {Ducout},
  {Dupac}, {Efstathiou}, {Elsner}, {En{\ss}lin}, {Eriksen}, {Falceta-Gon{\c
  c}alves}, {Falgarone}, {Ferri{\`e}re}, {Finelli}, {Forni}, {Frailis},
  {Fraisse}, {Franceschi}, {Frejsel}, {Galeotta}, {Galli}, {Ganga}, {Ghosh},
  {Giard}, {Gjerl{\o}w}, {Gonz{\'a}lez-Nuevo}, {G{\'o}rski}, {Gregorio},
  {Gruppuso}, {Gudmundsson}, {Guillet}, {Harrison}, {Helou}, {Hennebelle},
  {Henrot-Versill{\'e}}, {Hern{\'a}ndez-Monteagudo}, {Herranz}, {Hildebrandt},
  {Hivon}, {Holmes}, {Hornstrup}, {Huffenberger}, {Hurier}, {Jaffe}, {Jaffe},
  {Jones}, {Juvela}, {Keih{\"a}nen}, {Keskitalo}, {Kisner}, {Knoche}, {Kunz},
  {Kurki-Suonio}, {Lagache}, {Lamarre}, {Lasenby}, {Lattanzi}, {Lawrence},
  {Leonardi}, {Levrier}, {Liguori}, {Lilje}, {Linden-V{\o}rnle},
  {L{\'o}pez-Caniego}, {Lubin}, {Mac{\'{\i}}as-P{\'e}rez}, {Maino},
  {Mandolesi}, {Mangilli}, {Maris}, {Martin}, {Mart{\'{\i}}nez-Gonz{\'a}lez},
  {Masi}, {Matarrese}, {Melchiorri}, {Mendes}, {Mennella}, {Migliaccio},
  {Miville-Desch{\^e}nes}, {Moneti}, {Montier}, {Morgante}, {Mortlock},
  {Munshi}, {Murphy}, {Naselsky}, {Nati}, {Netterfield}, {Noviello}, {Novikov},
  {Novikov}, {Oppermann}, {Oxborrow}, {Pagano}, {Pajot}, {Paladini},
  {Paoletti}, {Pasian}, {Perotto}, {Pettorino}, {Piacentini}, {Piat},
  {Pierpaoli}, {Pietrobon}, {Plaszczynski}, {Pointecouteau}, {Polenta},
  {Ponthieu}, {Pratt}, {Prunet}, {Puget}, {Rachen}, {Reinecke}, {Remazeilles},
  {Renault}, {Renzi}, {Ristorcelli}, {Rocha}, {Rossetti}, {Roudier},
  {Rubi{\~n}o-Mart{\'{\i}}n}, {Rusholme}, {Sandri}, {Santos}, {Savelainen},
  {Savini}, {Scott}, {Soler}, {Stolyarov}, {Sudiwala}, {Sutton}, {Suur-Uski},
  {Sygnet}, {Tauber}, {Terenzi}, {Toffolatti}, {Tomasi}, {Tristram}, {Tucci},
  {Umana}, {Valenziano}, {Valiviita}, {Van Tent}, {Vielva}, {Villa}, {Wade},
  {Wandelt}, {Wehus}, {Ysard}, {Yvon}, \& {Zonca}}]{planck2016a}
---. 2016, \aap, 586, A138

\bibitem[{{Rao} {et~al.}(2009){Rao}, {Girart}, {Marrone}, {Lai}, \&
  {Schnee}}]{rao2009}
{Rao}, R., {Girart}, J.~M., {Marrone}, D.~P., {Lai}, S.-P., \& {Schnee}, S.
  2009, \apj, 707, 921

\bibitem[{{Redman} {et~al.}(2004){Redman}, {Keto}, {Rawlings}, \&
  {Williams}}]{redman2004}
{Redman}, M.~P., {Keto}, E., {Rawlings}, J.~M.~C., \& {Williams}, D.~A. 2004,
  \mnras, 352, 1365

\bibitem[{{Redman} {et~al.}(2002){Redman}, {Rawlings}, {Nutter},
  {Ward-Thompson}, \& {Williams}}]{redman2002}
{Redman}, M.~P., {Rawlings}, J.~M.~C., {Nutter}, D.~J., {Ward-Thompson}, D., \&
  {Williams}, D.~A. 2002, \mnras, 337, L17

\bibitem[{{Ridge} {et~al.}(2006){Ridge}, {Di Francesco}, {Kirk}, {Li},
  {Goodman}, {Alves}, {Arce}, {Borkin}, {Caselli}, {Foster}, {Heyer},
  {Johnstone}, {Kosslyn}, {Lombardi}, {Pineda}, {Schnee}, \&
  {Tafalla}}]{ridge2006}
{Ridge}, N.~A., {Di Francesco}, J., {Kirk}, H., {et~al.} 2006, \aj, 131, 2921

\bibitem[{{Roy} {et~al.}(2013){Roy}, {Martin}, {Polychroni}, {Bontemps},
  {Abergel}, {Andr{\'e}}, {Arzoumanian}, {Di Francesco}, {Hill}, {Konyves},
  {Nguyen-Luong}, {Pezzuto}, {Schneider}, {Testi}, \& {White}}]{roy2013}
{Roy}, A., {Martin}, P.~G., {Polychroni}, D., {et~al.} 2013, \apj, 763, 55

\bibitem[{{Roy} {et~al.}(2014){Roy}, {Andr{\'e}}, {Palmeirim}, {Attard},
  {K{\"o}nyves}, {Schneider}, {Peretto}, {Men'shchikov}, {Ward-Thompson},
  {Kirk}, {Griffin}, {Marsh}, {Abergel}, {Arzoumanian}, {Benedettini}, {Hill},
  {Motte}, {Nguyen Luong}, {Pezzuto}, {Rivera-Ingraham}, {Roussel}, {Rygl},
  {Spinoglio}, {Stamatellos}, \& {White}}]{roy2014}
{Roy}, A., {Andr{\'e}}, P., {Palmeirim}, P., {et~al.} 2014, \aap, 562, A138

\bibitem[{{Sadavoy} {et~al.}(2010){Sadavoy}, {Di Francesco}, \&
  {Johnstone}}]{sadavoy2010a}
{Sadavoy}, S.~I., {Di Francesco}, J., \& {Johnstone}, D. 2010, \apjl, 718, L32

\bibitem[{{Sadavoy} {et~al.}(2018){Sadavoy}, {Myers}, {Stephens}, {Tobin},
  {Kwon}, {Segura-Cox}, {Henning}, {Commer{\c{c}}on}, \&
  {Looney}}]{sadavoy2018}
{Sadavoy}, S.~I., {Myers}, P.~C., {Stephens}, I.~W., {et~al.} 2018, \apj, 869,
  115

\bibitem[{{Sadavoy} {et~al.}(2019){Sadavoy}, {Stephens}, {Myers}, {Looney},
  {Tobin}, {Kwon}, {Commer{\c{c}}on}, {Segura-Cox}, {Henning}, \&
  {Hennebelle}}]{sadavoy2019}
{Sadavoy}, S.~I., {Stephens}, I.~W., {Myers}, P.~C., {et~al.} 2019, \apjs, 245,
  2

\bibitem[{{Santos} {et~al.}(2017){Santos}, {Ade}, {Angil{\`e}}, {Ashton},
  {Benton}, {Devlin}, {Dober}, {Fissel}, {Fukui}, {Galitzki}, {Gandilo},
  {Klein}, {Korotkov}, {Li}, {Martin}, {Matthews}, {Moncelsi}, {Nakamura},
  {Netterfield}, {Novak}, {Pascale}, {Poidevin}, {Savini}, {Scott}, {Shariff},
  {Diego Soler}, {Thomas}, {Tucker}, {Tucker}, \& {Ward-Thompson}}]{santos2017}
{Santos}, F.~P., {Ade}, P.~A.~R., {Angil{\`e}}, F.~E., {et~al.} 2017, \apj,
  837, 161

\bibitem[{{Santos} {et~al.}(2019){Santos}, {Chuss}, {Dowell}, {Houde},
  {Looney}, {Lopez Rodriguez}, {Novak}, {Ward-Thompson}, {Berthoud}, {Dale},
  {Guerra}, {Hamilton}, {Hanany}, {Harper}, {Henning}, {Jones}, {Lazarian},
  {Michail}, {Morris}, {Staguhn}, {Stephens}, {Tassis}, {Trinh}, {Van Camp},
  {Volpert}, \& {Wollack}}]{santos2019}
{Santos}, F.~P., {Chuss}, D.~T., {Dowell}, C.~D., {et~al.} 2019, \apj, 882, 113

\bibitem[{{Schirmer} {et~al.}(2020){Schirmer}, {Abergel}, {Verstraete},
  {Ysard}, {Juvela}, {Jones}, \& {Habart}}]{schirmer2020}
{Schirmer}, T., {Abergel}, A., {Verstraete}, L., {et~al.} 2020, \aap, 639, A144

\bibitem[{{Schlafly} \& {Finkbeiner}(2011)}]{schlafly2011}
{Schlafly}, E.~F., \& {Finkbeiner}, D.~P. 2011, \apj, 737, 103

\bibitem[{{Schlegel} {et~al.}(1998){Schlegel}, {Finkbeiner}, \&
  {Davis}}]{schlegel1998}
{Schlegel}, D.~J., {Finkbeiner}, D.~P., \& {Davis}, M. 1998, \apj, 500, 525

\bibitem[{{Schnee} {et~al.}(2013){Schnee}, {Brunetti}, {Di Francesco},
  {Caselli}, {Friesen}, {Johnstone}, \& {Pon}}]{schnee2013}
{Schnee}, S., {Brunetti}, N., {Di Francesco}, J., {et~al.} 2013, \apj, 777, 121

\bibitem[{{Seifried} {et~al.}(2020){Seifried}, {Walch}, {Weis}, {Reissl},
  {Soler}, {Klessen}, \& {Joshi}}]{seifried2020}
{Seifried}, D., {Walch}, S., {Weis}, M., {et~al.} 2020, arXiv e-prints,
  arXiv:2003.00017

\bibitem[{{Seo} {et~al.}(2013){Seo}, {Hong}, \& {Shirley}}]{seo2013}
{Seo}, Y.~M., {Hong}, S.~S., \& {Shirley}, Y.~L. 2013, \apj, 769, 50

\bibitem[{{Serkowski}(1962)}]{serkowski1962}
{Serkowski}, K. 1962, Advances in Astronomy and Astrophysics, 1, 289

\bibitem[{{Soam} {et~al.}(2018){Soam}, {Pattle}, {Ward-Thompson}, {Lee},
  {Sadavoy}, {Koch}, {Kim}, {Kwon}, {Kwon}, {Arzoumanian}, {Berry}, {Hoang},
  {Tamura}, {Lee}, {Liu}, {Kim}, {Johnstone}, {Nakamura}, {Lyo}, {Onaka},
  {Kim}, {Furuya}, {Hasegawa}, {Lai}, {Bastien}, {Chung}, {Kim}, {Parsons},
  {Rawlings}, {Mairs}, {Graves}, {Robitaille}, {Liu}, {Whitworth}, {Eswaraiah},
  {Rao}, {Yoo}, {Houde}, {Kang}, {Doi}, {Choi}, {Kang}, {Coud{\'e}}, {Li},
  {Matsumura}, {Matthews}, {Pon}, {Di Francesco}, {Hayashi}, {Kawabata},
  {Inutsuka}, {Qiu}, {Franzmann}, {Friberg}, {Greaves}, {Kirk}, {Li},
  {Shinnaga}, {van Loo}, {Aso}, {Byun}, {Chen}, {Chen}, {Chen}, {Ching}, {Cho},
  {Chrysostomou}, {Drabek-Maunder}, {Eyres}, {Fiege}, {Friesen}, {Fuller},
  {Gledhill}, {Griffin}, {Gu}, {Hatchell}, {Holland}, {Inoue}, {Iwasaki},
  {Jeong}, {Kang}, {Kemper}, {Kim}, {Kim}, {Lacaille}, {Lee}, {Li}, {Liu},
  {Liu}, {Moriarty-Schieven}, {Nakanishi}, {Ohashi}, {Peretto}, {Pyo}, {Qian},
  {Retter}, {Richer}, {Rigby}, {Savini}, {Scaife}, {Tang}, {Tomisaka}, {Wang},
  {Wang}, {Yen}, {Yuan}, {Zhang}, {Zhang}, {Zhou}, {Zhu}, {Andr{\'e}},
  {Dowell}, {Falle}, {Tsukamoto}, {Kanamori}, {Kataoka}, {Kobayashi}, {Nagata},
  {Saito}, {Seta}, {Hwang}, {Han}, {Lee}, \& {Zenko}}]{soam2018}
{Soam}, A., {Pattle}, K., {Ward-Thompson}, D., {et~al.} 2018, \apj, 861, 65

\bibitem[{{Soam} {et~al.}(2019){Soam}, {Liu}, {Andersson}, {Lee}, {Liu},
  {Juvela}, {Li}, {Goldsmith}, {Zhang}, {Koch}, {Kim}, {Qiu}, {Evans},
  {Johnstone}, {Thompson}, {Ward-Thompson}, {Di Francesco}, {Tang},
  {Montillaud}, {Kim}, {Mairs}, {Sanhueza}, {Kim}, {Berry}, {Gordon},
  {Tatematsu}, {Liu}, {Pattle}, {Eden}, {McGehee}, {Wang}, {Ristorcelli},
  {Graves}, {Alina}, {Lacaille}, {Montier}, {Park}, {Kwon}, {Chung},
  {Pelkonen}, {Micelotta}, {Saajasto}, \& {Fuller}}]{soam2019}
{Soam}, A., {Liu}, T., {Andersson}, B.~G., {et~al.} 2019, \apj, 883, 95

\bibitem[{{Sohn} {et~al.}(2007){Sohn}, {Lee}, {Park}, {Lee}, {Myers}, \&
  {Lee}}]{sohn2007}
{Sohn}, J., {Lee}, C.~W., {Park}, Y.-S., {et~al.} 2007, \apj, 664, 928

\bibitem[{{Soler}(2019)}]{soler2019}
{Soler}, J.~D. 2019, \aap, 629, A96

\bibitem[{{Soler} \& {Hennebelle}(2017)}]{soler2017}
{Soler}, J.~D., \& {Hennebelle}, P. 2017, \aap, 607, A2

\bibitem[{{Soler} {et~al.}(2013){Soler}, {Hennebelle}, {Martin},
  {Miville-Desch{\^e}nes}, {Netterfield}, \& {Fissel}}]{soler2013}
{Soler}, J.~D., {Hennebelle}, P., {Martin}, P.~G., {et~al.} 2013, \apj, 774,
  128

\bibitem[{{Steinacker} {et~al.}(2016){Steinacker}, {Bacmann}, {Henning}, \&
  {Heigl}}]{steinacker2016}
{Steinacker}, J., {Bacmann}, A., {Henning}, T., \& {Heigl}, S. 2016, \aap, 593,
  A6

\bibitem[{{Tafalla} {et~al.}(2004){Tafalla}, {Myers}, {Caselli}, \&
  {Walmsley}}]{tafalla2004}
{Tafalla}, M., {Myers}, P.~C., {Caselli}, P., \& {Walmsley}, C.~M. 2004, \aap,
  416, 191

\bibitem[{{Tang} {et~al.}(2019){Tang}, {Koch}, {Peretto}, {Novak},
  {Duarte-Cabral}, {Chapman}, {Hsieh}, \& {Yen}}]{tang2019}
{Tang}, Y.-W., {Koch}, P.~M., {Peretto}, N., {et~al.} 2019, \apj, 878, 10

\bibitem[{{Vrba}(1977)}]{vrba1977}
{Vrba}, F.~J. 1977, \aj, 82, 198

\bibitem[{{Vrba} {et~al.}(1976){Vrba}, {Strom}, \& {Strom}}]{vrba1976}
{Vrba}, F.~J., {Strom}, S.~E., \& {Strom}, K.~M. 1976, \aj, 81, 958

\bibitem[{{Walker} {et~al.}(1988){Walker}, {Lada}, {Young}, \&
  {Margulis}}]{walker1988}
{Walker}, C.~K., {Lada}, C.~J., {Young}, E.~T., \& {Margulis}, M. 1988, \apj,
  332, 335

\bibitem[{{Wang} {et~al.}(2019){Wang}, {Lai}, {Eswaraiah}, {Pattle}, {Di
  Francesco}, {Johnstone}, {Koch}, {Liu}, {Tamura}, {Furuya}, {Onaka},
  {Ward-Thompson}, {Soam}, {Kim}, {Lee}, {Lee}, {Mairs}, {Arzoumanian}, {Kim},
  {Hoang}, {Hwang}, {Liu}, {Berry}, {Bastien}, {Hasegawa}, {Kwon}, {Qiu},
  {Andr{\'e}}, {Aso}, {Byun}, {Chen}, {Chen}, {Chen}, {Ching}, {Cho}, {Choi},
  {Chrysostomou}, {Chung}, {Coud{\'e}}, {Doi}, {Dowell}, {Drabek-Maunder},
  {Duan}, {Eyres}, {Falle}, {Fanciullo}, {Fiege}, {Franzmann}, {Friberg},
  {Friesen}, {Fuller}, {Gledhill}, {Graves}, {Greaves}, {Griffin}, {Gu}, {Han},
  {Hatchell}, {Hayashi}, {Holland}, {Houde}, {Inoue}, {Inutsuka}, {Iwasaki},
  {Jeong}, {Kanamori}, {Kang}, {Kang}, {Kang}, {Kataoka}, {Kawabata}, {Kemper},
  {Kim}, {Kim}, {Kim}, {Kim}, {Kirk}, {Kobayashi}, {Konyves}, {Kwon},
  {Lacaille}, {Lee}, {Lee}, {Lee}, {Lee}, {Li}, {Li}, {Li}, {Liu}, {Liu},
  {Lyo}, {Matsumura}, {Matthews}, {Moriarty-Schieven}, {Nagata}, {Nakamura},
  {Nakanishi}, {Ohashi}, {Park}, {Parsons}, {Pascale}, {Peretto}, {Pon}, {Pyo},
  {Qian}, {Rao}, {Rawlings}, {Retter}, {Richer}, {Rigby}, {Robitaille},
  {Sadavoy}, {Saito}, {Savini}, {Scaife}, {Seta}, {Shinnaga}, {Tang},
  {Tomisaka}, {Tsukamoto}, {van Loo}, {Wang}, {Whitworth}, {Yen}, {Yoo},
  {Yuan}, {Yun}, {Zenko}, {Zhang}, {Zhang}, {Zhang}, {Zhou}, \&
  {Zhu}}]{wang2019}
{Wang}, J.-W., {Lai}, S.-P., {Eswaraiah}, C., {et~al.} 2019, \apj, 876, 42

\bibitem[{{Ward-Thompson} {et~al.}(2000){Ward-Thompson}, {Kirk}, {Crutcher},
  {Greaves}, {Holland}, \& {Andr{\'e}}}]{wardthompson2000}
{Ward-Thompson}, D., {Kirk}, J.~M., {Crutcher}, R.~M., {et~al.} 2000, \apjl,
  537, L135

\bibitem[{Ward-Thompson {et~al.}(1994)Ward-Thompson, Scott, Hills, \&
  Andr\'{e}}]{wardthompson1994}
Ward-Thompson, D., Scott, P.~F., Hills, R.~E., \& Andr\'{e}, P. 1994, \mnras,
  268, 276

\bibitem[{{Ward-Thompson} {et~al.}(2017){Ward-Thompson}, {Pattle}, {Bastien},
  {Furuya}, {Kwon}, {Lai}, {Qiu}, {Berry}, {Choi}, {Coud{\'e}}, {Di Francesco},
  {Hoang}, {Franzmann}, {Friberg}, {Graves}, {Greaves}, {Houde}, {Johnstone},
  {Kirk}, {Koch}, {Kwon}, {Lee}, {Li}, {Matthews}, {Mottram}, {Parsons}, {Pon},
  {Rao}, {Rawlings}, {Shinnaga}, {Sadavoy}, {van Loo}, {Aso}, {Byun},
  {Chakali}, {Chen}, {Chen}, {Chen}, {Ching}, {Cho}, {Chrysostomou}, {Chung},
  {Doi}, {Drabek-Maunder}, {Eyres}, {Fiege}, {Friesen}, {Fuller}, {Gledhill},
  {Griffin}, {Gu}, {Hasegawa}, {Hatchell}, {Hayashi}, {Holland}, {Inoue},
  {Inutsuka}, {Iwasaki}, {Jeong}, {Kang}, {Kang}, {Kang}, {Kawabata}, {Kemper},
  {Kim}, {Kim}, {Kim}, {Kim}, {Kim}, {Kim}, {Lacaille}, {Lee}, {Lee}, {Li},
  {Li}, {Liu}, {Liu}, {Liu}, {Liu}, {Lyo}, {Mairs}, {Matsumura},
  {Moriarty-Schieven}, {Nakamura}, {Nakanishi}, {Ohashi}, {Onaka}, {Peretto},
  {Pyo}, {Qian}, {Retter}, {Richer}, {Rigby}, {Robitaille}, {Savini}, {Scaife},
  {Soam}, {Tamura}, {Tang}, {Tomisaka}, {Wang}, {Wang}, {Whitworth}, {Yen},
  {Yoo}, {Yuan}, {Zhang}, {Zhang}, {Zhou}, {Zhu}, {Andr{\'e}}, {Dowell},
  {Falle}, \& {Tsukamoto}}]{wardthompson2017}
{Ward-Thompson}, D., {Pattle}, K., {Bastien}, P., {et~al.} 2017, \apj, 842, 66

\bibitem[{{Wenger} {et~al.}(2000){Wenger}, {Ochsenbein}, {Egret}, {Dubois},
  {Bonnarel}, {Borde}, {Genova}, {Jasniewicz}, {Lalo{\"e}}, {Lesteven}, \&
  {Monier}}]{simbad}
{Wenger}, M., {Ochsenbein}, F., {Egret}, D., {et~al.} 2000, A\&AS, 143, 9

\bibitem[{{Whittet} {et~al.}(2008){Whittet}, {Hough}, {Lazarian}, \&
  {Hoang}}]{whittet2008}
{Whittet}, D.~C.~B., {Hough}, J.~H., {Lazarian}, A., \& {Hoang}, T. 2008, \apj,
  674, 304

\bibitem[{{Wilking} {et~al.}(2008){Wilking}, {Gagn{\'e}}, \&
  {Allen}}]{wilking2008}
{Wilking}, B.~A., {Gagn{\'e}}, M., \& {Allen}, L.~E. 2008, in Handbook of Star
  Forming Regions, Volume II, ed. B.~{Reipurth} ({Astronomical Society of the
  Pacific Monograph Publications}), 351

\bibitem[{{Wootten}(1989)}]{wootten1989}
{Wootten}, A. 1989, \apj, 337, 858

\bibitem[{{Xu} {et~al.}(2020){Xu}, {Li}, {Dai}, {Goldsmith}, \&
  {Fuller}}]{xu2020}
{Xu}, X., {Li}, D., {Dai}, Y.~S., {Goldsmith}, P.~F., \& {Fuller}, G.~A. 2020,
  arXiv e-prints, arXiv:2006.04309

\bibitem[{{Ysard} {et~al.}(2013){Ysard}, {Abergel}, {Ristorcelli}, {Juvela},
  {Pagani}, {K{\"o}nyves}, {Spencer}, {White}, \& {Zavagno}}]{ysard2013}
{Ysard}, N., {Abergel}, A., {Ristorcelli}, I., {et~al.} 2013, \aap, 559, A133

\bibitem[{{Zhang} {et~al.}(2020){Zhang}, {Andre}, {Menshchikov}, \&
  {Wang}}]{zhang2020}
{Zhang}, G.~Y., {Andre}, P., {Menshchikov}, A., \& {Wang}, K. 2020, arXiv
  e-prints, arXiv:2002.05984

\end{thebibliography}
\end{document}